\documentclass[a4paper,11pt]{article}
\usepackage[a4paper]{geometry}
\usepackage{amssymb,amsmath,graphicx,epsfig,amscd,array,color}
\usepackage{amsthm}
\usepackage[round]{natbib}
\usepackage{mathrsfs}
\usepackage{enumerate}
\usepackage{authblk} 
\usepackage{subfigure}    % use subfigures

\DeclareMathOperator{\Cov} {Cov}

\title{Multivariate Gaussian Random Fields with Oscillating Covariance Functions using Systems of Stochastic Partial Differential Equations}
\author[1]{Xiangping Hu\footnote{Corresponding author. Email: \texttt{Xiangping.Hu@math.ntnu.no}}}
\author[2]{Finn Lindgren}
\author[1]{Daniel Simpson}
\author[1]{H\aa{}vard Rue}
\affil[1]{Department of Mathematical Sciences, Norwegian University of Science and Technology, N-7491 
Trondheim, Norway}
\affil[2]{Department of Mathematical Sciences, University of Bath, BA2 7AY, United Kingdom}

\date{July 4, 2013}
\begin{document}

\maketitle

\begin{abstract}
In this paper we propose a new approach for constructing \emph{multivariate} Gaussian random fields (GRFs) with oscillating covariance functions through systems of stochastic partial differential equations (SPDEs).
We discuss how to build systems of SPDEs that introduces oscillation characteristics in the covariance functions of the multivariate GRFs. By choosing different parametrization of 
the equations, some GRFs can be made with oscillating covariance functions but other fields can have Mat\'ern covariance functions or close to Mat\'ern covariance functions. The multivariate GRFs constructed 
by solving the systems of SPDEs automatically fulfill the hard requirement of nonnegative definiteness for the covariance functions.
The approximate weak solutions to the systems of SPDEs are used to represent the multivariate GRFs by multivariate Gaussian \emph{Markov} random fields (GMRFs). 
Since the multivariate GMRFs have sparse precision matrices (inverse of the covariance matrices), numerical algorithms for sparse matrices can be applied to the precision matrices for sampling and inference.
Thus from a computational point of view, 
the \emph{big-n} problem can be partially solved with these types of models.  Another advantage of the method is that the oscillation  in the covariance function can be controlled directly by the parameters 
in the system of SPDEs. We show how to use this proposed approach with simulated data and real data examples.

\end{abstract}

\section{Introduction} \label{sec: osci_introduction}
Statistics for spatial data appeared from hundreds of years ago, but spatial models for this type of data appeared much later \citep{cressie1993statistics}. The spatial models have been widely used to
model the spatial data in many areas. Gaussian random fields (GRFs) are some of the most commonly used models in spatial statistics. Since the normalizing constant can be computed explicitly, the GRFs are convenient to be used in many
applications, such as geo-statistical data, environmental and atmospheric data, longitudinal and survival data \citep{cressie1993statistics,stein1999interpolation,rue2005gaussian,gelfand2010handbook}.
GRFs also have other good properties, such as the fact that a GRF can be explicitly specified through a mean  $\boldsymbol{\mu}(\cdot)$ and a covariance function $\boldsymbol{C}(\cdot,\cdot)$. 
Let the coordinates members of $\mathbb{R}^d$. If $(x(\boldsymbol{s}_1), x(\boldsymbol{s}_2), \dots, x(\boldsymbol{s}_n))$ is Gaussian for every selection of points $(\boldsymbol{s}_1^T, \boldsymbol{s}_2^T, \dots, \boldsymbol{s}_n^T)$ for every $n \geq 1$, 
then we call $x(\boldsymbol{s})$ a continuously indexed GRF. The covariance matrix of a collection $(x(\boldsymbol{s}_1), x(\boldsymbol{s}_2), \dots, x(\boldsymbol{s}_n))$ is given by $\boldsymbol{\Sigma} = [C(\boldsymbol{s}_i, \boldsymbol{s}_j)]$.
However, there is a hard nonnegative definite requirement that must be fulfilled for the function $\boldsymbol{C}(\cdot, \cdot)$. This is one of the main
concerns when we build models with GRFs using covariance function based approaches.

A widely used class of covariance functions is the Mat\'ern family which was introduced by Mat\'ern \citep{matern1986spatial}. This family of covariance functions captures the most common form of the empirical behavior of 
stationary covariance functions, namely that the correlation between the locations $\boldsymbol{m}$ and $\boldsymbol{n}$ should decrease when the Euclidean distance $\| \boldsymbol{h} \| = \| \boldsymbol{m} - \boldsymbol{n} \|$ increases 
\citep{diggle2006model}. This family of covariance functions is isotropic and usually written as 
$\sigma^2 M(\boldsymbol{m,n}|\nu,\kappa)$, where $M(\boldsymbol{m,n}|\nu, \kappa)$ is the Mat\'ern correlation function between the spatial locations $\boldsymbol{m}, \boldsymbol{n} \in \mathbb{R}^d$. 
The Mat\'ern correlation function is a two-parameter family with the form 
\begin{equation} \label{eq: osci_matern_cocariance}
M(\boldsymbol{h} |\nu, \kappa) = \frac{2^{1-\nu}}{\Gamma(\nu)}(\kappa \| \boldsymbol{h} \|)^\nu K_\nu(\kappa \| \boldsymbol{h} \|),
\end{equation}
where $\sigma^2$ is the marginal variance, $K_\nu$ is the modified Bessel function of second kind and $\kappa > 0$ is the scaling parameter. 
The order $\nu $ is a smoothness parameter and must be positive. The smoothness parameter $\nu$ is of critical concern in spatial
statistics since it defines the differentiability of the sample paths and the Hausdorff dimension \citep{goff1988stochastic,handcock1993bayesian, gneitingmatern}. It is known that the smoothness parameter $\nu$ is poorly identifiable from 
the data and hence it is usually fixed \citep{diggle2006model,lindgren2011explicit,hu2012multivariate}. The Mat\'ern family contains the commonly used models with exponential covariance function 
$M(\boldsymbol{h} |\frac{1}{2}, \kappa) = \exp(-\kappa \| \boldsymbol{h} \|)$. 
The Mat\'ern covariance function is the key factor in the explicit link between the GRFs and GMRFs through the stochastic partial differential equations (SPDEs) discussed by \citet{lindgren2011explicit}. 
\citet{gneitingmatern} presented one direct approach for constructing the multivariate GRFs by using matrix-valued covariance functions, and all components in the matrix-valued covariance function are Mat\'ern covariance functions.
\citet{hu2012multivariate} discussed the important role of Mat\'ern covariance function for constructing stationary and isotropic multivariate GRFs with systems of SPDEs.

In spatial statistics oscillating models usually deal with ocean waves, and we usually work entirely with 
their spectra only rarely with covariance functions \citep{georg2010secondstationary}.
For time series there are plenty of applications with oscillating models in discrete time since the $p$-order auto-regressive (AR($p$)) processes can result in oscillating 
models \citep{wei2006time}. However, it is less common for continuous time models. 
\citet{lindgren2011explicit} have discussed an approach for constructing univariate GRFs with oscillating covariance functions.
In their approach they have chosen a coupled system of SPDEs to construct two independent random fields with the same precision matrix. Their discussion was focused on the case $\alpha = 2$.
Since it is important for our approach for constructing multivariate GRFs with oscillating covariance functions, we give an overview of their  approach in Section \ref{sec: osci_univariate_oscillation}.

In this paper we focus on the methodology for constructing multivariate GRFs with oscillating covariance functions through systems of SPDEs. 
This work is an extension of the discussion given by \citet{lindgren2011explicit} for the univariate case.
It is also an extension of the approach discussed by \citet{hu2012multivariate} to construct larger class of useful models.
In our approach the GRFs are constructed by solving systems of stochastic partial differential equations.
One of the main advantages of this approach is that we do not need to consider the notorious nonnegative definite requirement for the covariance functions. This requirement is fulfilled automatically because we are working on the processes directly. 
During the computational stage the GRFs are represented by the Gaussian \emph{Markov} random fields (GMRFs) by GMRF approximations.
A GMRF $x(\boldsymbol{s})$ is a discretely indexed GRF with some Markov property. The full conditionals $\{ \pi(x_i|\boldsymbol{x}_{-i}); i = 1, \dots, n \}$ of a GMRF depend only on a set of neighbors of each site $i$. Denote the neighbors of the 
node $i$ by $\partial i$. The Markov property implies that $Q_{ij} \neq  0$ if and only if $j \in \partial i \cup i$, where $\boldsymbol{Q}$ denotes the precision matrix of the GMRF.
Consistency requirement implies that if $i \in \partial j$, then $j \in \partial i$. 
The precision matrix $\boldsymbol{Q}$ for the GMRF is sparse which enables us to use numerical algorithms for sparse matrices.
Thus the \emph{big-n} problem \citep{banerjee2004hierarchical} can be partially solved in our case with these types of models. We refer to \citet[Section 2.1]{rue2009approximate} for a condensed overview of the theory of GMRFs.
Detailed discussions about GMRFs are given in \citet[Chapter 2]{rue2005gaussian}.

\section{State-of-the-art and preliminaries}
We review the state-of-the-art research on GRFs with SPDE approach in Section \ref{sec: osci_SPDEsunivariate} and the methodologies for constructing multivariate GRFs in Section \ref{sec: osci_multivariateGRFs}. 
The GMRF approximation for representing a GRF with a GMRF is introduced in Section \ref{sec: osci_GMRFs2GRFs} since it is crucial for computations. 

\subsection{GRFs through the SPDE approach} \label{sec: osci_SPDEsunivariate}
As mentioned in Section \ref{sec: osci_introduction}, \citet{lindgren2011explicit} proposed a novel approach for constructing GRFs by using the SPDE
\begin{equation} \label{eq: osci_spde_simple}
 (\kappa^2 - \Delta)^{\alpha/2} {x}(\boldsymbol{s}) = \mathcal{W}(\boldsymbol{s}),
\end{equation}
where $(\kappa^2 - \Delta)^{\alpha/2}$ is a pseudo-differential operator, $\kappa$ is the scaling parameter, $\alpha$ is related to the smoothness parameters $\nu > 0 $ and $\alpha = \nu + d/2$. 
$\Delta$ is the Laplacian with definition
\begin{displaymath}
 \Delta = \sum_{i=1}^{d}{\frac{\partial^2}{\partial x_i^2}}.
\end{displaymath}
A range parameter $\rho$ connects the scaling parameter $\kappa$ and the smoothness parameter $\nu$. The simple and empirically derived relationship 
$\rho = \sqrt{8 \nu}/\kappa$ is commonly used \citep{lindgren2011explicit,hu2012multivariate}. It corresponds to correlation near $0.1$ at distance $\rho$, with 
parameters $\kappa$ and $\nu$. $\mathcal{W}(\boldsymbol{s})$ is the innovation process which is a standard spatial Gaussian white noise. 
The most important result given by \citet{whittle1954stationary, whittle1963stochastic}, and used by 
 \citet{lindgren2011explicit} extensively, is that the solution $x(\boldsymbol{s})$ to SPDE \eqref{eq: osci_spde_simple} is a GRF with the Mat\'ern covariance function given in Equation \eqref{eq: osci_matern_cocariance}.
We follow the terminology used by \citet{lindgren2011explicit} and call GRFs with Mat\'ern covariance functions Mat\'ern random fields. \citet{lindgren2011explicit} commented that
their approach can be extended in many directions. 

\citet{fuglstad2011spatial} has extended the SPDE approach to include a diffusion matrix in Equation \eqref{eq: osci_spde_simple}
which provides a way of controlling the covariance structures of the GRF. Using the diffusion matrix $\boldsymbol{H}$ in SPDE \eqref{eq: osci_spde_simple}, it is not only possible to construct homogeneous isotropic fields,
but also anisotropic fields. \citet{fuglstad2011spatial} showed that it is possible to construct inhomogeneous fields. The SPDE discussed by \citet{fuglstad2011spatial} has the form
\begin{equation} \label{eq: osci_spde_diffusion}
 \kappa^2 x(\boldsymbol{s}) - \nabla \cdot \boldsymbol{H}(\boldsymbol{s}) \nabla x(\boldsymbol{s}) = \mathcal{W}(\boldsymbol{s}),
\end{equation}
where $\boldsymbol{H}$ is a $2 \times 2$ matrix-valued function, $\nabla$ is the gradient operator and $\mathcal{W}(\boldsymbol{s})$ is a standard spatial Gaussian white noise process. The main contribution of 
\citet{fuglstad2011spatial} is that he introduced the matrix $\boldsymbol{H}$ to \eqref{eq: osci_spde_simple} to control the structure of covariance matrix. However, he focused on discussing the univariate GRFs with $\alpha = 2$.

\citet{bolin2011spatial} discussed how to use nested SPDEs to construct stationary and non-stationary GRFs.
The equation chosen by \citet{bolin2011spatial} has the form
\begin{equation} \label{eq: osci_nestedspde}
 \mathcal{L}_1 x(\boldsymbol{s}) = \mathcal{L}_2 \mathcal{W}(\boldsymbol{s})
\end{equation}
for some linear differential operators $\mathcal{L}_1$ and $\mathcal{L}_2$. $\mathcal{W}(\boldsymbol{s})$ is a standard spatial Gaussian white noise process. The SPDE \eqref{eq: osci_nestedspde} may not exist in the common sense since the operator 
$\mathcal{L}_2$ may contain some differentiation of the noise process $\mathcal{W}(\boldsymbol{s})$. In this case SPDE \eqref{eq: osci_nestedspde} can then be interpreted as the following nested system of SPDEs
\begin{equation}
 \begin{split}
  \mathcal{L}_1 x_0(\boldsymbol{s}) = \mathcal{W}(\boldsymbol{s}),  \\
  x(\boldsymbol{s}) = \mathcal{L}_2 x_0(\boldsymbol{s}),
 \end{split}
\end{equation}
when $\mathcal{L}_1$ and $\mathcal{L}_2$ are commutative operators.
As pointed out by \citet{bolin2011spatial} this interpretation not only avoids the apparent problem with the differentiation of the Gaussian white noise process, but also gives an interpretation of the consequence of the additional 
differential operator $\mathcal{L}_2$. The random field $x(\boldsymbol{s})$ is obtained by applying $\mathcal{L}_2$ to the solution of \eqref{eq: osci_nestedspde} with $\mathcal{L}_2 = \boldsymbol{I}$.

\citet{fuglstad2010approximating} discussed the approximated solutions to a SPDE for constructing space-time GMRFs. The equation discussed by him has the form
\begin{equation} \label{eq: osci_approximating}
 \frac{\partial}{\partial t} x(s,t) - \nabla \cdot \nabla x(s,t) = \tau \mathcal{W}(s,t), \hspace{10mm} (s,t) \in [0,L] \times [0,T],
\end{equation}
where, $\nabla = \frac{\partial}{\partial x}$, $\tau > 0$ is a constant and $\mathcal{W}(s,t)$ is Gaussian space-time white noise. \citet{fuglstad2010approximating} only discussed $s \in \mathbb{R}$ and
argued that this equation has real physical meaning since the prototype of \eqref{eq: osci_approximating} is the stochastic heat equation
\begin{equation} \label{eq: osci_heatequation}
 \frac{\partial}{\partial x} q(s,t) + \nabla F(s,t) = f(s,t).
\end{equation}
The heat equation relates the change of the size $q$ in time to the spatial divergences of the flux $F(s,t)$  and the source term $f(s,t)$. \citet{fuglstad2010approximating} chose a finite volume method for solving
Equation \eqref{eq: osci_approximating}, and claimed that the finite volume method gave the correct distribution of the total energy of the solution to SPDE \eqref{eq: osci_approximating}.

\citet{lindgren2011explicit} discussed the methodology for constructing non-stationary GRFs and non-separable space-time models. 
They claimed that if the parameters $\kappa$ and $\tau$ depend on the coordinate $\boldsymbol{s}$, then we can construct a non-stationary GRF.
 The SPDE then has the form
\begin{equation}
 (\kappa^2 (\boldsymbol{s}) - \Delta)^{\alpha/2} \{ \tau({\boldsymbol{s})} x(\boldsymbol{s}) \} = \mathcal{W}(\boldsymbol{s}).
\end{equation}
The non-separable space-time models have interaction between space and time in the covariance structure. In general it is difficult to construct this kind of model through a covariance function based approach. However, the SPDE approach can
be used. One of the  non-separable SPDEs which can result in this kind of model is
\begin{equation}
 \left\{ \frac{\partial}{\partial t} + (\kappa^2 + \boldsymbol{m} \cdot \nabla -\nabla \cdot \boldsymbol{H} \nabla) \right\} x(\boldsymbol{s}, t) = \varepsilon (\boldsymbol{s}, t),
\end{equation}
where $\boldsymbol{m}$ is a transport vector, $\boldsymbol{H}$ is a positive definite diffusion matrix and $\varepsilon (\boldsymbol{s}, t)$ is a stochastic space-time noise process.
We refer to \citet[Section $3.5$]{lindgren2011explicit} for detailed discussion on this topic.

\subsection{Multivariate GRFs} \label{sec: osci_multivariateGRFs}
A multivariate GRF with $p$ components 
$\boldsymbol{x}(\boldsymbol{s}) = ({x}_1(\boldsymbol{s}), {x}_2(\boldsymbol{s}), \cdots, {x}_p(\boldsymbol{s}) )^T$, $\boldsymbol{s} \in \mathbb{R}^d$, is a collection of continuously indexed multivariate normal random vectors such that 
\begin{displaymath}
 \boldsymbol{x}(\boldsymbol{s}) \sim \mbox{MVN}(\boldsymbol{\mu}, \boldsymbol{\Sigma}),
\end{displaymath}
where $\boldsymbol{\mu}$ is the mean of the random field and $\boldsymbol{\Sigma}$ is the covariance matrix.
Assume, at the current stage, that the process is second-order stationary with mean zero.
One approach for constructing stationary and isotropic multivariate GRFs using covariance-based models was proposed by \citet{gneitingmatern}. 
The covariance function $\boldsymbol{C}$ in their approach is given by
\begin{align} \label{eq: CovarianceMatrix_M}
 \ \boldsymbol{C}(\boldsymbol{h})  = \left( \begin{array}{cccc}
  C_{11}(\boldsymbol{h}) & C_{12}(\boldsymbol{h}) & \cdots & C_{1p}(\boldsymbol{h}) \\
  C_{21}(\boldsymbol{h}) & C_{22}(\boldsymbol{h}) & \cdots & C_{2p}(\boldsymbol{h}) \\
  \vdots & \vdots & \ddots &  \vdots \\
  C_{p1}(\boldsymbol{h}) & C_{p2}(\boldsymbol{h}) & \cdots & C_{pp}(\boldsymbol{h}) \\
         \end{array} \right),
\end{align}
where $\{ C_{ii}(\boldsymbol{h}) = \sigma_{ii} M(\boldsymbol{h} |\nu_{ii}, \kappa_{ii}); i = 1, \dots, p \} $ are the marginal covariance functions
and $\{ C_{ij}(\boldsymbol{h}) = \rho_{ij} \sigma_{i} \sigma_{j} M(\boldsymbol{h} |\nu_{ij}, \kappa_{ij}); i, j = 1, \dots, p, i \neq j \}$ are the cross-covariance functions.
$\{ C_{ii}(\boldsymbol{h}) = \mathbb{E}(x_i(\boldsymbol{s}+\boldsymbol{h}) x_i(\boldsymbol{s})); i = 1, 2, \dots, p \}$ give information about the covariance structures within the fields $\{x_i (\boldsymbol{s})\}$.
$\{ C_{ij}(\boldsymbol{h}) = \mathbb{E}(x_i(\boldsymbol{s}+\boldsymbol{h}) x_j(\boldsymbol{s})); i, j = 1, 2, \dots, p, i \neq j \}$ describes the covariance structure between fields $\{x_i(\boldsymbol{s})\}$ and $\{x_j(\boldsymbol{s})\}$. 
$\{ \rho_{ij} \}$ are the co-located correlation coefficients. $\{\sigma_{ii} \geq 0\}$ are the marginal variances, and $\{\sigma_{i}\}$ and $\{\sigma_{j}\}$ are the corresponding standard deviations. 
They satisfy the relationships throughout this paper $\sigma_{ii} = \sigma_i^2 , \sigma_{ij} = \sigma_i \sigma_j$.
The main difficulty in constructing useful multivariate models using this kind of approach is the nonnegative definiteness requirement for the covariance functions. 
\citet{gneitingmatern} proposed a way to  specify valid parametric models through the covariance functions given in Equation \eqref{eq: CovarianceMatrix_M} directly. 
Several theorems were presented in order to ensure the matrix-valued covariance function to be symmetric and nonnegative definite. 

\citet{hu2012multivariate} proposed to use a system of SPDEs to construct a multivariate GRF. They claimed that the notorious requirement nonnegative definiteness for the covariance function is 
automatically fulfilled with their approach. \citet{hu2012multivariate} also discussed the link between the system of SPDEs approach and the covariance function based approach discussed by \citet{gneitingmatern}.
The system of SPDEs which has been used for constructing the multivariate GRFs by \citet{hu2012multivariate} is
\begin{align} \label{eq: osci_SPDEs_system}
\begin{pmatrix}
\mathcal{L}_{11} & \mathcal{L}_{12} & \ldots & \mathcal{L}_{1p}\\
\mathcal{L}_{21} & \mathcal{L}_{22} & \ldots & \mathcal{L}_{2p}\\
\vdots & \vdots & \ddots&\vdots\\
\mathcal{L}_{p1} & \mathcal{L}_{p2} &\ldots & \mathcal{L}_{pp}
\end{pmatrix}
\begin{pmatrix}
{x}_1(\boldsymbol{s}) \\ {x}_2(\boldsymbol{s})\\ \vdots \\ {x}_p(\boldsymbol{s})
\end{pmatrix}
=
\begin{pmatrix}
\varepsilon_1(\boldsymbol{s}) \\ \varepsilon_2(\boldsymbol{s}) \\ \vdots \\ \varepsilon_p(\boldsymbol{s})
\end{pmatrix},
\end{align}
where $\{ \mathcal{L}_{ij} = b_{ij}(\kappa_{ij}^2 - \Delta)^{\alpha_{ij}/2}; i = j = 1, 2,\dots, p \}$ are differential operators with $\{\alpha_{ij} = 0 \text{ or } 2\}$,
$\{ \kappa_{ij} \}$ and $\{\nu_{ij} \}$ are scaling parameters and smoothness parameters.
$\{ \varepsilon_i(\boldsymbol{s}); i = 1,2, \dots, p \}$ are independent but not necessarily identically distributed noise processes, and
$\{ b_{ij} \}$ are the parameters related to the marginal variances of the fields and the cross-covariances between the fields. 
\citet{hu2012multivariate} pointed out that the GMRF approximation can be applied to the GRFs. Hence they can use computationally efficient GMRFs for sampling and inference. 
However, the constructed GRFs are always stationary and isotropic, and the covariance functions are not oscillating. 

\subsection{GMRFs approximation to GRFs} \label{sec: osci_GMRFs2GRFs}
Generally speaking, GRFs are commonly used in statistical modelling because of their good theoretical properties. However, the GRFs have a bottle-neck on the computational side. The computational cost for factorizing a dense covariance
matrix $\boldsymbol{\Sigma}$ with dimension $n \times n$ is $\mathcal{O}(n^3)$. Even though the computational power is at an all time high, 
it seems that in many situations it is infeasible to do the computations in reasonable time. \citet[Appendix A.5]{banerjee2004hierarchical} informally call this situation ``the big $n$ problem''. 

There are many different approaches trying to avoid or overcome ``the big $n$ problem'', such as covariance tapering \citep{furrer2006covariance, zhang2008covariance, kaufman2008covariance, shaby2012tapered}, 
likelihood approximations \citep{vecchia1988estimation, stein2004approximating}, and fixed rank kriging and fixed rank filtering \citep{cressie2008fixed,cressie2010fixed} .
The approach which has been chosen in this paper is based on the GMRF approximation to GRFs.
The sparsity of the precision matrix $\boldsymbol{Q}$ enables the numerical algorithms for sparse matrix for fast inference with large datasets \citep{rue2001fast,rue2005gaussian,lindgren2011explicit}.
The general cost for factorizing the sparse matrix $\boldsymbol{Q}$ is $\mathcal{O} (n)$, $\mathcal{O} (n^{3/2})$ and $\mathcal{O} (n^2)$ in one dimension, two dimensions and three dimensions, respectively \citep{rue2005gaussian}. 
\citet{hartman2008fast} proposed to use the GMRFs for GRFs for spatial prediction with Kriging, due to the pleasant computational properties of GMRFs. 

In this paper we only give an overview of GMRF approximation to univariate GRF and refer to \citet[Section $2.3$]{hu2012multivariate} for detailed discussion on GMRF approximation to multivariate GRF.
In order to find a GMRF approximation of a GRF on a triangulated lattice, we at first need to find the stochastic weak formulation of SPDE \eqref{eq: osci_spde_simple} \citep{kloeden1992numerical}. 
In this paper we use Delaunay triangulation. We refer to \citet{hjelle2006triangulations} for more information about triangulations. 
Denote the inner product of functions $h$ and $g$ as
\begin{equation} \label{eq: osci_inner}
 \langle h, g \rangle = \int h(\boldsymbol{s})g(\boldsymbol{s})d(\boldsymbol{s}),
\end{equation}
where the integration is within the region of interest. The stochastic weak solution of SPDE \eqref{eq: osci_spde_simple} is found by requiring 
\begin{equation} \label{eq: soci_fem_spdesimple}
\left\{ \langle \phi_i, (\kappa^2 - \Delta)^{\alpha/2} x \rangle; i = 1, \dots, M \right\} \stackrel{d}{=} \left\{ \langle \phi_i, \mathcal{W} \rangle; i = 1, \dots, M \right\},
\end{equation}
where $M$ is the number of test functions $\{ \phi_i (\boldsymbol{s}) \}$ and  ``$\stackrel{d}{=}$'' denotes equality in distribution.

Then we need to find the finite element representation of the solution to the SPDE. 
The finite element representation of the solution is
\begin{equation} \label{eq: osci_spde_representing}
x(\boldsymbol{s}) = \sum_{i = 1}^{N} {\psi_i(\boldsymbol{s})\omega_i} 
\end{equation}
with basis functions $\{ \psi_i(\boldsymbol{s}); i = 1,2, \dots, N \}$ and Gaussian distributed weights $\{ \omega_i;  i = 1, 2, \dots, N\}$. $N$ is the number of vertexes in the triangulation. 
We refer to \citet{zienkiewicz2005finite} and \citet{brenner2008mathematical} for more information and theoretical background of finite element methods.
The approach given by \citet{lindgren2011explicit} for choosing the basis functions is used in this paper.
With $M = N$ we choose each basis function $\psi_i(\boldsymbol{s})$ to be piecewise linear on each triangle with $\psi_i (\boldsymbol{s}) = 1$ at vertex $i$ and $\psi_i(\boldsymbol{s}) = 0$
at other vertexes. This choice of basis functions means that the local interpolation on a triangle is linear.
\citet{lindgren2011explicit} pointed out that other methods, such as kernel method, are useful in theory but not necessary in practice.
When  $\alpha_{ij} = 1$ the \emph{least squares} approximation is chosen, $\phi_k = (\kappa^2 - \Delta)^{\frac{1}{2}} \psi_k$.
When $\alpha_{ij} = 2$ the \emph{Galerkin} solution is chosen, $\phi_k = \psi_k$.
When  $\alpha_{ij} \geq 3$ the \emph{recursive Galerkin} formulation is used. 
We refer to \citet[Section $2.3$]{lindgren2011explicit} for more information about the recursive Galerkin formation.

\subsection{Outline of the paper}
The structure of the rest of the paper is organized as follows. Section \ref{sec: osci_Model_constr} gives the detailed discussion on how to construct multivariate GRFs 
with oscillating covariance functions through the systems of SPDEs approach. Examples with simulated data and real data are given in Section \ref{sec: osci_exampl_applications}. 
Discussion and future work in Section \ref{sec: osci_discussion} ends this paper.

\section{Model formulation} \label{sec: osci_Model_constr}
GRFs with oscillating covariance functions can be used in many situations, for example, for modelling global pressure \citep{lindgren2011explicit} and ocean waves \citep{georg2010secondstationary}. First, an overview 
for constructing the \emph{univariate} GRFs with oscillating covariance function is given since it is needed for constructing \emph{multivariate} GRFs with oscillating covariance functions. Next, we introduce a 
general approach for constructing the multivariate GRFs with oscillating covariance functions. Then explicit approach for constructing the \emph{bivariate} GRFs is discussed. At last we discuss the  procedure for sampling the 
multivariate GRFs with oscillating covariance functions. 

\subsection{Univariate GRFs with oscillating covariance functions} \label{sec: osci_univariate_oscillation}
\citet[Section $3.3$]{lindgren2011explicit} discussed how to construct an univariate GRF with oscillating covariance function using a SPDE with complex number. For the case $\alpha = 2$ the SPDE has the form
\begin{equation} \label{eq: osci_univariate_complex}
 \left\{ \kappa^2 \exp(i \pi\omega) - \Delta \right\} x(\boldsymbol{s})  = \mathcal{W} (\boldsymbol{s}) ,
\end{equation}
where $\omega \in [0,1]$ is the oscillation parameter, $x(\boldsymbol{s})  = x_1(\boldsymbol{s}) + i x_2(\boldsymbol{s})$, and $\mathcal{W} (\boldsymbol{s}) = \mathcal{W}_1 (\boldsymbol{s}) + i \mathcal{W}_2 (\boldsymbol{s})$.
The innovation processes $\mathcal{W}_1(\boldsymbol{s})$ and $\mathcal{W}_2(\boldsymbol{s})$ are independent standard Gaussian white noise processes. 
\citet{lindgren2011explicit} pointed out that the real and imaginary parts, $x_1(\boldsymbol{s})$ and $x_2(\boldsymbol{s})$, of the stationary solution $x(\boldsymbol{s})$ are independent with identical spectrum densities
\begin{equation} \label{osci_spectrum_univariate}
 R(\boldsymbol{k}) = \frac{1}{(2\pi)^{d}\left( \kappa^4 +2 \cos(\pi \omega) \kappa^2  \| \boldsymbol{k} \|^2 + \| \boldsymbol{k} \|^4 \right)}, \hspace{2mm} \boldsymbol{k} \in \mathbb{R}^d.
\end{equation}
With this approach the common isotropic stationary Mat\'ern random fields can be obtained by setting $\omega = 0$.  We can notice that $\omega = 1$ generates intrinsic stationary random fields. 
We refer to \citet[Chapter 3]{rue2005gaussian} for more information on the intrinsic random fields. When $\omega \in (0,1)$, the constructed GRFs have covariance functions with oscillation. 
The oscillation is increasing with larger value of $\omega$. The closed form of the precision matrix $\boldsymbol{Q}$ for the stationary GRFs with oscillation can be obtained from \eqref{osci_spectrum_univariate},
\begin{equation} \label{osci_covariance_univariate}
 \boldsymbol{Q}(\kappa^2, \omega) = \kappa^4 \boldsymbol{C} +2 \cos(\pi \omega) \kappa^2 \boldsymbol{G} + \boldsymbol{G} \boldsymbol{C}^{-1} \boldsymbol{G}.
\end{equation}
The matrices $\boldsymbol{C}$ and $\boldsymbol{G}$ in Equation \eqref{osci_covariance_univariate} are defined through
\begin{equation}
 \begin{split}
  C_{ij} = \langle \psi_i, 1 \rangle , \\
  G_{ij} = \langle \nabla \psi_i, \nabla \psi_j \rangle ,
 \end{split}
\end{equation}
with basis functions $\{ \psi_i; i = 1, 2, \dots, n \}$. 
We use $C_{ij} = \langle \psi_i, 1 \rangle$, instead of $C_{ij} = \langle \psi_i, \psi_j \rangle,$ in order to make the precision matrix sparse.
This setting yields a Markov approximation to the FEM solution. \citet{bolin2009wavelet} studied the effects of the Markov approximation and claimed that
the difference between the Markov approximation and exact FEM representation is negligible.

\citet{lindgren2011explicit} pointed out that this complex-valued version of SPDE \eqref{eq: osci_univariate_complex} can be rewritten as a special case 
of the coupled systems of SPDEs
\begin{align} \label{eq: osci_coupledSPDE}
\begin{pmatrix}
h_{1} - \Delta& -h_{2} \\
h_{2}         & h_{1} - \Delta  \\
\end{pmatrix}
\begin{pmatrix}
{x}_1(\boldsymbol{s}) \\ {x}_2(\boldsymbol{s}) 
\end{pmatrix}
=
\begin{pmatrix}
\mathcal{W}_1(\boldsymbol{s}) \\ \mathcal{W}_2(\boldsymbol{s})
\end{pmatrix},
\end{align}
where $h_1 = \kappa^2 \cos(\pi\omega)$ and $h_2 = \kappa^2 \sin(\pi\omega)$. 
The random fields $x_1(\boldsymbol{s})$ and $x_2(\boldsymbol{s})$ from Equation \eqref{eq: osci_coupledSPDE} have the same precision matrix $\boldsymbol{Q}$ given in Equation \eqref{osci_covariance_univariate}. 
\citet{lindgren2011explicit} commented that it is surprising that these two fields from the coupled system of SPDEs \eqref{eq: osci_coupledSPDE} are always independent regardless of 
the choices of parameters. Additionally, the univariate GRFs with oscillating covariance functions from Equation \eqref{eq: osci_univariate_complex} are always isotropic. However, it is possible to construct non-isotropic GRFs by slightly 
modifying the coupled system of SPDEs \eqref{eq: osci_coupledSPDE}. We are not going to discuss this issue here and we focus only on the isotropic GRFs. We refer to \citet[Appendix C.4]{lindgren2011explicit} for 
more information about the oscillating and non-isotropic cases.

\subsection{Multivariate GRFs with oscillating covariance functions} \label{sec: osci_multiGRFs}
The multivariate GRFs with oscillating covariance functions, in this paper, all have the assumption that the mean is zero, i.e.,  $\boldsymbol{x}(\boldsymbol{s}) \sim \text{MVN} (\boldsymbol{0}, \boldsymbol{\Sigma})$.
\citet{hu2012multivariate} proposed to construct the multivariate GRFs using the system of SPDEs given in \eqref{eq: osci_SPDEs_system}. In their approach the multivariate GRFs are always isotropic and stationary.
The covariance functions cannot be oscillating. They argued that, under some conditions, it is possible to construct the multivariate GRFs with Mat\'ern covariance functions as discussed by \citet{gneitingmatern}.
In this section we are going to discuss how to construct multivariate GRFs where some components of the random fields have oscillating covariance functions. 
The main idea is to replace the noise processes by noise processes with oscillating covariance functions.  With this approach the systems of SPDEs have the same
form as given in \eqref{eq: osci_SPDEs_system}, but the noise processes are different. Even though the system of SPDEs in \eqref{eq: osci_SPDEs_system} is theoretically general, we recommend to 
use the triangular system of SPDEs in applications. The triangular system of SPDEs is 

\begin{align} \label{eq: osci_SPDEs_system_triangular}
\begin{pmatrix}
\mathcal{L}_{11} &  &  &  \\
\mathcal{L}_{21} & \mathcal{L}_{22} &  & \\
\vdots & \vdots & \ddots & \\
\mathcal{L}_{p1} & \mathcal{L}_{p2} &\ldots & \mathcal{L}_{pp}
\end{pmatrix}
\begin{pmatrix}
{x}_1(\boldsymbol{s}) \\ {x}_2(\boldsymbol{s})\\ \vdots \\ {x}_p(\boldsymbol{s})
\end{pmatrix}
=
\begin{pmatrix}
\varepsilon_1(\boldsymbol{s}) \\ \varepsilon_2(\boldsymbol{s}) \\ \vdots \\ \varepsilon_p(\boldsymbol{s})
\end{pmatrix},
\end{align}
where $\{ \mathcal{L}_{ij}; i, j = 1,2, \dots, p, i \geq j \}$ are differential operators as defined in \eqref{eq: osci_SPDEs_system}, $\{\mathcal{L}_{ij} = 0; i, j = 1,2, \dots, p, i < j\} $ and 
$\{\varepsilon_i(\boldsymbol{s}); i = 1,2, \dots, p\}$ are noise processes where some of them have oscillating covariance functions. We recommend to use as fewer noise processes with oscillating 
covariance functions as possible.
This system has many advantages, such as interpretation of the properties of the fields. For example, we know which components of the random field must have
non-oscillating covariance functions and have oscillating covariance functions. However, there are some components of the random field which might have oscillating covariance functions.
We divide the random fields into three categories $\boldsymbol{x}_\text{m}$, $\boldsymbol{x}_\text{o}$ and $\boldsymbol{x}_\text{p}$, where $\boldsymbol{x}_\text{m}$ denotes the random fields with non-oscillating covariance functions,  
$\boldsymbol{x}_\text{o}$ denotes the random fields with oscillating covariance functions and $\boldsymbol{x}_\text{p}$ denotes the random fields with covariance functions which might be oscillating.
Assume that only $\{\varepsilon_i(\boldsymbol{s}); i = 1,2, \dots, p\}$ is the noise process with oscillating covariance function and other noise processes $\{\varepsilon_j(\boldsymbol{s}); j = 1,2, \dots, p, j \neq i\}$ are noise processes
with non-oscillating covariance functions, and then we can obtain the following results.

\begin{itemize}
 \item If only the covariance function for the noise process $\varepsilon_i(\boldsymbol{s})$ is oscillating, the covariances functions for all the random fields $x_j(\boldsymbol{s})  (j < i)$ are non-oscillating,
       $x_j(\boldsymbol{s}) \in \boldsymbol{x}_\text{n}(\boldsymbol{s})$;
 \item If only the covariance function for the noise process $\varepsilon_i(\boldsymbol{s})$ is oscillating, the random field $x_j(\boldsymbol{s}) (j = i)$ has an oscillating covariance function, 
       $x_j(\boldsymbol{s}) \in \boldsymbol{x}_\text{o}(\boldsymbol{s})$
 \item If only the covariance function for the noise process $\varepsilon_i(\boldsymbol{s})$ is oscillating, the random fields $x_j(\boldsymbol{s}) (j > i)$ belong to $\boldsymbol{x}_\text{p}(\boldsymbol{s})$, which means that
       the covariance functions for these random fields might be oscillating.
\end{itemize}

This result is rather intuitive since we can obtain it by checking the system of SPDEs \eqref{eq: osci_SPDEs_system_triangular} directly. However, these results are important in the real-world application since it gives information for how to build
models in a reasonable way. For instance, we can get information about how to choose the order of the random fields. 

\subsection{Bivariate GRFs with oscillating covariance functions} \label{sec: osci_bivariateGRFs}
The methodology for constructing non-oscillating and isotropic bivariate GRFs explicitly has been studied by \citet{gneitingmatern} and \citet{hu2012multivariate}. 
In this section we discuss the  approach for constructing bivariate GRFs explicitly with oscillating covariance functions using systems of SPDEs.
We start the investigation with random fields constructed by the full system of SPDEs
\begin{equation}
\begin{split}
 \label{eq: osci_bivariate_SPDE_explict}
 b_{11}(\kappa_{11}^2 - \Delta)^{\alpha_{11}/2} {x}_1(\boldsymbol{s}) +  b_{12}(\kappa_{12}^2 - \Delta)^{\alpha_{12}/2} {x}_2(\boldsymbol{s}) & = \varepsilon_1(\boldsymbol{s}),   \\
 b_{22}(\kappa_{22}^2 - \Delta)^{\alpha_{22}/2} {x}_2(\boldsymbol{s}) +  b_{21}(\kappa_{21}^2 - \Delta)^{\alpha_{21}/2} {x}_1(\boldsymbol{s}) & = \varepsilon_2(\boldsymbol{s}),
\end{split}
\end{equation}
where $b_{ij}$ and $\{ \mathcal{L}_{ij}(\boldsymbol{s}); i,j = 1, 2 \}$ are the same as in \eqref{eq: osci_SPDEs_system}, and
$\{ \varepsilon_1 (\boldsymbol{s}); i = 1,2 \}$ are noise processes which can have oscillating covariance functions. By changing the
properties of the noise processes we can construct more interesting random fields. Use the matrix notion and define the operator matrix as
\begin{align} \label{eq: osci_bivariate_SPDE_operator_full}
 \mathscr{L}(\boldsymbol{\theta}) = \begin{pmatrix}
 \mathcal{L}_{11}(\theta_{11})&  \mathcal{L}_{12}(\theta_{12}) \\
 \mathcal{L}_{21}(\theta_{21}) & \mathcal{L}_{22}(\theta_{22})
\end{pmatrix},
\end{align}
and let $\boldsymbol{\varepsilon}(\boldsymbol{s}) = \left( \varepsilon_1(\boldsymbol{s}), \varepsilon_2(\boldsymbol{s}) \right)^{\mbox{H}}$, where $\mbox{H}$ denotes the Hermitian transpose of a vector or a matrix. 
$\theta_{ij} = \left\{ \alpha_{ij}, \kappa_{ij}, b_{ij} \right\}$ is defined as the collection of parameters for $\mathcal{L}_{ij}$.
The system of equations \eqref{eq: osci_bivariate_SPDE_explict} can then be written in a compact matrix form as
\begin{equation} \label{eq: osci_bivariate_SPDE_compact}
 \mathscr{L}(\boldsymbol{\theta}) \boldsymbol{x}(\boldsymbol{s}) = \boldsymbol{\varepsilon}(\boldsymbol{s}),
\end{equation}
where $\boldsymbol{\theta} = \{ \theta_{ij}, i, j = 1,2\}$.
With \eqref{eq: osci_bivariate_SPDE_compact} we can obtain the power spectrum $\boldsymbol{S}_{\boldsymbol{x}} = \mathbb{E}(\hat{\boldsymbol{x}} \cdot \hat{\boldsymbol{x}}^{\mbox{H}})$ by
\begin{equation} \label{eq: osci_spectrum_bivariate}
\boldsymbol{S}_{\boldsymbol{x}} = \mathscr{H}^{\mbox{H}} \boldsymbol{S}_{\boldsymbol{\varepsilon}} \mathscr{H}^{-{\mbox{H}}},
\end{equation}
where $-\mbox{H}$ denotes the inverse of the complex conjugate of the matrix. $\hat{x}_{ij}$ is the Fourier transform of $x_{ij}$, $\hat{x}_{ij} = \mathscr{F} (x_{ij})$, and $\mathscr{H}$ is the Fourier transform of the operator matrix
$\mathscr{L}$,
\begin{align} \label{eq: osci_Hmatrix_bivariate}
\mathscr{H}(\boldsymbol{\theta}) = \begin{pmatrix}
 \mathcal{H}_{11}(\theta_{11}) & \mathcal{H}_{12}(\theta_{12})\\
 \mathcal{H}_{21}(\theta_{21}) & \mathcal{H}_{22}(\theta_{22})
\end{pmatrix}.
\end{align}
$\boldsymbol{S}_{\boldsymbol{\varepsilon}}(\boldsymbol{k}) = \mathbb{E}\left( \hat{\boldsymbol{{\varepsilon}}} \hat{\boldsymbol{{\varepsilon}}}^{\mbox{H}} \right)$ is the power spectrum matrix for the independent noise processes

\begin{align} \label{eq: osci_noise_spectrum}
\boldsymbol{S}_{\boldsymbol{\varepsilon}} (\boldsymbol{k}) = \begin{pmatrix}
 S_{\varepsilon_1}(\boldsymbol{k}) & 0 \\
 0  & S_{\varepsilon_2}(\boldsymbol{k})
\end{pmatrix},
\end{align}
where $\boldsymbol{k}$ is the frequency. Since the noise processes are mutually independent, the power spectrum 
matrix of noise processes is a diagonal matrix.  Using Equation \eqref{eq: osci_spectrum_bivariate} - Equation \eqref{eq: osci_Hmatrix_bivariate}, the elements in the power spectrum matrix of the 
bivariate fields from the full system of SPDEs in \eqref{eq: osci_bivariate_SPDE_explict} can be obtained,
\begin{equation} \label{eq: osci_Sx_separate}
 \begin{split}
 S_{x_{11}}(\boldsymbol{k}) & = \frac{S_{\varepsilon_1} |\mathcal{H}_{22}^2| + S_{\varepsilon_2} |\mathcal{H}_{12}^2|}{|(\mathcal{H}_{11}\mathcal{H}_{22}-\mathcal{H}_{12}\mathcal{H}_{21})^2|}, \\
 S_{x_{12}}(\boldsymbol{k}) & =-\frac{\mathcal{H}_{22} S_{\varepsilon_1} |\mathcal{H}_{21}^2|\mathcal{H}_{11} + \mathcal{H}_{12} S_{\varepsilon_2} 
                                |\mathcal{H}_{11}^2|\mathcal{H}_{21}}{|(\mathcal{H}_{11}\mathcal{H}_{22}-\mathcal{H}_{12}\mathcal{H}_{21})^2|\mathcal{H}_{21}\mathcal{H}_{11}}, \\
 S_{x_{21}}(\boldsymbol{k}) & =-\frac{\mathcal{H}_{21} S_{\varepsilon_1} |\mathcal{H}_{22}^2|\mathcal{H}_{12} + \mathcal{H}_{11} S_{\varepsilon_2} |\mathcal{H}_{12}^2|\mathcal{H}_{22}}
                                 {|(\mathcal{H}_{11}\mathcal{H}_{22}-\mathcal{H}_{12}\mathcal{H}_{21})^2|\mathcal{H}_{22}\mathcal{H}_{12}}, \\
 S_{x_{22}}(\boldsymbol{k}) & = \frac{S_{\varepsilon_1} |\mathcal{H}_{21}^2| + S_{\varepsilon_2} |\mathcal{H}_{11}^2|}{|(\mathcal{H}_{11}\mathcal{H}_{22}-\mathcal{H}_{12}\mathcal{H}_{21})^2|}.
\end{split}
\end{equation}
Define poles and zeros as the roots of the denominators and numerators of the power spectrum elements $\{S_{x_{ij}}; i, j = 1,2 \}$, respectively.
From \eqref{eq: osci_Sx_separate}, we can see that the poles of the power spectrum, in general, are the same for both the fields, 
but zeros of the power spectrum will be different. It gives us a possibility to construct bivariate GRFs with oscillating 
covariance functions by carefully re-parametrization of system of SPDEs \eqref{eq: osci_bivariate_SPDE_explict}.
However, we will not discuss this approach in this paper, but leave it for future research. The approach we have chosen here is to change the noise process at the right hand of system of SPDEs \eqref{eq: osci_bivariate_SPDE_explict}.

Theoretically, we could choose the full version of the system of SPDEs given in \eqref{eq: osci_bivariate_SPDE_explict} and give more flexibility for constructing bivariate random fields. However, we choose to simplify the model.
\citet{hu2012multivariate} used the triangular version of the SPDEs system to construct bivaraite GRFs and this suggestion is followed in this paper. In the following sections, 
we focus on a special form of the triangular system of the SPDEs discussed in Section \ref{sec: osci_multiGRFs}.
In the special form the operator matrix is
\begin{equation} \label{eq: osci_bivariate_SPDE1_operator1}
  \mathscr{L}_1(\boldsymbol{\theta}) = \begin{pmatrix}
 b_{11} \left( h_{11} -\Delta \right) &  0 \\
 b_{21} & b_{22} \left( h_{22} -\Delta \right)
\end{pmatrix},
\end{equation}
where the subscript ``$1$'' in $\mathscr{L}_1$ is used to denote the first operator matrix we use for constructions. Some other operator matrices are discussed in Appendix A.
We can rewrite the system of SPDEs with matrix notation as 
\begin{equation} \label{eq: osci_bivariate_SPDE1_compact}
\begin{split}
 \mathscr{L}_1(\boldsymbol{\theta}) \boldsymbol{x}(\boldsymbol{s}) = \boldsymbol{\varepsilon}(\boldsymbol{s}),
\end{split}
\end{equation}
and Equation \eqref{eq: osci_bivariate_SPDE1_compact} can be written down explicitly as 
\begin{equation} \label{eq: osci_bivariate_SPDE1_explicit}
 \begin{split}
 b_{11} (h_{11} - \Delta) x_1(\boldsymbol{s}) & = \varepsilon_1(\boldsymbol{s}),   \\
 b_{21} x_1(\boldsymbol{s}) + b_{22} (h_{22} - \Delta) x_2(\boldsymbol{s}) & = \varepsilon_2(\boldsymbol{s}). 
 \end{split}
\end{equation}
In this form both random fields can have oscillating covariance function. 
The following discussion are based on the system of SPDEs \eqref{eq: osci_bivariate_SPDE1_explicit}.
Let $\varepsilon_1(\boldsymbol{s})$ be a noise process with non-oscillating covariance function, such as a white noise process or noise process with Mat\'ern covariance function, and $\varepsilon_2(\boldsymbol{s})$ be a noise 
process with oscillating covariance function generated from the complex-valued SPDEs \eqref{eq: osci_univariate_complex}. 
We can then conclude that the first field $x_1(\boldsymbol{s})$ is a stationary and isotropic random field with non-oscillating covariance function,
and that $x_2(\boldsymbol{s})$ is a random field with oscillating covariance function. 
On the other hand, if $\varepsilon_1(\boldsymbol{s})$ has an oscillating covariance function and $\varepsilon_2(\boldsymbol{s})$ has a
non-oscillating covariance function, then the covariance functions for both random fields $x_1(\boldsymbol{s})$ and $x_2(\boldsymbol{s})$ are oscillating given that $b_{21} \neq 0$.

We use the power spectra $\{S_{x_{ii}}(\boldsymbol{k}); i = 1,2 \}$ of the random fields together with the cross spectrum $S_{x_{21}}(\boldsymbol{k})$ to investigate the properties of the random fields,
\begin{equation} \label{eq: osci_bivariate_SPDE1_spectrum}
\begin{split}
 S_{x_{11}}(\boldsymbol{k}) & = \frac{S_{\varepsilon_1}}{b_{11}^2 ( h_{11} + \| \boldsymbol{k} \|^2 )^2}, \\
 S_{x_{21}}(\boldsymbol{k}) & = -\frac{b_{21}S_{\varepsilon_1}}{b_{22}(h_{22} + \| \boldsymbol{k} \|^2) b_{11}^2 (h_{11} + \| \boldsymbol{k} \|^2)^2 }, \\
 S_{x_{22}}(\boldsymbol{k}) & = \frac{b_{21}^2 S_{\varepsilon_1} + b_{11}^2 (h_{11} +\| \boldsymbol{k} \|^2)^2 S_{\varepsilon_2} }{b_{11}^2 ( h_{11} + \| \boldsymbol{k} \|^2 )^2 b_{22}^2 ( h_{22} + \| \boldsymbol{k} \|^2 )^2}.
\end{split}
\end{equation}
The following results can be obtained from \eqref{eq: osci_bivariate_SPDE1_spectrum}.
\begin{enumerate}
 \item[$\bullet$] The marginal variance of the first random field $x_1(\boldsymbol{s})$ is related only to the parameters $b_{11}$ and $h_{11}$. 
 \item[$\bullet$] The marginal variance of the second random field $x_2(\boldsymbol{s})$ is related only to $\{b_{ij}, h_{ij}; i, j = 1,2\}$.
 \item[$\bullet$] The sign of  $b_{11}$ is irrelevant to the sign of the cross-correlation between the two fields. Since the marginal variance of the first random field $x_1(\boldsymbol{s})$ is only related to $b_{11}$ and $h_{11}$, and 
                  there is a requirement $h_{11} > 0$, we can set $b_{11} > 0$;
 \item[$\bullet$] The sign of the correlation between the two fields only depends on the sign of the product of $b_{22}$ and $b_{21}$. We recommend to set $b_{22} > 0$, and then the sign of the correlation between  
                  the fields will be related only to the sign of $b_{21}$. If $b_{21} >0$, the two random fields are negatively correlated. If $b_{21} < 0$, the random fields are positively correlated. 
\end{enumerate}

\subsection{Sampling the bivariate GRFs} \label{sec: osci_sampling}
The common approach for sampling GRFs uses the covariance matrix $\boldsymbol{\Sigma}$ or precision matrix $\boldsymbol{Q}$. Since the bivariate GRFs from the systems of SPDEs are represented by the GMRFs, the 
precision matrices $\boldsymbol{Q}$ are sparse. Therefore, the direct approach for sampling a (multivariate)
GMRF is usually through the Cholesky triangle $\boldsymbol{L}$, where $\boldsymbol{Q} = \boldsymbol{L} \boldsymbol{L}^T$. The commonly used  procedure for getting a sample from the GMRF $\boldsymbol{x} \sim \mathcal{N}(\boldsymbol{\mu}, \boldsymbol{Q}^{-1})$
is through the following steps
\begin{enumerate}[I.]
 \item Use the Cholesky factorization to find the Cholesky triangle $\boldsymbol{L}$ of the precision matrix $\boldsymbol{Q}$. We usually do the Cholesky factorization with standard libraries.
 \item Get a sample $\boldsymbol{z} \sim \mathcal{N}(\boldsymbol{0}, \boldsymbol{I})$. $\boldsymbol{I}$ is an identity matrix and has the same dimensions as the precision matrix $\boldsymbol{Q}$.
 \item Solve a linear system of equations with Cholesky triangle $\boldsymbol{L} \boldsymbol{v} = \boldsymbol{z}$. Thus $\boldsymbol{v}$ has the correct covariance matrix $\boldsymbol{Q}^{-1}$ 
      since $\Cov(\boldsymbol{v}) = \Cov(\boldsymbol{L}^{-T}\boldsymbol{z}) = (\boldsymbol{L}\boldsymbol{L}^T)^{-1} = \boldsymbol{Q}^{-1}$.       
 \item Correct the mean by $\boldsymbol{x}=\boldsymbol{\mu}+\boldsymbol{v}$, and then $\boldsymbol{x}$ has the correct mean $\boldsymbol{\mu}$ and covariance matrix $\boldsymbol{Q}^{-1}$, $\boldsymbol{x} \sim \mathcal{N}(\boldsymbol{\mu}, \boldsymbol{Q}^{-1})$. 
\end{enumerate}
If the precision matrix $\boldsymbol{Q}$ is a band matrix, the Cholesky triangle $\boldsymbol{L}$ will be also a band matrix. 
The corresponding algorithm for finding the Cholesky triangle when $\boldsymbol{Q}$ is a band matrix can be found in \citet[Algorithm $2.9$]{rue2005gaussian}.
We also refer to \citet[Chapter $2$]{rue2005gaussian} for detailed discussion about different sampling algorithms for GMRFs with different kinds of parametrization. \citet{hu2012specifying} showed that 
it is possible to find a sparser triangular matrix $\tilde{\boldsymbol{L}}$ with incomplete orthogonal factorization for sampling the GMRF, but they pointed out that it needs longer computation time for
finding the sparse triangular matrix.

In the following two examples, we choose all values of parameters to be equal. However, we set $\varepsilon_1(\boldsymbol{s})$ to be a noise process with a non-oscillating covariance function and $\varepsilon_2(\boldsymbol{s})$ 
to be a noise process with an oscillating covariance function in the first example. In the second example, we simply switch the order of the noise processes, i.e., we set $\varepsilon_1(\boldsymbol{s})$ to be a noise process 
with an oscillating covariance function and $\varepsilon_2(\boldsymbol{s})$ to be a noise process with a non-oscillating covariance function.

\begin{figure}[tbp]
    \centering
    \subfigure[]{\includegraphics[width=0.48\textwidth,height=0.48\textwidth]{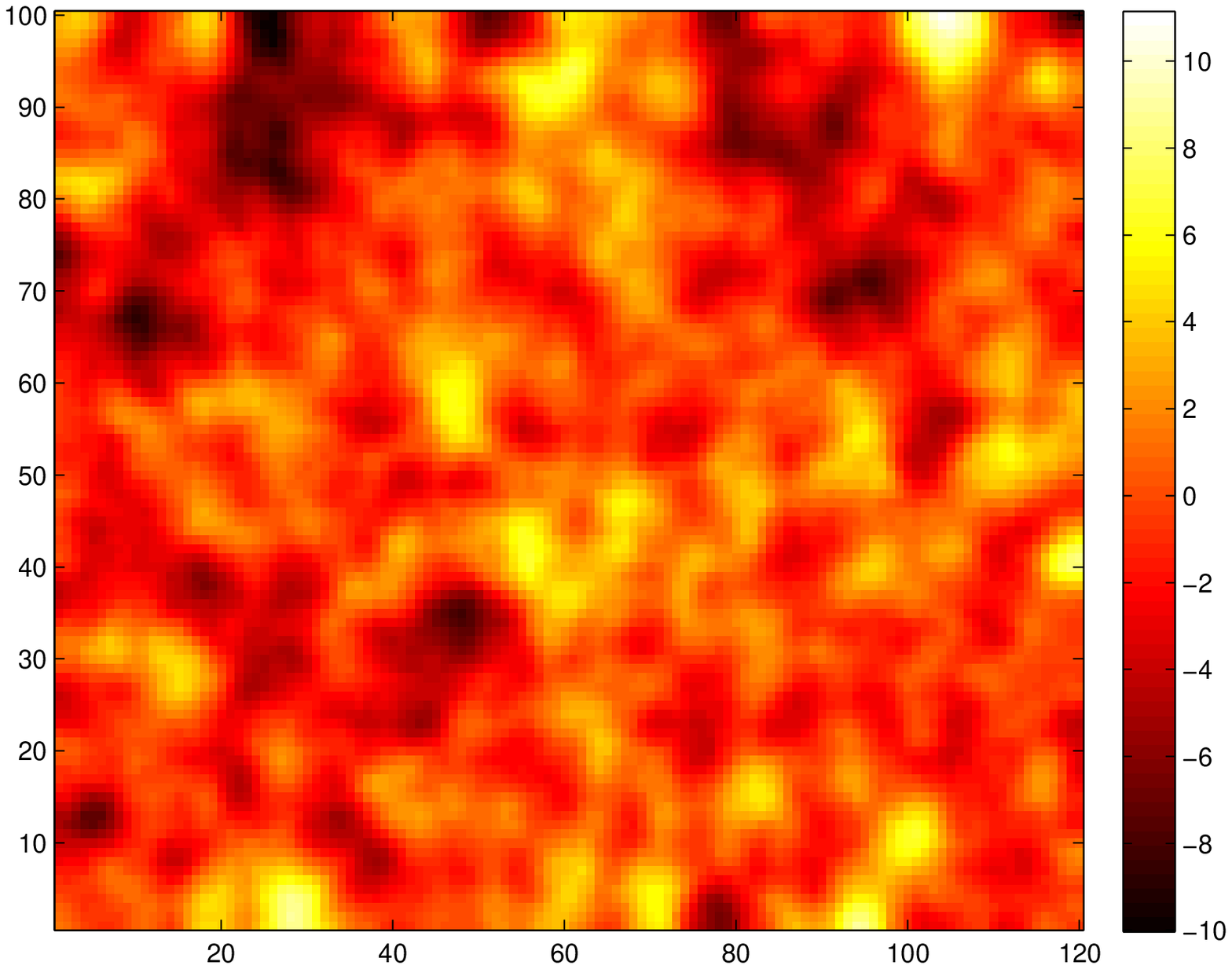} \label{fig: osci_sampling_case1_field1}} 
    \subfigure[]{\includegraphics[width=0.48\textwidth,height=0.48\textwidth]{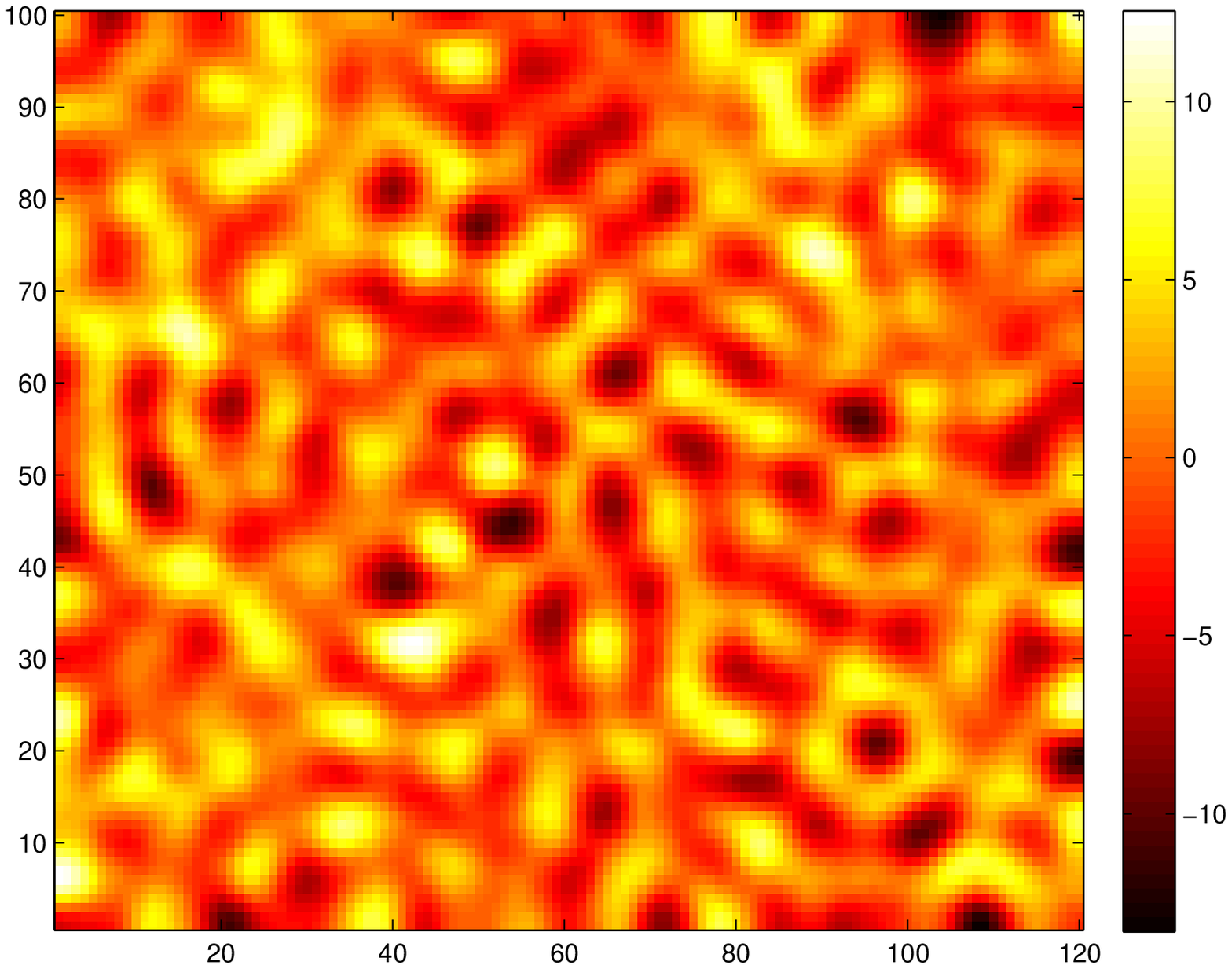}  \label{fig: osci_sampling_case1_field2}} \\
    \caption{A realization of the bivariate random field with parameters given in Table \ref{tab: simulated_parameters_1}.
             In this example the first noise process has a non-oscillating covariance function and the second noise process has an oscillating covariance function. } 
\label{fig: osci_bivariate_matern_oscillation}
 \end{figure}

\begin{figure}[tbp]
    \centering
    \includegraphics[width=0.8\textwidth,height=0.6\textwidth]{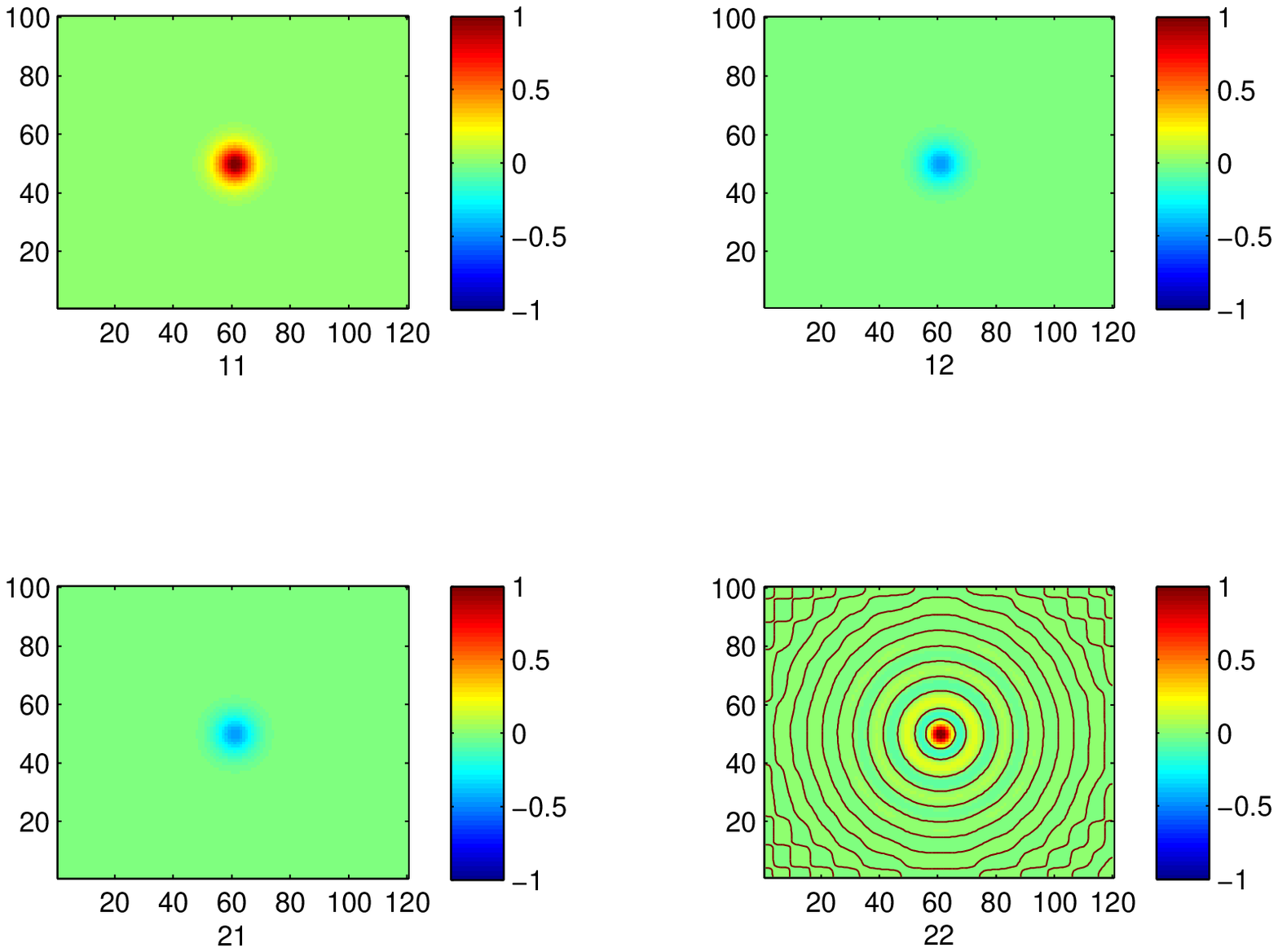} 
     \caption{Correlations and cross-correlation functions for the bivariate random field with parameters given in Table \ref{tab: simulated_parameters_1}.
             In this example the first noise process has a non-oscillating covariance function and the second noise process has an oscillating covariance function.
             We see that only the second random field has an oscillating covariance function.}
\label{fig: osci_sampling_cov1}
 \end{figure}

\begin{figure}[tbp]
    \centering
    \subfigure[]{\includegraphics[width=0.48\textwidth,height=0.48\textwidth]{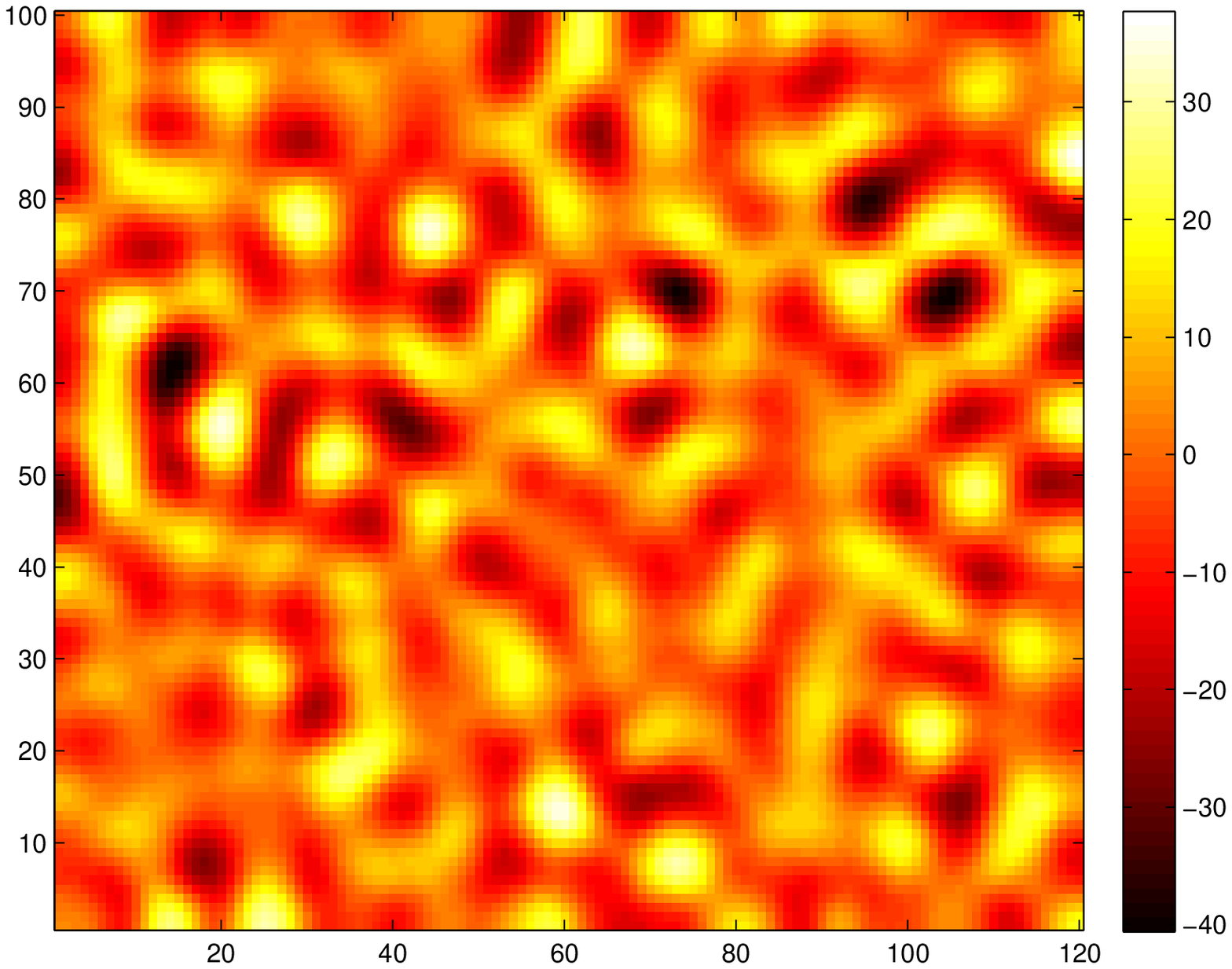} \label{fig: osci_sampling_case2_field1}}
    \subfigure[]{\includegraphics[width=0.48\textwidth,height=0.48\textwidth]{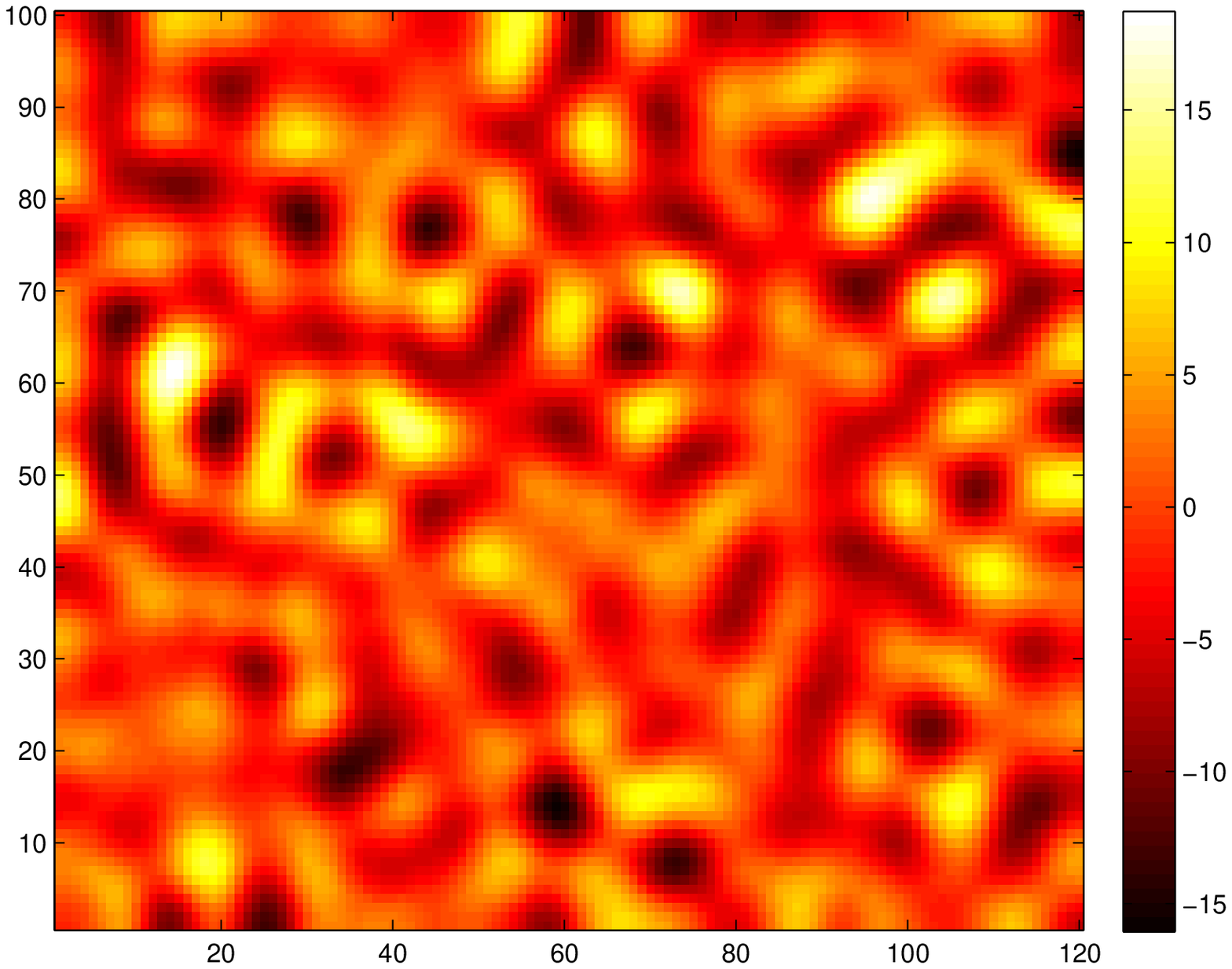} \label{fig: osci_sampling_case2_field2}} 
    \caption{A realization of the bivariate random field with the parameters given in Table \ref{tab: simulated_parameters_1}.
             In this example the first noise process has an oscillating covariance function and the second noise process has a non-oscillating covariance function.} 
\label{fig: osci_bivariate_oscillation2}
 \end{figure}

\begin{figure}[tbp]
    \centering
    \includegraphics[width=0.8\textwidth,height=0.6\textwidth]{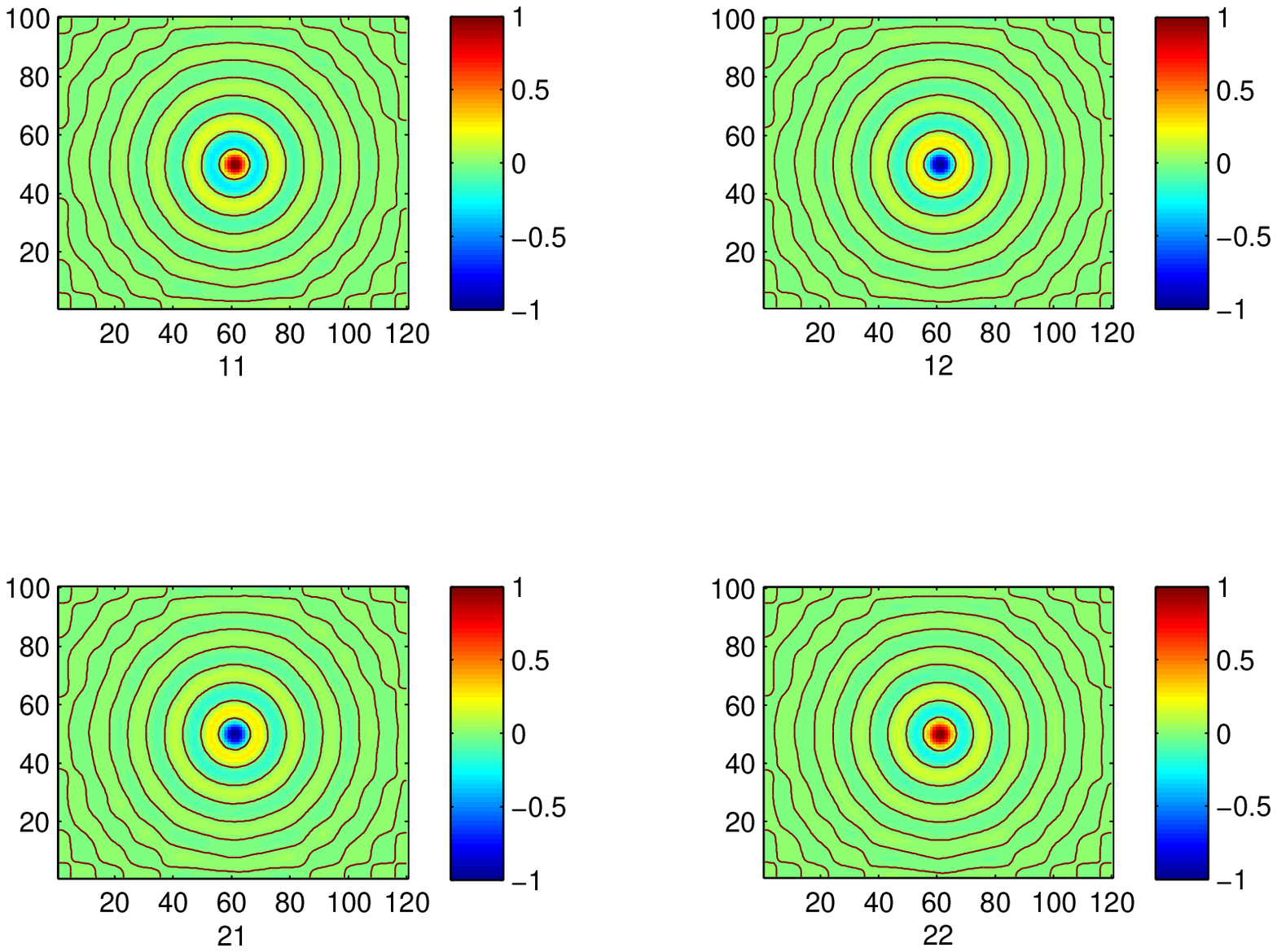}
    \caption{Correlations and cross-correlation functions for the bivariate random field with parameters given in Table \ref{tab: simulated_parameters_1}.
             In this example the first noise process has an oscillating covariance function and the second noise process has a non-oscillating covariance function.
             We see that both the random fields have oscillating covariance functions.}
\label{fig: osci_sampling_cov2}
 \end{figure}

One sample from the GMRF in the first example is shown in Fig. \ref{fig: osci_bivariate_matern_oscillation} and one sample in the second example is shown in Fig. \ref{fig: osci_bivariate_oscillation2}. 
The corresponding correlation matrices are shown in Fig. \ref{fig: osci_sampling_cov1} and Fig. \ref{fig: osci_sampling_cov2}. 
In these figures, the red lines indicate that the correlation is $0$.
In the first example the random field $x_1(\boldsymbol{s})$ has a non-oscillating covariance function and the second random field $x_2(\boldsymbol{s})$ has an oscillating covariance function.
In the second example both the fields have oscillating covariance functions. These two examples verify the conclusion in Section \ref{sec: osci_bivariateGRFs}.

\begin{table}
\centering
\caption{Parameters for sampling the bivariate GRFs}
 \begin{tabular}{c|c|c}
\hline
\hline
\multicolumn{3}{c}{Parameters}  \\
\hline
$\boldsymbol{\alpha}$     &  $\boldsymbol{\kappa}$         &   $\boldsymbol{b}$  \\
\hline
$\alpha_{11} = 2 $    &  $h_{11}  = 0.25$          & $b_{11} = 0.5$   \\
$\alpha_{12} = 0 $    &  $h_{22} = 0.36$           & $b_{12} = 0$     \\
$\alpha_{21} = 2 $    &  $\kappa_{n_1} = 0.5$      & $b_{21} = 0.25$   \\
$\alpha_{22} = 2 $    &  $\kappa_{n_2} = 0.6$      & $b_{22} = 1$     \\
$\alpha_{n_1} = 2$    &  $\omega = 0.95$           &                  \\
$\alpha_{n_2} = 2$    &                            &                  \\
\hline
 \end{tabular}
  \label{tab: simulated_parameters_1}
\end{table}

  \begin{figure}[tbp]
    \centering
    \includegraphics[width=0.8\textwidth,height=0.6\textwidth]{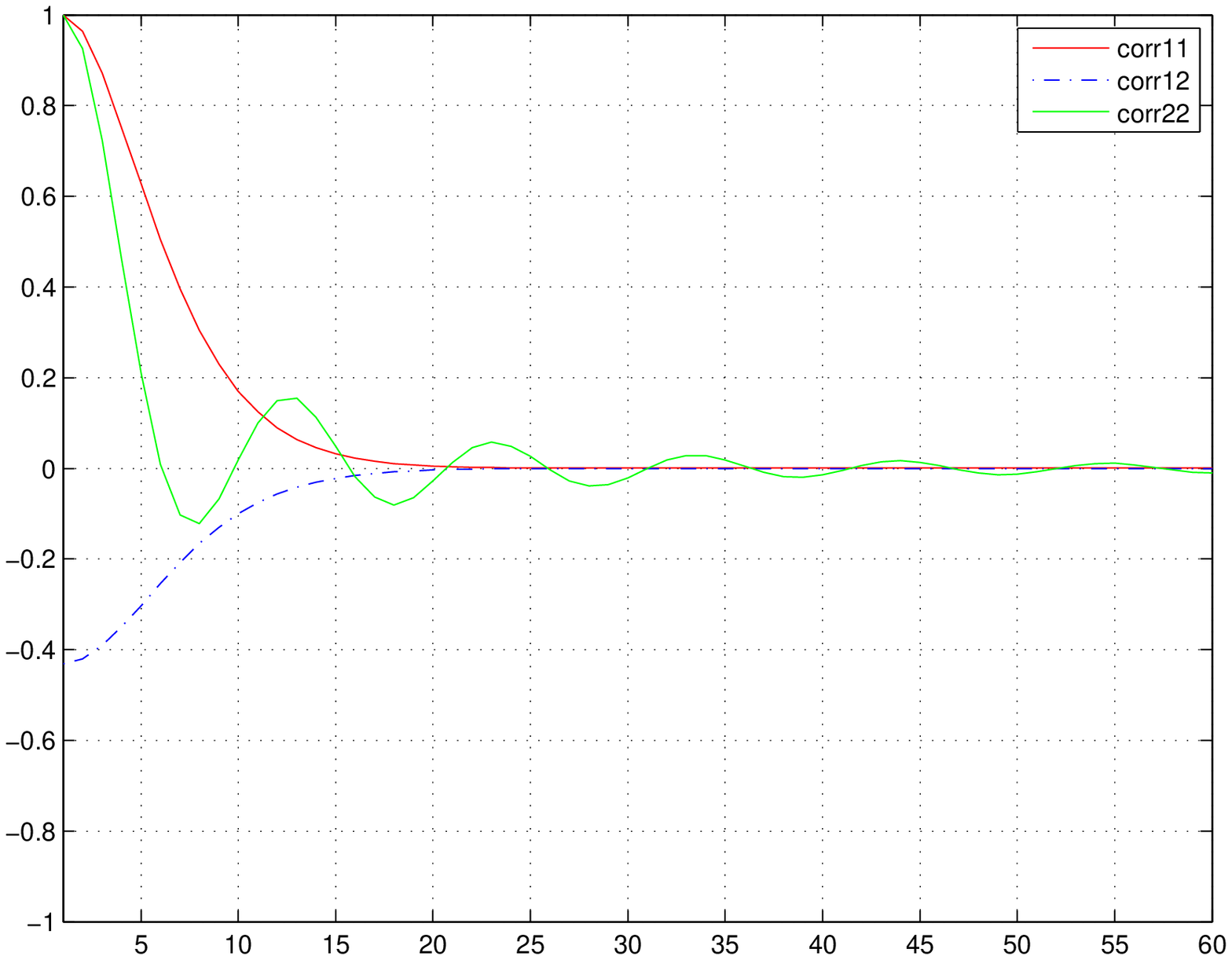} 
     \caption{Correlations and cross-correlation functions for the bivariate random field with parameters given in Table \ref{tab: simulated_parameters_1}.
             In this example the first noise process has a non-oscillating covariance function and the second noise process has an oscillating covariance function.
             We see that only the second random field has an oscillating covariance function.}
\label{fig: osci_sampling_cov1_1d}
 \end{figure}
 
  \begin{figure}[tbp]
    \centering
    \includegraphics[width=0.8\textwidth,height=0.6\textwidth]{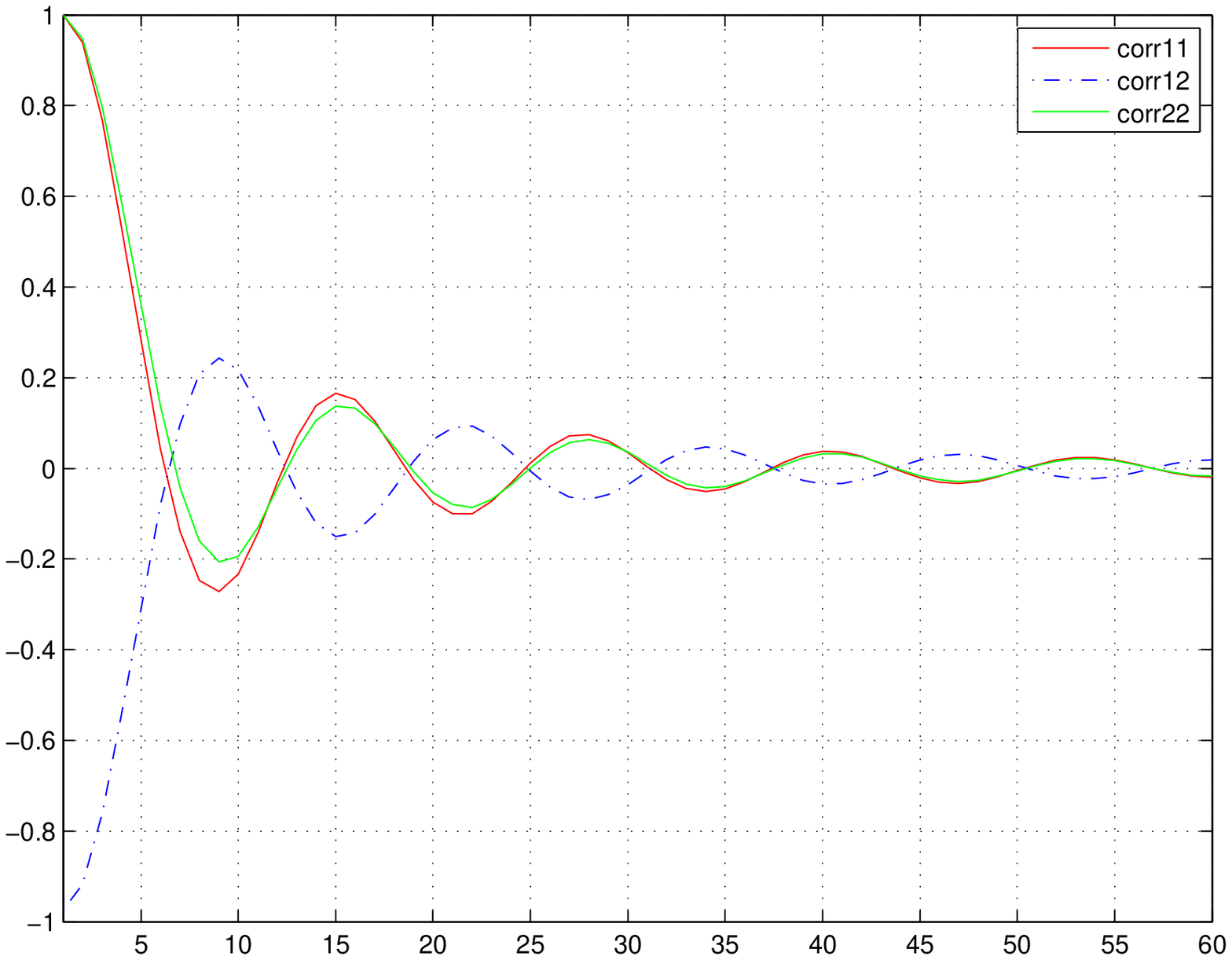}
    \caption{Correlations and cross-correlation functions for the bivariate random field with parameters given in Table \ref{tab: simulated_parameters_1}.
             In this example the first noise process has an oscillating covariance function and the second noise process has a non-oscillating covariance function.
             We see that both the random fields have oscillating covariance functions.}
\label{fig: osci_sampling_cov2_1d}
 \end{figure}

\section{Inference with simulated data and real data} \label{sec: osci_exampl_applications}
In this section we illustrate how to use our approach with some simulated data examples and one real data example. In the first example, the covariance function of the  first random field is non-oscillating, but the 
second random field has an oscillating covariance function. In the second example, both the random fields have oscillating covariance functions. The third and the forth examples 
show that if the two fields are independent, the inferences give indications about this, no matter which field is oscillating. 
One real data example in the end shows that our approach can be applied in practical applications. 
As pointed out by \citet[Chapter 5]{diggle1998model} and \citet[Section 2]{lindgren2011explicit}, the smoothness parameter $\{\nu_{ij}; i,j = 1,2\}$ are poorly identifiable. Therefore, we 
fix the values of $\{\alpha_{ij}\}$ in the simulated data examples and in the real data example.

\subsection{Posterior for the hyper-parameters} \label{sec: osci_posterior}
The first step of the inference is usually to derive the (log-) posterior distribution $\pi(\boldsymbol{\theta}|\boldsymbol{y})$ of $\boldsymbol{\theta}$. 
The well known Bayesian formula \eqref{eq: osci_baysian_y & theta} is at the core of Bayesian inference,

\begin{equation} \label{eq: osci_baysian_y & theta}
\begin{split}
 \pi(\boldsymbol{y}, \boldsymbol{\theta}) & = \frac{\pi(\boldsymbol{\theta}, \boldsymbol{x}, \boldsymbol{y})}{\pi(\boldsymbol{x} | \boldsymbol{\theta}, \boldsymbol{y})} \\
                                      & = \frac{\pi(\boldsymbol{\theta}) \pi(\boldsymbol{x}|\boldsymbol{\theta}) \pi(\boldsymbol{y|\boldsymbol{x},\boldsymbol{\theta}})}{\pi(\boldsymbol{x}|\boldsymbol{y}, \boldsymbol{\theta})}.
\end{split}
\end{equation}
where $\pi(\boldsymbol{\theta})$ is the prior distribution of the hyper-parameters, and we return to this topic in Section \ref{sec: osci_prior}, $\pi(\boldsymbol{x}|\boldsymbol{\theta})$ is the density for the bivariate random fields, 
$\pi(\boldsymbol{y|\boldsymbol{x},\boldsymbol{\theta}})$ is the density for the observations given the random field and the parameters
and $\pi(\boldsymbol{x}|\boldsymbol{y}, \boldsymbol{\theta})$ is the full conditional of the random fields given the observations and parameters.

Assume that there are $N$ triangles in the domain for each of the random field $\{x_i(\boldsymbol\theta); i = 1,2\}$. With the bivariate random fields $\boldsymbol{x} = (x_1, x_2)^T$, $2N$ triangles are used and hence the probability density
of the bivariate random field                                                                                                                                                                                                                                                                                                                                                                                                   has the form

\begin{equation} \label{eq: osci_bivariate_x|theta}
 \pi(\boldsymbol{x} | \boldsymbol{\theta}) = \left(\frac{1}{2\pi} \right)^{2N} |\boldsymbol{Q}(\boldsymbol\theta)|^{1/2} \exp \left(-\frac{1}{2} \boldsymbol{x}^T \boldsymbol{Q}(\boldsymbol{\theta}) \boldsymbol{x} \right), 
\end{equation}
where $\boldsymbol{Q}(\boldsymbol{\theta)}$ is the precision matrix for the bivariate field. We assume that the length of the data is $t = k_1+k_2$, where $\{k_i; i = 1,2\}$ are the length of the observations for each field.  
Then $\pi(\boldsymbol{y|\boldsymbol{x},\boldsymbol{\theta}})$ has the form   

\begin{equation} \label{eq: osci_bivariate_y|x,theta}
 \pi(\boldsymbol{y|\boldsymbol{x},\boldsymbol{\theta}}) = \left(\frac{1}{2\pi} \right)^{t}|\boldsymbol{Q}_n|^{1/2}
               \exp \left(-\frac{1}{2} (\boldsymbol{y}-\boldsymbol{Ax})^T \boldsymbol{Q}_n (\boldsymbol{y}-\boldsymbol{Ax}) \right),
\end{equation}
where $\boldsymbol{Q}_n$ is the precision matrix for the measurement errors with dimension $t \times 2k$, and $\boldsymbol{A}$ is a matrix with dimension $t \times 2N$
that links the sparse observations to the dense random fields. One thing we want to point out is that the length of the observations for 
each field can be different and they are not necessarily observed at the same locations.
We used the notation 
$\pi(\boldsymbol{y|\boldsymbol{x}})$ instead of $\pi(\boldsymbol{y|\boldsymbol{x},\boldsymbol{\theta}}) $ since this function is independent of $\boldsymbol{\theta}$. The full conditional 
$\pi(\boldsymbol{x}|\boldsymbol{y}, \boldsymbol{\theta})$ can be obtained,

\begin{equation} \label{eq: osci_bivariate_x|y,theta}
\begin{split}
 \pi(\boldsymbol{x}|\boldsymbol{y}, \boldsymbol{\theta}) & \propto \pi({\boldsymbol{x}, \boldsymbol{y} | \boldsymbol{\theta}})   \\
          & = \pi(\boldsymbol{x}|\boldsymbol{\theta}) \pi(\boldsymbol{y}|\boldsymbol{x}, \boldsymbol{\theta}) \\
          & \propto \exp \left( -\frac{1}{2} \left[ x^T (\boldsymbol{Q}(\boldsymbol{\theta}) + \boldsymbol{A}^T \boldsymbol{Q}_n \boldsymbol{A}) \boldsymbol{x} - 2\boldsymbol{x}^T \boldsymbol{A}^T\boldsymbol{Q}_n \boldsymbol{y} \right] \right).
\end{split}
\end{equation}
Denote $\boldsymbol{\mu}_c (\boldsymbol{\theta}) =  \boldsymbol{Q}_c^{-1}(\boldsymbol{\theta}) \boldsymbol{A}^T \boldsymbol{Q}_n \boldsymbol{y} $, 
 and $\boldsymbol{Q}_c (\boldsymbol{\theta}) = \boldsymbol{Q}(\boldsymbol{\theta}) + \boldsymbol{A}^T \boldsymbol{Q}_n \boldsymbol{A}$. Then we have 
\begin{equation} \label{eq: osci_bivariate_x|y,theta_canonical}
 {\boldsymbol{x}|\boldsymbol{y}, \boldsymbol{\theta}} \sim \mathcal{N} \left( \boldsymbol{\mu}_c (\boldsymbol{\theta}), \boldsymbol{Q}_c^{-1} (\boldsymbol{\theta}) \right).
\end{equation}
Thus ${\boldsymbol{x}|\boldsymbol{y}, \boldsymbol{\theta}}$ is a $2N$-dimensional multivariate Gaussian distribution. We can write \eqref{eq: osci_bivariate_x|y,theta_canonical} in 
the canonical form ${\boldsymbol{x}|\boldsymbol{y}, \boldsymbol{\theta}} \sim \mathcal{N}_c \left( \boldsymbol{A}^T \boldsymbol{Q}_n \boldsymbol{y}, \boldsymbol{Q}_c(\boldsymbol{\theta}) \right)$. 
For more information about canonical form of the GMRFs, we refer to, for example, \citet[Chapter 2]{rue2005gaussian}.

With a given prior $\pi(\boldsymbol{\theta})$, together with \eqref{eq: osci_bivariate_x|theta} to \eqref{eq: osci_bivariate_x|y,theta_canonical}, the posterior distribution of $\boldsymbol{\theta}$ becomes

\begin{equation} \label{eq: osci_bivariate_theta|y}
 \begin{split}
  \pi(\boldsymbol{\theta} | \boldsymbol{y}) & \propto \pi(\boldsymbol{\theta}) \frac{|\boldsymbol{Q}(\boldsymbol{\theta})|^{1/2} |\boldsymbol{Q_n}|^{1/2}}{|\boldsymbol{Q}_c(\boldsymbol{\theta})|^{1/2}} 
                          \exp \left( -\frac{1}{2} \boldsymbol{x}^T \boldsymbol{Q}(\boldsymbol{\theta}) \boldsymbol{x} \right) \\
                          & \times \exp \left(-\frac{1}{2} (\boldsymbol{y}-\boldsymbol{Ax})^T \boldsymbol{Q}_n (\boldsymbol{\theta}) (\boldsymbol{y}-\boldsymbol{Ax}) \right) \\
                          & \times \exp \left( \frac{1}{2} (\boldsymbol{x}-\boldsymbol{\mu}_c (\boldsymbol{\theta}))^T \boldsymbol{Q}_c (\boldsymbol{\theta}) (\boldsymbol{x}-\boldsymbol{\mu}_c (\boldsymbol{\theta})) \right).
 \end{split}
\end{equation}

\noindent And hence the logarithm of the posterior distribution is
\begin{equation} \label{eq: osci_bivariate_log(theta|y)}
\begin{split}
 \log(\pi(\boldsymbol{\theta}|\boldsymbol{y})) = & \text{ Const} + \log(\pi(\boldsymbol{\theta})) + \frac{1}{2}\log(|\boldsymbol{Q}(\boldsymbol{\theta})|) \\
             & - \frac{1}{2}\log(|\boldsymbol{Q}_c(\boldsymbol{\theta})|) + \frac{1}{2} \boldsymbol{\mu}_c(\boldsymbol{\theta})^T \boldsymbol{Q}_c(\boldsymbol{\theta}) \boldsymbol{\mu}_c(\boldsymbol{\theta}).
\end{split}
\end{equation}

\subsection{Priors for the parameters}\label{sec: osci_prior}
The prior distribution is important in Bayesian inference, and choosing the priors is an important part of inference. Two common approaches for choosing
the prior distribution are the conjugate prior approach and the non-informative prior approach. There is no unique way for choosing priors. We refer to \citet[Chapter 3]{robert2007bayesian} for detailed discussion about
the prior information and prior distribution.

General speaking, it is hard to specify an informative prior for the hyper-parameters in our system of SPDEs approach. Therefore, the non-informative approach has been chosen. The following choice for the priors of the parameters 
are recommended with the bivariate random fields.

\begin{itemize}
 \item $b_{11}$ and $b_{22}$ should be positive values. So log-normal distributions are used for these two parameters. Gamma distribution can also be considered;
 \item Because of the requirement on the systems of SPDEs that $\{h_{ij}; i, j = 1,2\}$ and $\{\kappa_{n_i}; i = 1,2\}$ should be positive values, we can use log-normal or gamma distributions;
 \item $b_{21}$ is related to the sign of the correlation of the two random fields and it can be either positive or negative. Therefore, a Gaussian distribution can be used;
 \item The oscillation parameter $\omega$ should fulfill the requirement $\omega \in [0,1]$ and hence a beta distribution can be used.
\end{itemize}

\subsection{Inference with simulated data} \label{sec: osci_exampl_applications_simulated_data}
Four simulated data examples are presented in this section to illustrate how to use our proposed approach. The datasets are divided into $2$ groups. In the first group we use the correlated random fields given in Section \ref{sec: osci_sampling}. 
In the second group the fields are independent. We want our model to capture these features, and to return whether $b_{21} = 0$ or not. 
However, if the first noise process is generated from the univariate SPDE given in Equation \eqref{eq: osci_spde_simple}, $\kappa_{n_1}^2$ and $h_{11}$ are not identifiable. See Appendix B for more 
information. We use the setting $\kappa_{n_1}^2 = h_{11} $ in this situation. 
It is our experience that $\omega$ is likely to be in the range of $(0.5, 1)$ if we have empirical knowledge that the random field has an oscillating covariance function,
 and hence we recommend to use a beta distribution with negative skew.
 In all of our simulated data examples, we use the following priors for the parameters (if they are needed to be estimated) following the discussion given in Section \ref{sec: osci_prior}.

\begin{itemize}
 \item $b_{11}, b_{22}, h_{11}, h_{22}, \kappa_{n_1} \text{ and } \kappa_{n_2}$ have the log-normal distributions with $\mu = 0$ and $\sigma^2 = 100$;
 \item $b_{21}$ has a normal distribution with $\mu = 0$ and $\sigma^2 = 100$, $b_{21} \sim \mathcal{N}(0, 100)$
 \item $\omega$ has a beta distribution with $\alpha = 1$ and $\beta = 1$, $\omega \sim \text{Beta}(1,1)$, i.e., it is a uniform distribution. 
\end{itemize}

\begin{figure}[tbp]
    \centering
    \subfigure[]{\includegraphics[width=0.48\textwidth,height=0.48\textwidth]{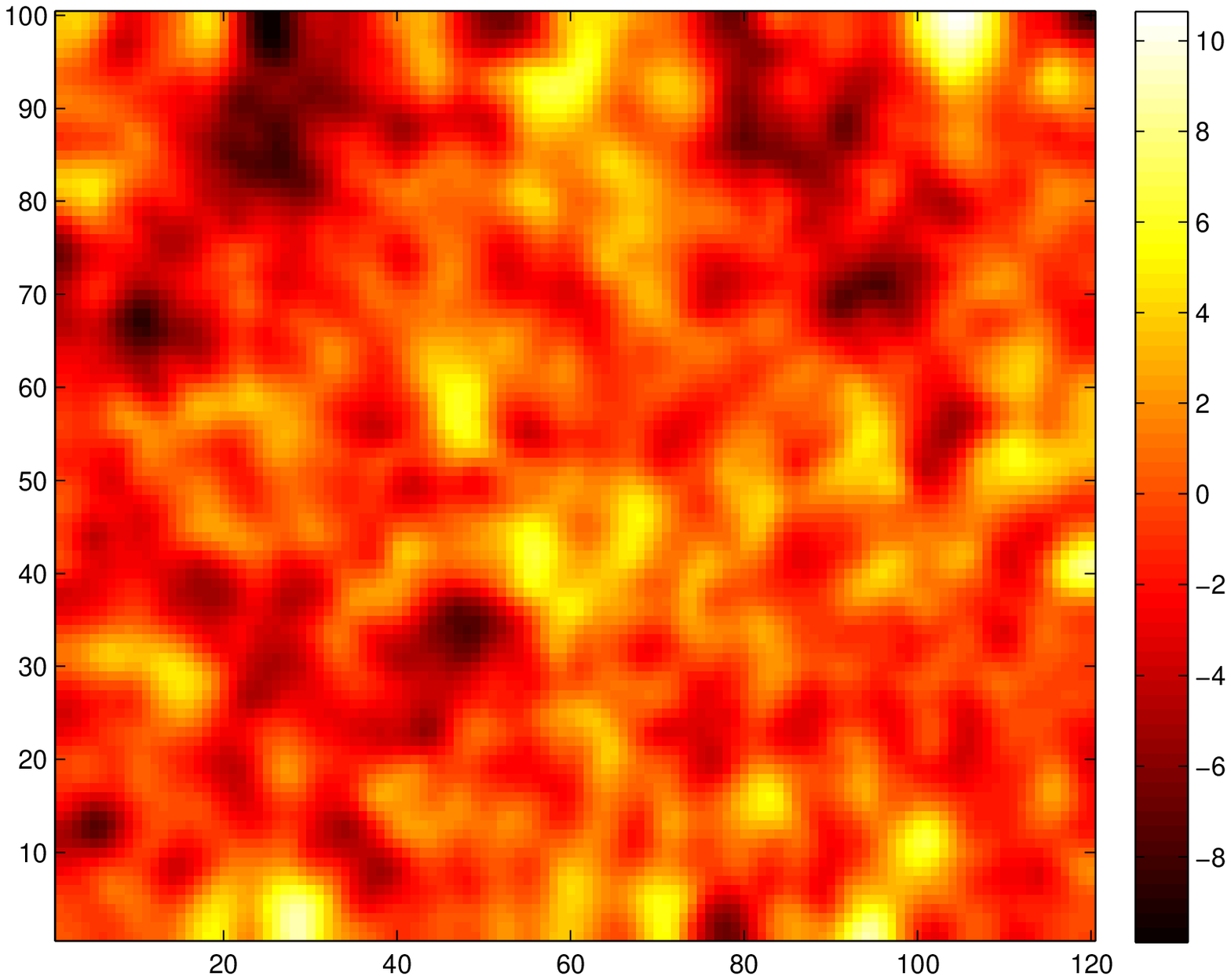} \label{fig: osci_estimated_case1_field1}} 
    \subfigure[]{\includegraphics[width=0.48\textwidth,height=0.48\textwidth]{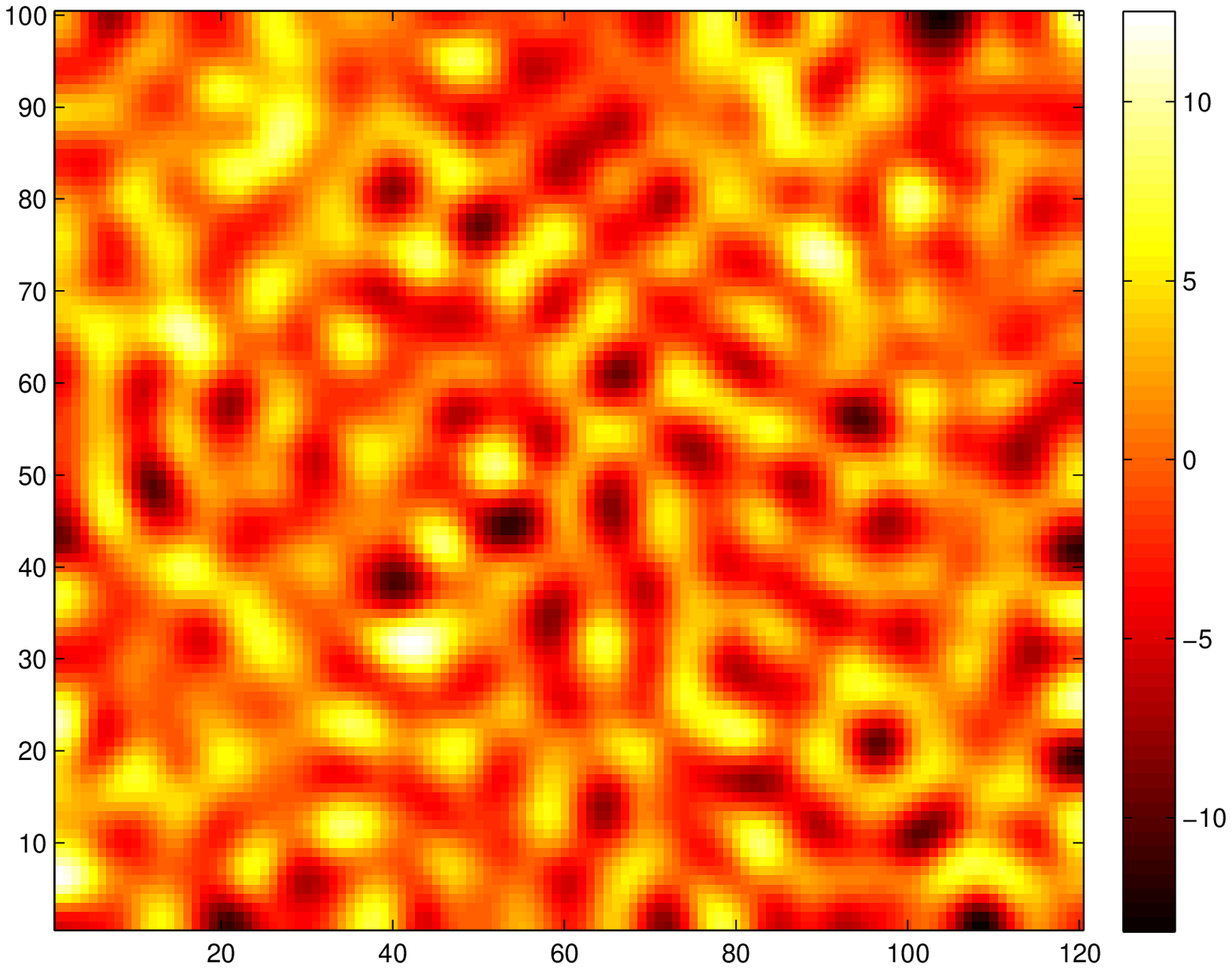} \label{fig: osci_estimated_case1_field2}} \\
 %   \subfigure[]{\includegraphics[width=0.8\textwidth,height=0.6\textwidth]{correlation_oscillation_case1} \label{fig: osci_sampling_cov1}}
    \caption{Estimated conditional mean of the bivariate random field for the dataset 1.} 
\label{fig: osci_bivariate_matern_oscillation_estimated}
 \end{figure}

\begin{figure}[tbp]
    \centering
    \subfigure[]{\includegraphics[width=0.48\textwidth,height=0.48\textwidth]{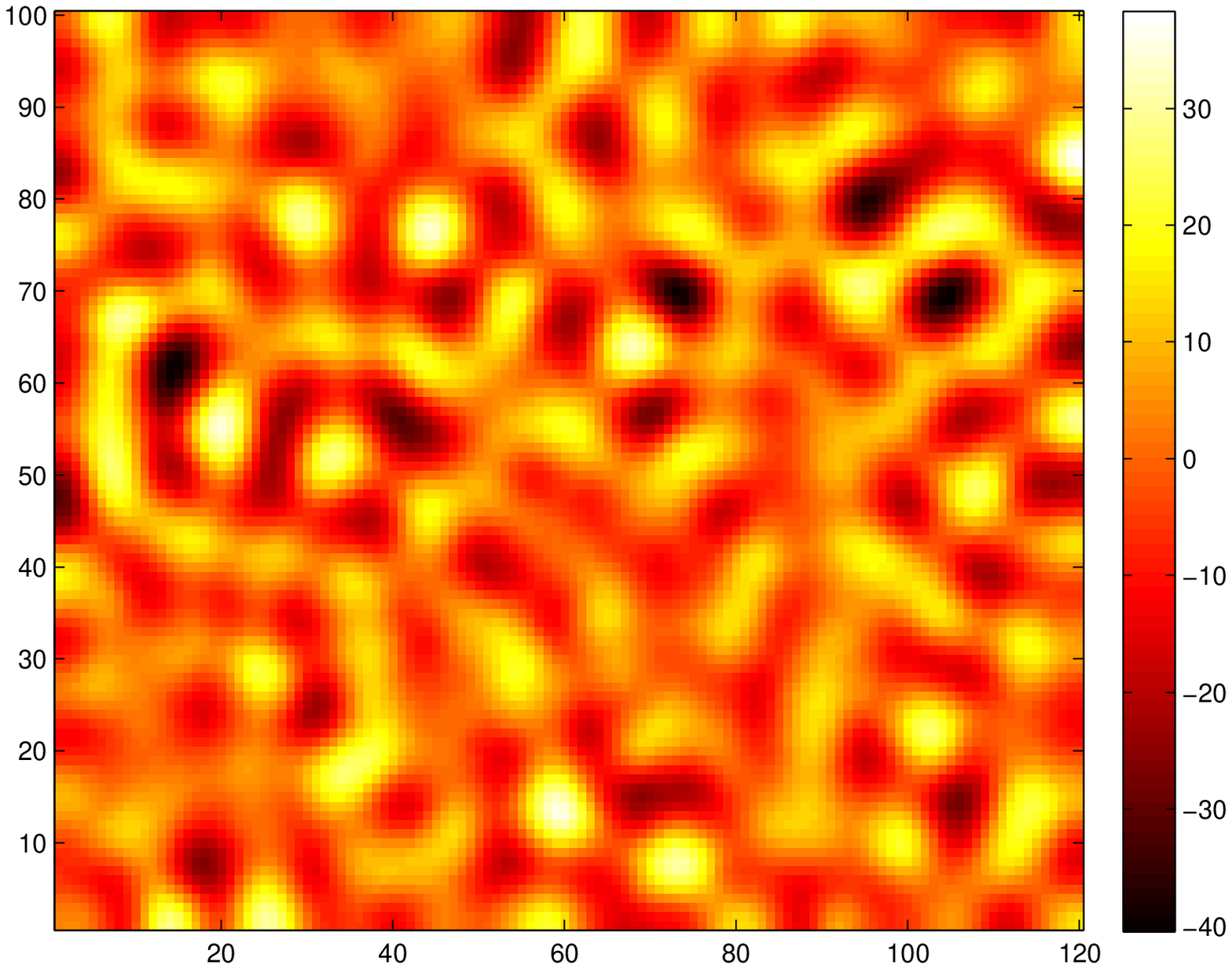} \label{fig: osci_estimated_case2_field1}} 
    \subfigure[]{\includegraphics[width=0.48\textwidth,height=0.48\textwidth]{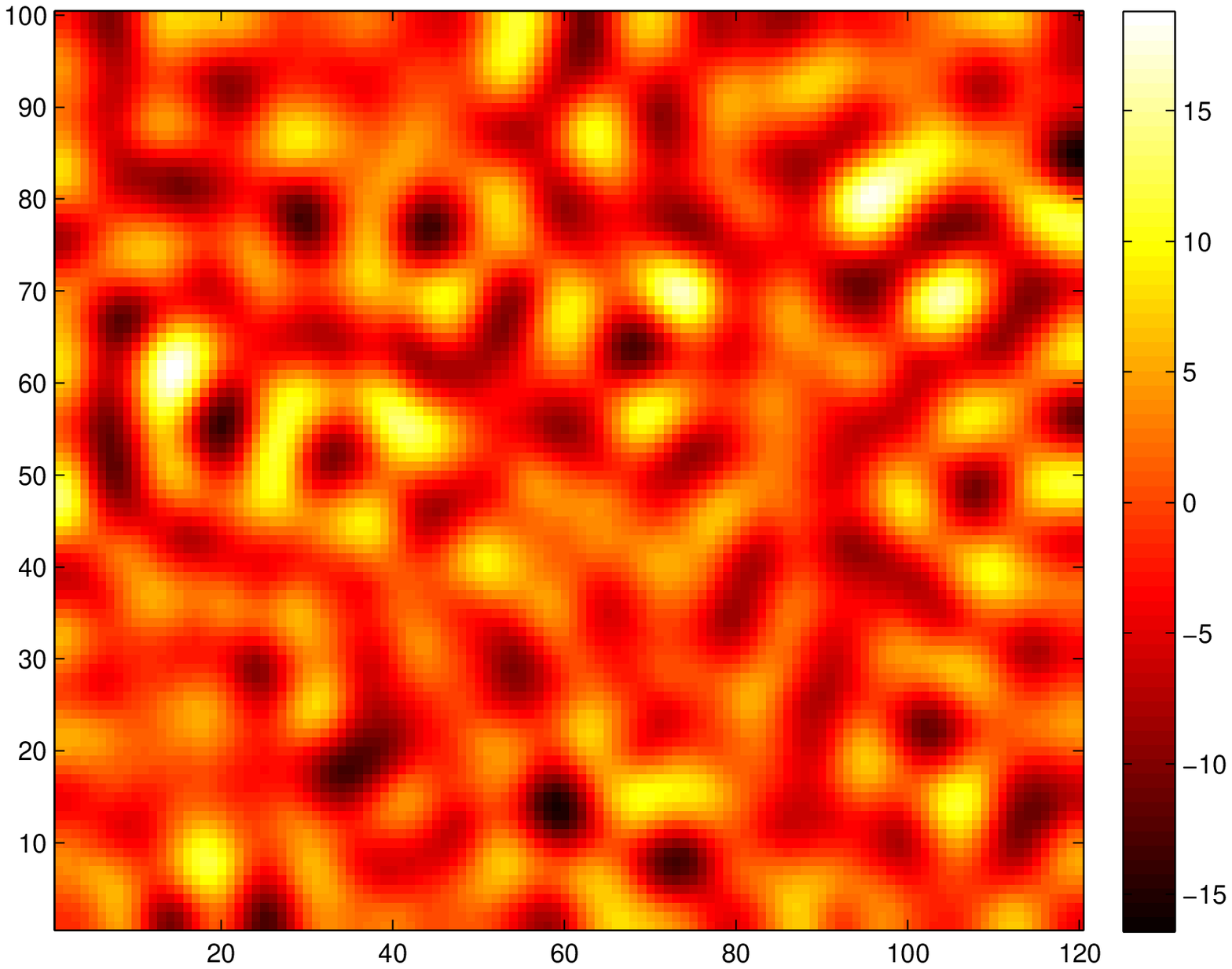} \label{fig: osci_estimated_case2_field2}} \\
 %   \subfigure[]{\includegraphics[width=0.8\textwidth,height=0.6\textwidth]{correlation_oscillation_case1} \label{fig: osci_sampling_cov1}}
    \caption{Estimated conditional mean of the bivariate random field for the dataset 2.} 
\label{fig: osci_bivariate_oscillation_estimated}
 \end{figure}

The results for the first and second simulated datasets are given in Table \ref{tab: simulated_result_1} and Table \ref{tab: simulated_result_2}, respectively. 
We can notice that the estimates are quite precise. Most of the true values are
within $1$ standard derivation away from the estimates. None of the true values are $2$ standard deviations away from the estimates.  
The estimated conditional mean of the bivariate fields for these two datasets are given in 
Fig. \ref{fig: osci_bivariate_matern_oscillation_estimated} and Fig. \ref{fig: osci_bivariate_oscillation_estimated}. Compare with the true random fields given in 
Fig. \ref{fig: osci_sampling_case1_field1} - Fig. \ref{fig: osci_sampling_case1_field2} and Fig. \ref{fig: osci_sampling_case2_field1} - Fig. \ref{fig: osci_sampling_case2_field2}. There is no large difference between them.

\begin{table}
\centering
\caption{Inference for the simulated dataset $1$}
\begin{tabular}{c|c|c|c}
  \hline
  \hline
Parameters     &   True values   &  Estimates      & Standard deviations \\
\hline
$b_{11}$  &     0.5       &   0.495         &     0.013    \\
$b_{21}$  &     0.25      &   0.248         &     0.017    \\
$b_{22}$  &     1         &   1.027         &     0.032    \\ 
$h_{11}$  &     0.25      &   0.248         &     0.010    \\
$h_{22}$  &     0.36      &   0.355         &     0.029    \\
$\kappa_{n_2}$   &   0.6  &   0.601         &     0.004     \\
$\omega$         &  0.95  &   0.953         &     0.092     \\     
\hline
 \end{tabular}
 \label{tab: simulated_result_1} 
\end{table}
 
\begin{table}
\centering
\caption{Inference for the simulated dataset $2$}
\begin{tabular}{c|c|c|c}
  \hline
  \hline
Parameters     &   True values   &  Estimates      & Standard deviations \\
\hline
$b_{11}$  &     0.5       &   0.497         &     0.014    \\
$b_{21}$  &     0.25      &   0.234         &     0.012    \\
$b_{22}$  &     1         &   0.964         &     0.029    \\ 
$h_{11}$  &     0.25      &   0.269         &     0.024    \\
$h_{22}$  &     0.36      &   0.339         &     0.022    \\
$\kappa_{n_1}$   &   0.5  &   0.496         &     0.005     \\
$\kappa_{n_2}$   &   0.6  &   0.636         &     0.049      \\
$\omega$         &  0.95  &   0.956         &     0.113      \\     
\hline
 \end{tabular}
 \label{tab: simulated_result_2} 
\end{table}

Similarly, the results for the second group are given in Table \ref{tab: parametes_simudata_3} and Table \ref{tab: parameters_simudata_4}. In both examples,
the estimates are precise and they are within $2$ standard derivations from the true values. We can notice that if the fields are independent, i.e., $b_{21} = 0$, our model captures this characteristic
since $b_{21}$ is small and $0$ is within the $95\%$  credible interval.

\begin{table}
\centering
\caption{Inference for the simulated dataset $3$}
\begin{tabular}{c|c|c|c}
  \hline
  \hline
Parameters     &   True value   &  Estimated      & Standard deviations \\
\hline
$b_{11}$  &     0.5       &   0.491               &    0.012     \\
$b_{21}$  &     0         &   0.012               &    0.010     \\
$b_{22}$  &     0.3       &   0.301               &    0.010     \\ 
$h_{11}$  &     0.25      &   0.247               &    0.009     \\
$h_{22}$  &     0.36      &   0.374               &    0.033     \\
$\kappa_{n_2}$   &   0.6  &   0.596               &    0.004      \\
$\omega$         &  0.95  &   0.951               &    0.092      \\     
\hline
 \end{tabular}
 \label{tab: parametes_simudata_3} 
\end{table}

\begin{table}
\centering
\caption{Inference for the simulated dataset $4$}
\begin{tabular}{c|c|c|c}
  \hline
  \hline
Parameters     &   True value   &  Estimated      & Standard deviations \\
\hline
$b_{11}$  &     0.5       &   0.487               &     0.015     \\
$b_{21}$  &     0         &   0.001               &     0.002    \\
$b_{22}$  &     0.3       &   0.308               &     0.009    \\ 
$h_{11}$  &     0.25      &   0.284               &     0.026    \\
$h_{22}$  &     0.36      &   0.359               &     0.122    \\
$\kappa_{n_1}$   &  0.5   &   0.502               &     0.004     \\
$\kappa_{n_2}$   &  0.6   &   0.599               &     0.102     \\
$\omega$         &  0.95  &   0.949               &     0.107     \\     
\hline
 \end{tabular}
 \label{tab: parameters_simudata_4} 
\end{table}

\begin{figure}[tbp]
    \centering
    \subfigure[]{\includegraphics[width=0.48\textwidth,height=0.48\textwidth]{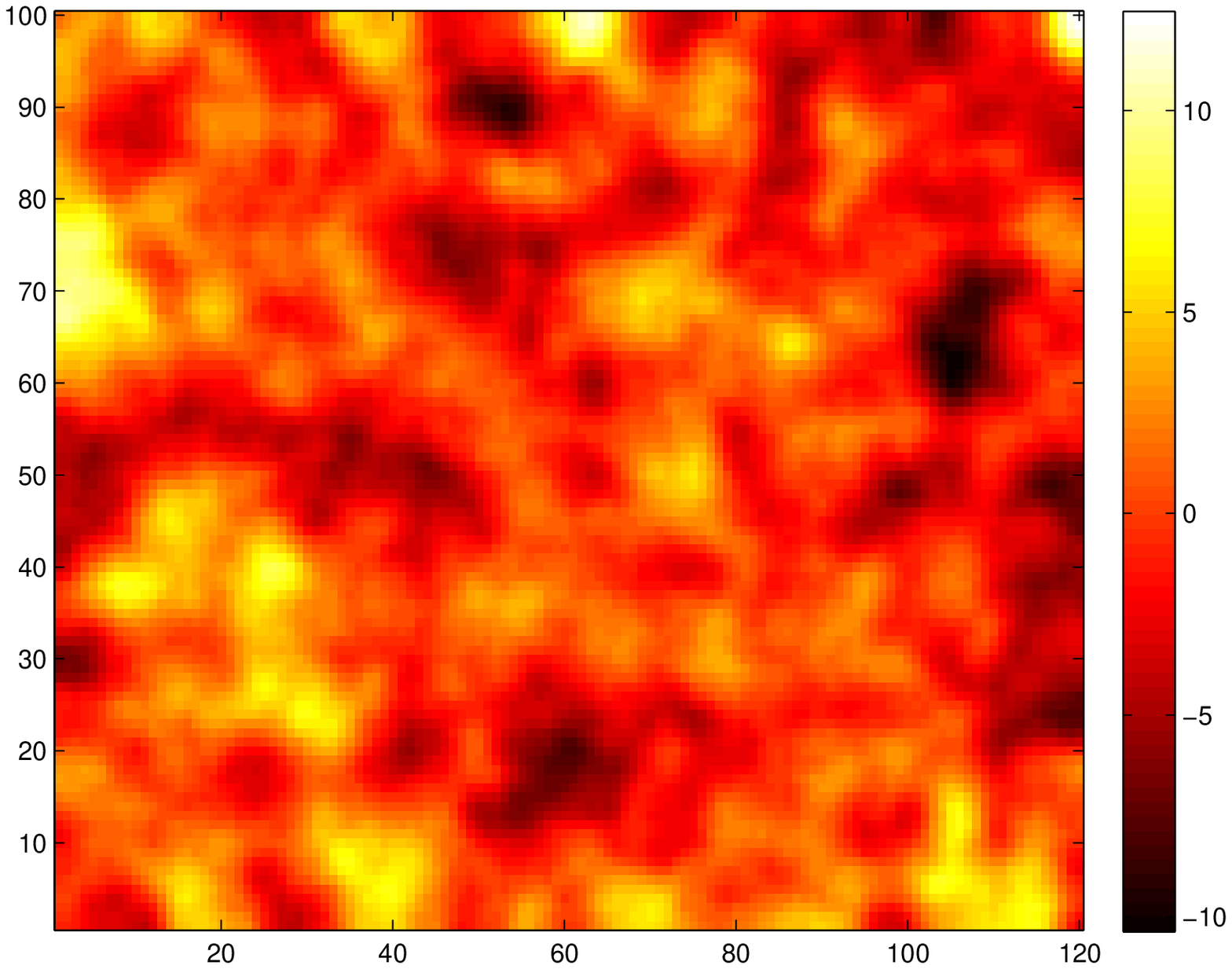}} 
    \subfigure[]{\includegraphics[width=0.48\textwidth,height=0.48\textwidth]{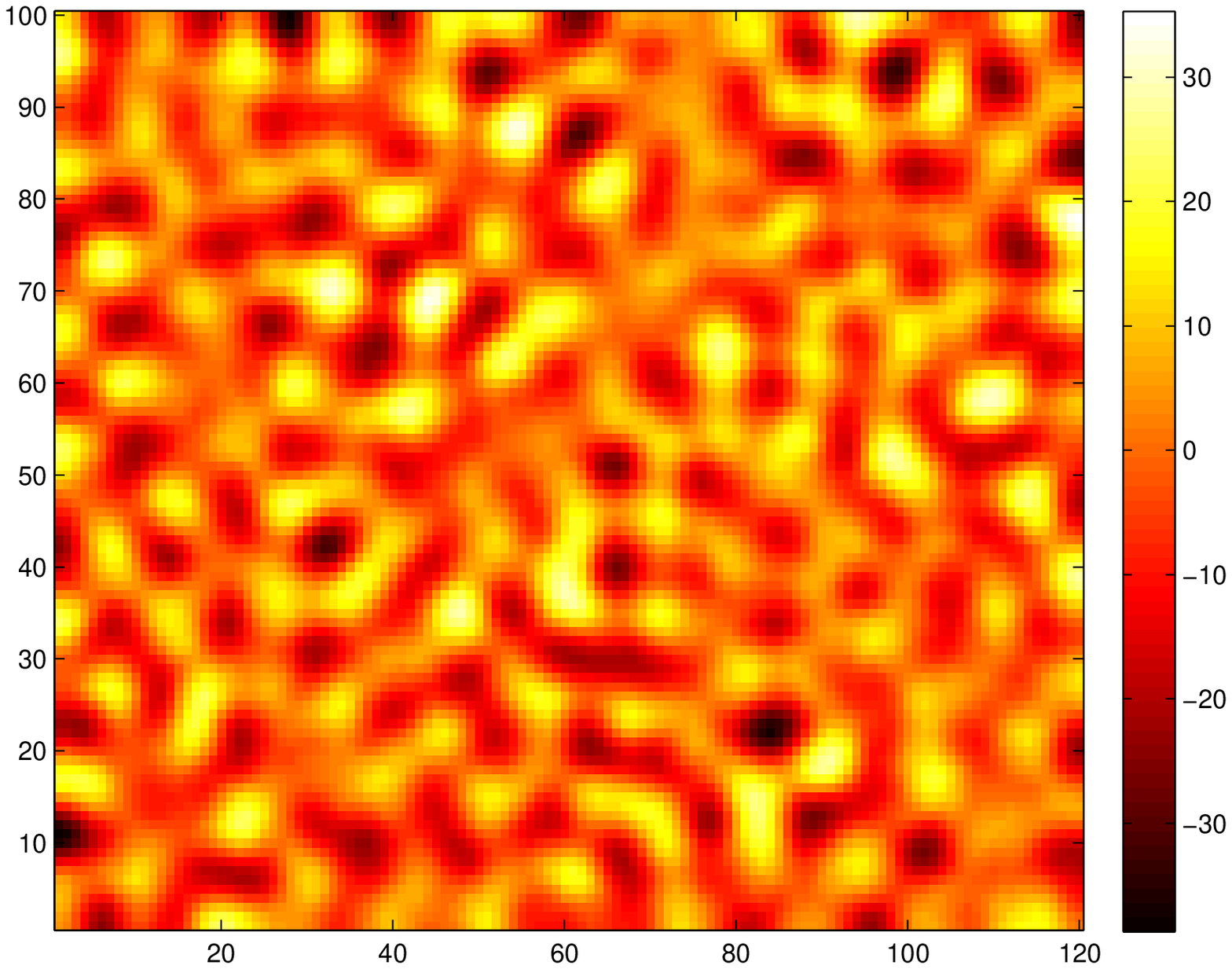}}
    \caption{One realization of the bivariate random field with parameters given in Table \ref{tab: parametes_simudata_3}. The first field random field has a non-oscillating covariance function and the second field has an oscillating 
             covariance. The two random fields are independent.} 
\label{fig: osci_bivariate_matern_oscillation_independent_sample}
 \end{figure}

 \begin{figure}[tbp]
    \centering
    \includegraphics[width=0.8\textwidth,height=0.6\textwidth]{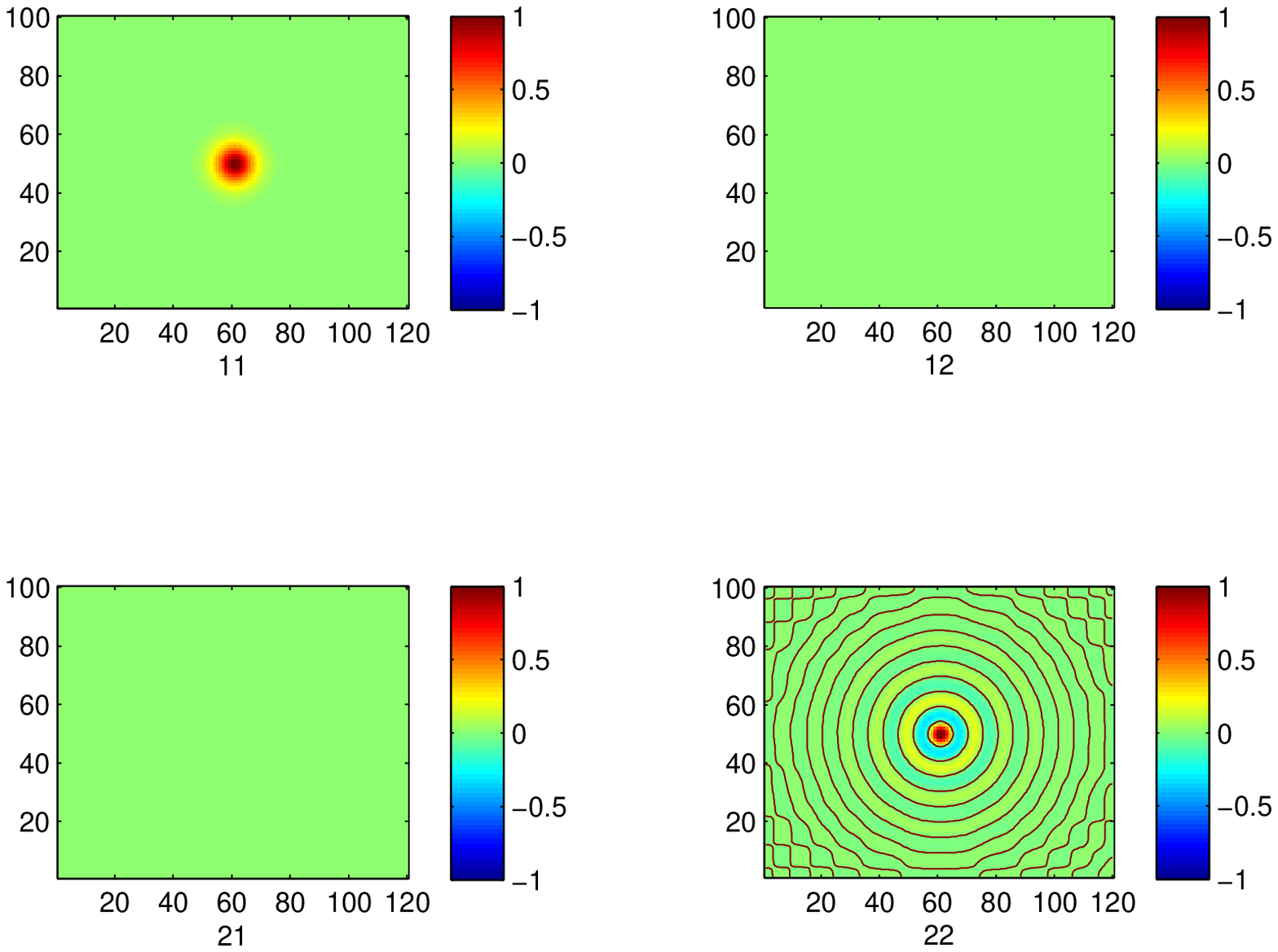}
    \caption{The correlation and cross correlation functions with parameters given in Table \ref{tab: parametes_simudata_3}. The first field random field has a non-oscillating covariance function and the second field has an oscillating 
             covariance. The two random fields are independent.} 
\label{fig: osci_sampling_cov3}
 \end{figure}
 
  \begin{figure}[tbp]
    \centering
    \includegraphics[width=0.8\textwidth,height=0.6\textwidth]{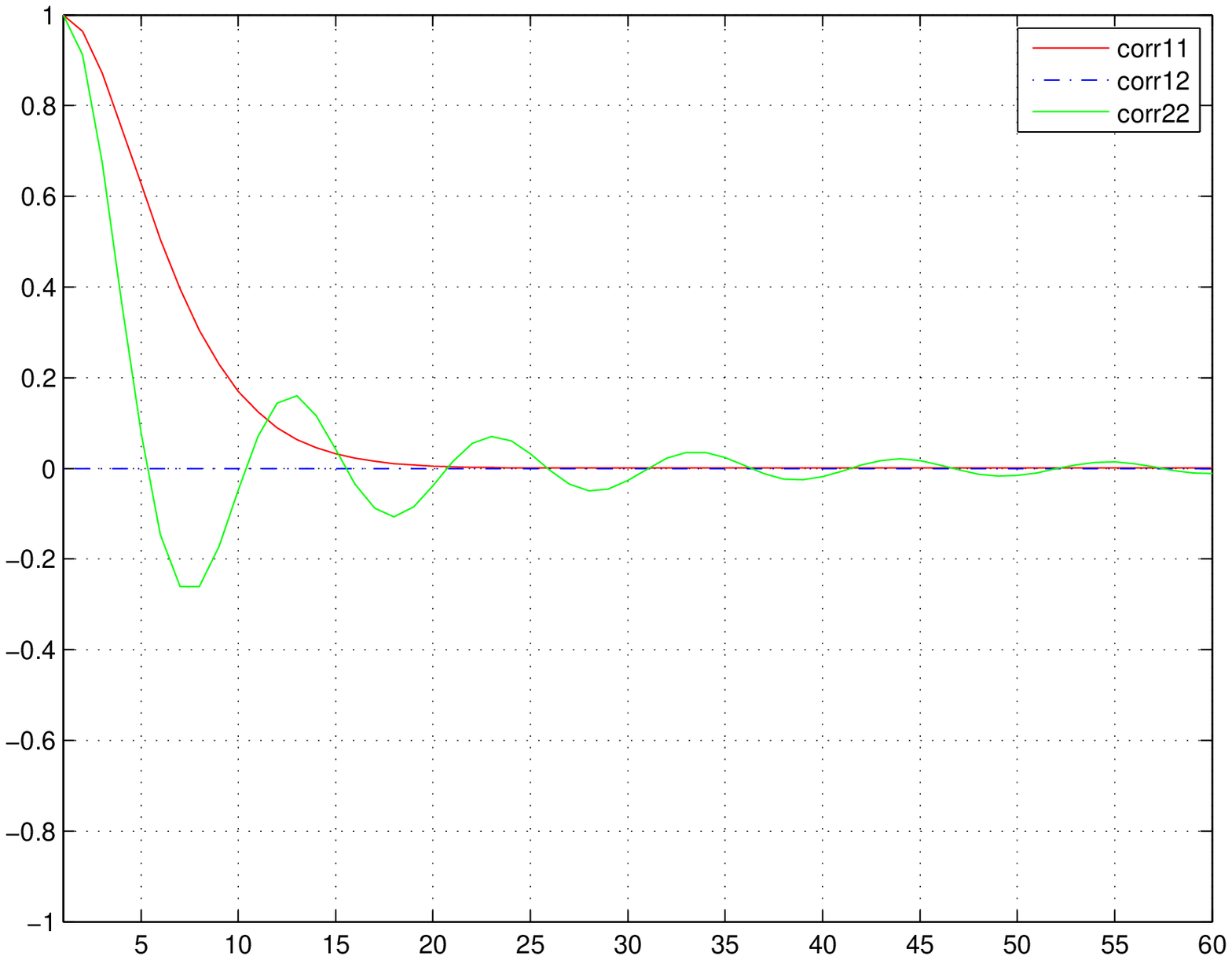}
    \caption{The correlation and cross correlation functions with parameters given in Table \ref{tab: parametes_simudata_3}. The first field random field has a non-oscillating covariance function and the second field has an oscillating 
             covariance. The two random fields are independent.} 
\label{fig: osci_sampling_cov3_1d}
 \end{figure}

\begin{figure}[tbp]
    \centering
    \subfigure[]{\includegraphics[width=0.48\textwidth,height=0.48\textwidth]{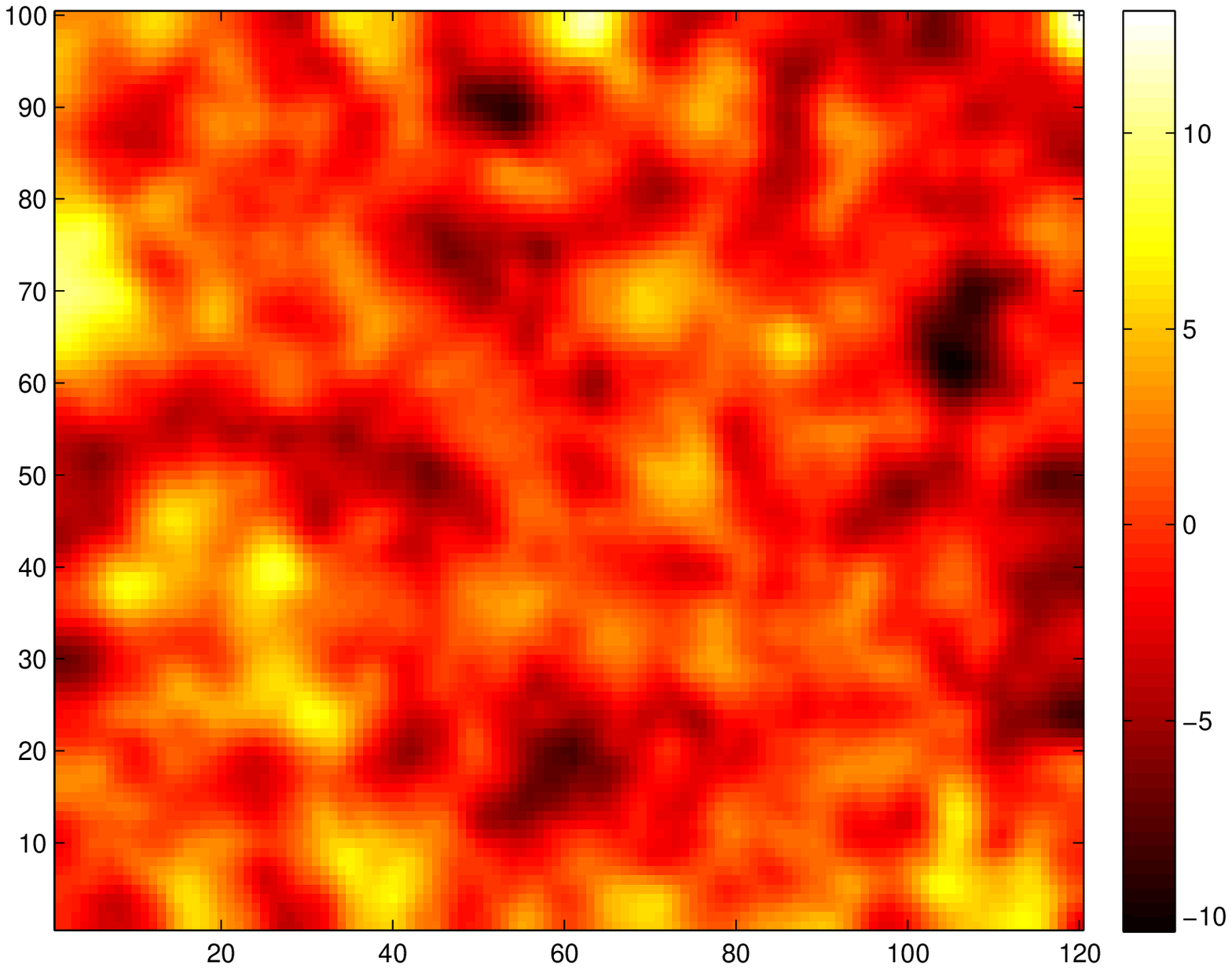}} 
    \subfigure[]{\includegraphics[width=0.48\textwidth,height=0.48\textwidth]{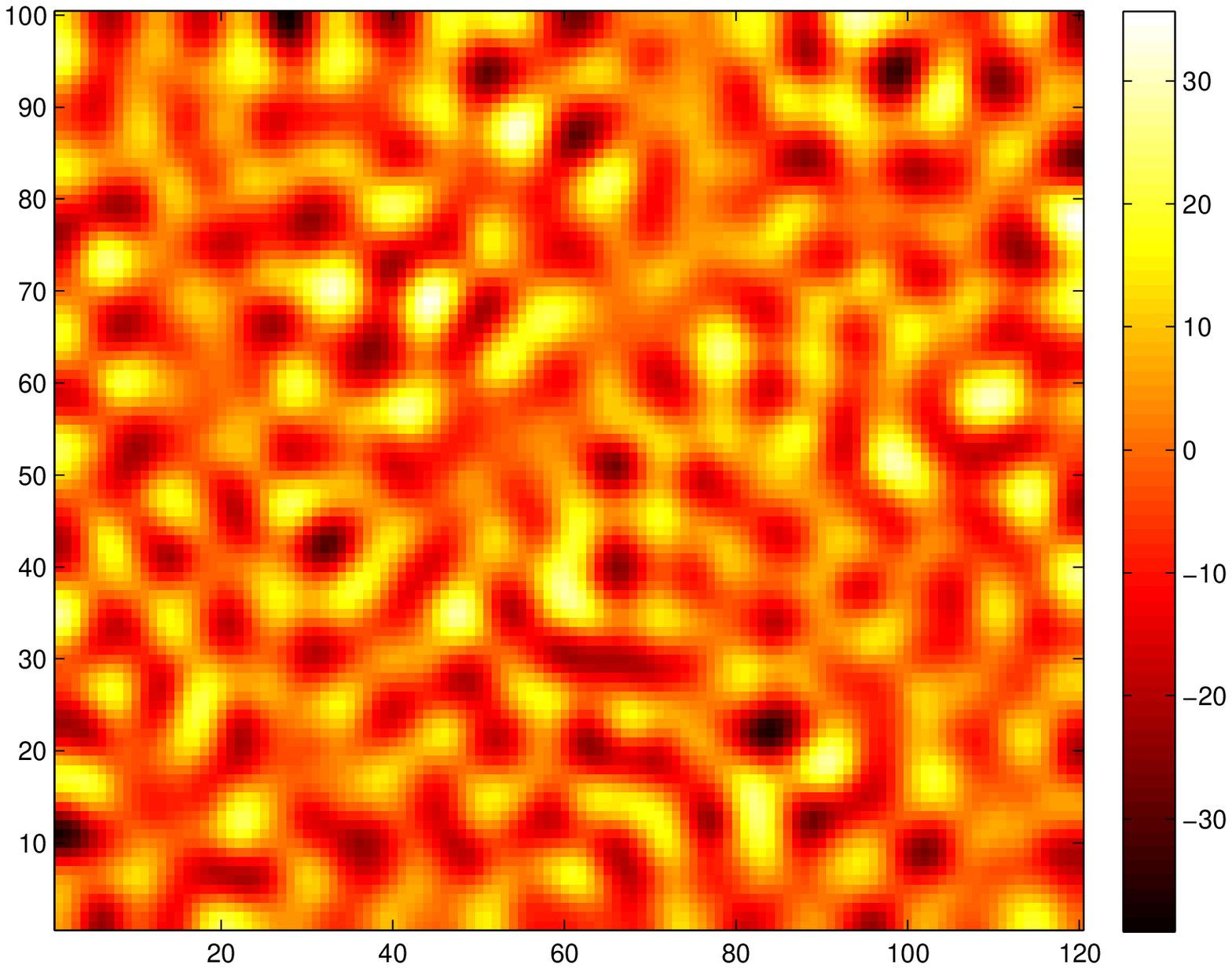}}
    \caption{Estimated conditional mean of the bivariate random fields for dataset $3$.} 
\label{fig: osci_bivariate_matern_oscillation_independent_estimated}
 \end{figure}

\begin{figure}[t]
    \centering
    \subfigure[]{\includegraphics[width=0.48\textwidth,height=0.48\textwidth]{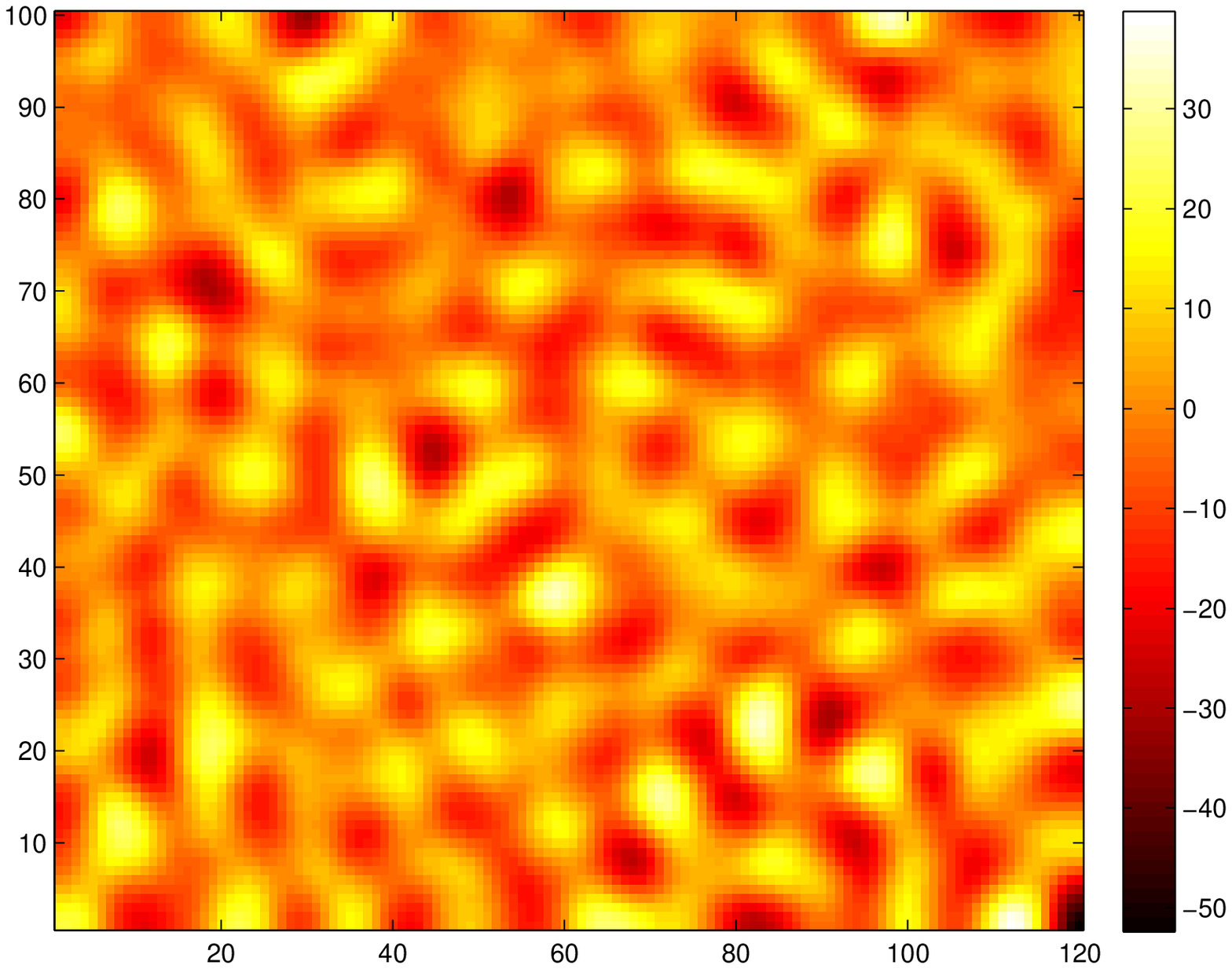}} 
    \subfigure[]{\includegraphics[width=0.48\textwidth,height=0.48\textwidth]{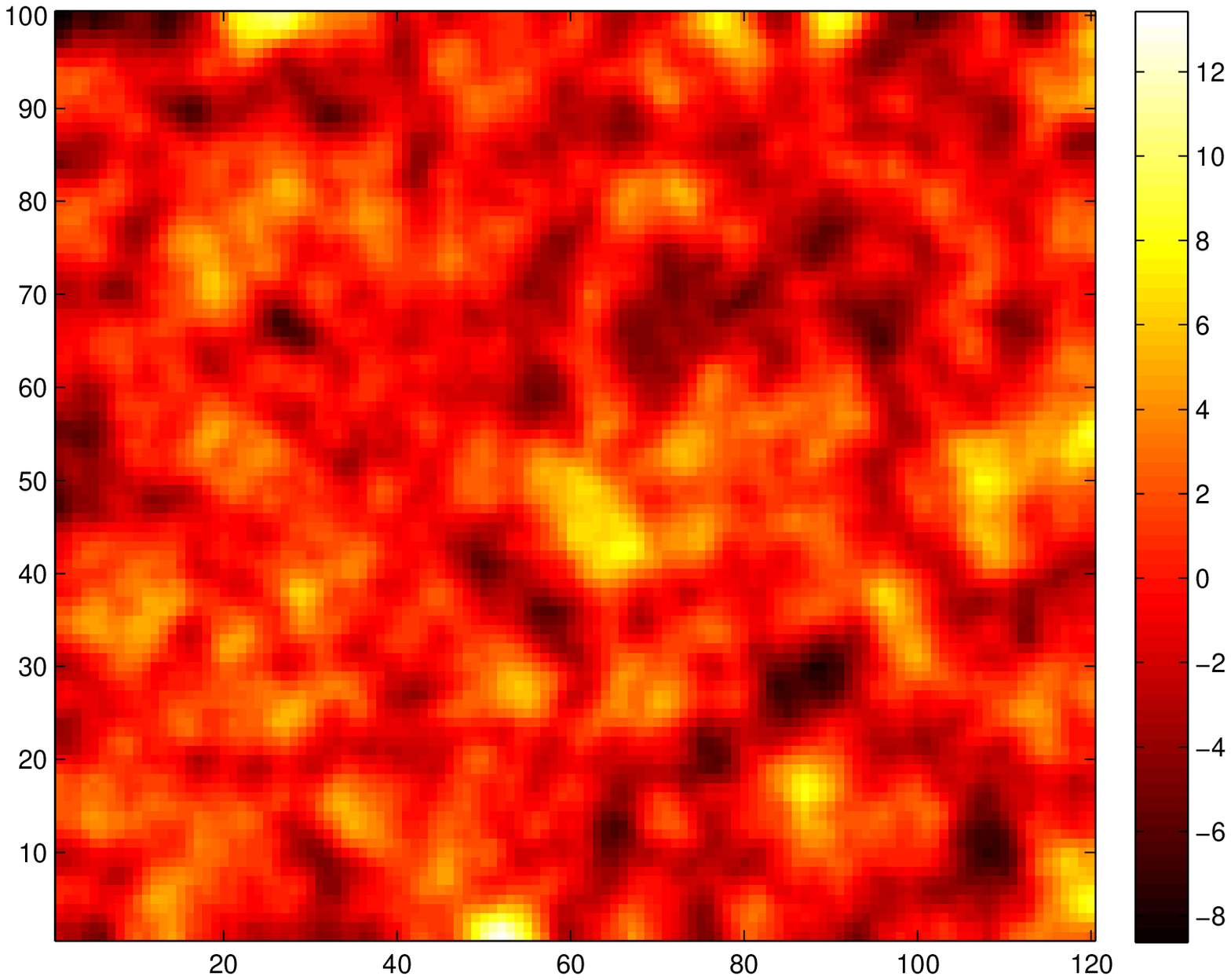}} 
    \caption{One realization of the bivariate random fields with parameters given in Table \ref{tab: parameters_simudata_4}. The field random field has an oscillating covariance function and the second field has a non-oscillating covariance.
             The two random fields are independent.} 
\label{fig: osci_bivariate_oscillation_independent_sample}
 \end{figure}
 
 \begin{figure}[tbp]
    \centering
   \includegraphics[width=0.8\textwidth,height=0.6\textwidth]{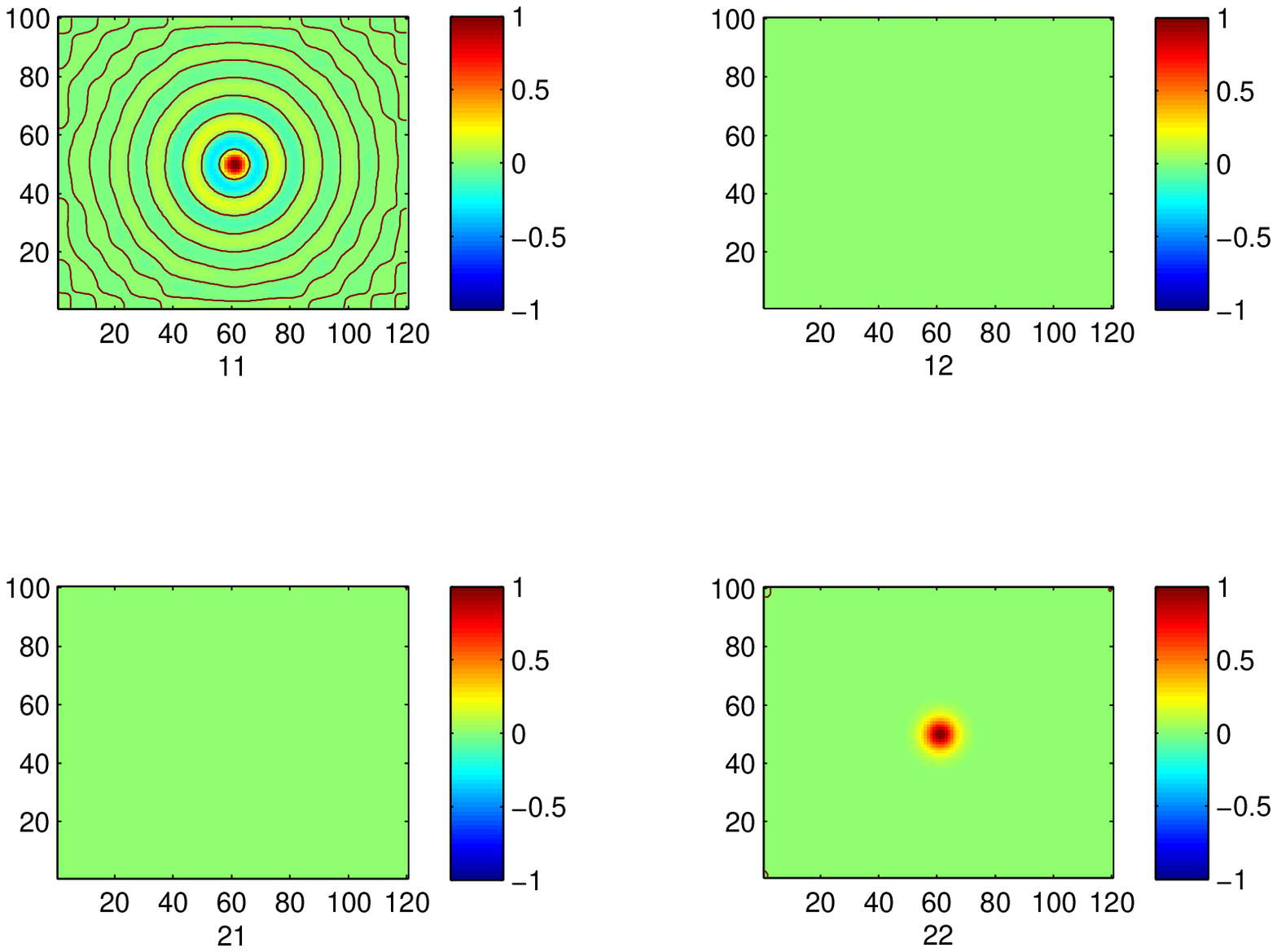}
    \caption{The correlation functions and cross correlation functions with parameters given in Table \ref{tab: parameters_simudata_4}. The field random field has an oscillating covariance function and the second field has a non-oscillating covariance.
             The two random fields are independent.} 
\label{fig: osci_sampling_cov4}
 \end{figure}

  \begin{figure}[tbp]
    \centering
   \includegraphics[width=0.8\textwidth,height=0.6\textwidth]{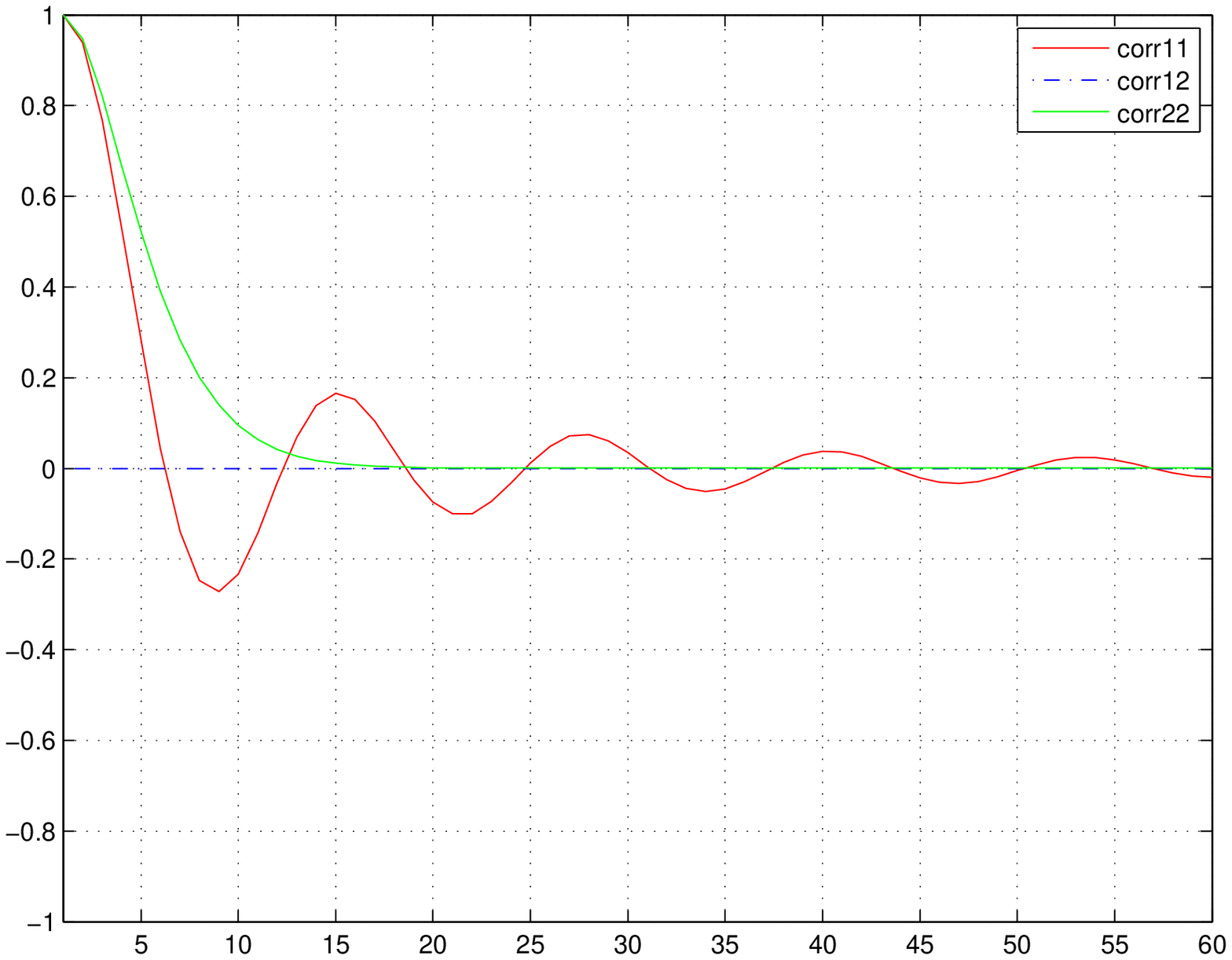}
    \caption{The correlation functions and cross correlation functions with parameters given in Table \ref{tab: parameters_simudata_4}. The field random field has an oscillating covariance function and the second field has a non-oscillating covariance.
             The two random fields are independent.} 
\label{fig: osci_sampling_cov4_1d}
 \end{figure}

\begin{figure}[tbp]
    \centering
    \subfigure[]{\includegraphics[width=0.48\textwidth,height=0.48\textwidth]{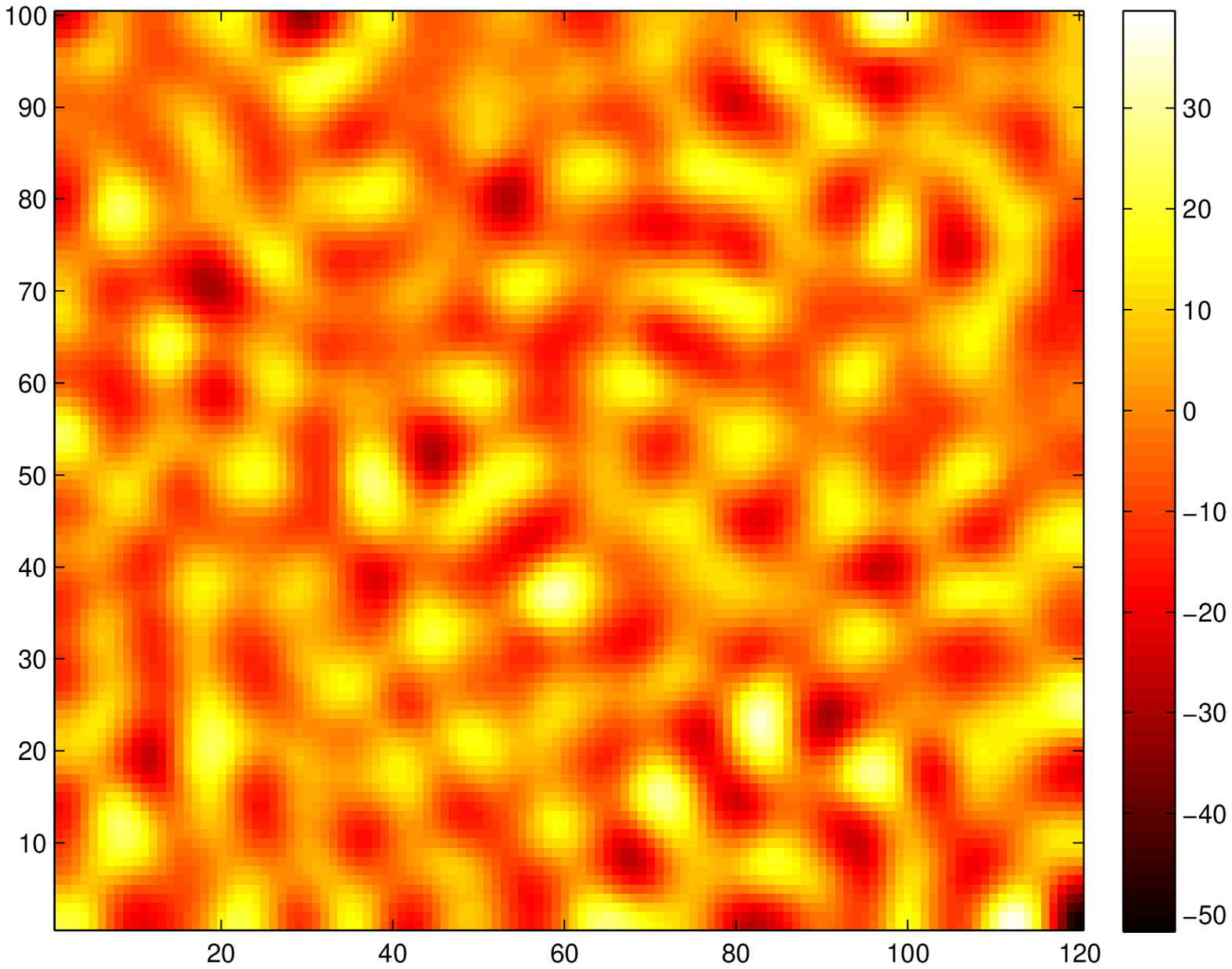}} 
    \subfigure[]{\includegraphics[width=0.48\textwidth,height=0.48\textwidth]{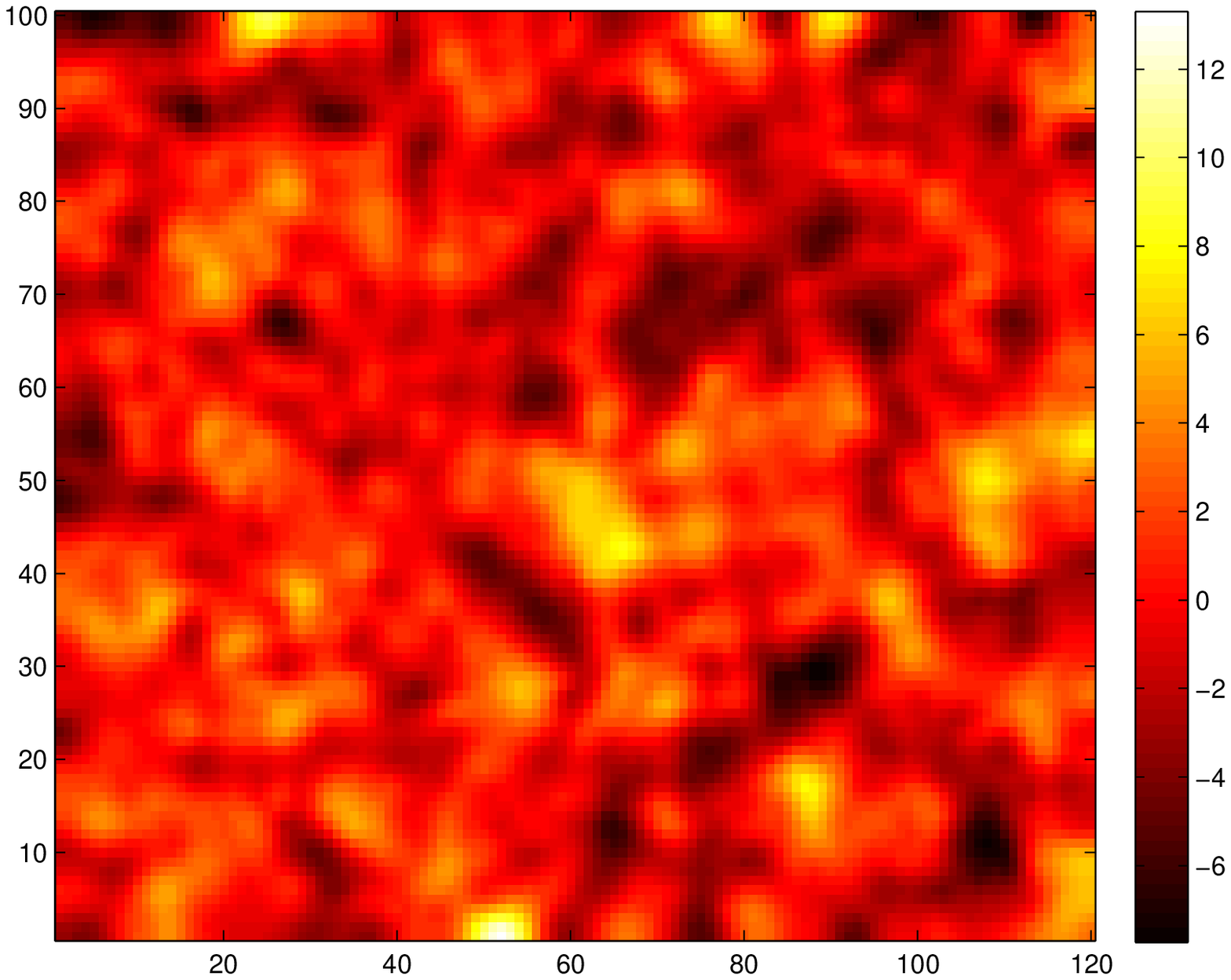}} 
    \caption{Estimated conditional mean of the bivariate random fields for dataset $4$.} 
\label{fig: osci_bivariate_oscillation_independent_estimated}
 \end{figure}

\subsection{Inference with real data}
In this section a dataset has been chosen in order to illustrate how to use our approach in real-world applications. This dataset is from the ERA $40$ database and can be downloaded from the ERA $40$ project homepage. 
The dataset contains the temperature and pressure on the whole globe on $4$th of September, $2002$. The main objective for this section is to illustrate how to use our model for a big dataset.
 All the results are only from the prediction point of view.  The dataset contains $10368$ observations both for temperature and pressure, and the observations are on the grid. The grid is
constructed with the latitude and longitude. The latitudes are from $90^\text{o}$ to $-90^\text{o}$ and longitudes are from $0^\text{o}$ to $357.5^\text{o}$, with increments of $2.5^\text{o}$ for both axes. The dataset
contains the temperatures in Kelvin and the mean sea level pressure in Pascal. We have subtracted the monthly mean for the temperature and pressure, respectively. 

Since the dataset is on the entire globe, we need to construct our model on the sphere.
\citet{jones1963stochastic} discussed how to construct stochastic processes on a sphere using the spectral representations for spherically symmetric and the axially symmetric cases. Another 
approach is to consider the sphere as a surface in $\mathbb{R}^3$. However, this has the disadvantage that the correlation between points are determined by the chordal distances \citep{lindgren2011explicit}.  
\citet{gneiting1998simple} pointed out that the random fields constructed on the plane were not suitable for this kind of dataset since the great circle distances in the original covariance function would not work in general.
\citet{jun2007approach} discussed an approach for constructing space-time covariance functions on spheres using a sum of independent processes. The main idea for their approach is to sum independent processes where each process
is obtained by applying the first-order differential operations to a fully symmetric processes on sphere $\times$ time. We refer to \citet{jun2007approach} for more information on the fully symmetric processes on sphere $\times$ time.

In this paper we follow the approach discussed by \citet{lindgren2011explicit} to construct the GRFs on the sphere. 
They claimed that using the SPDE approach for constructing GRFs on the sphere is similar to constructing the GRFs on $\mathbb{R}^d$. 
By reinterpreting the SPDE defined on $\mathbb{S}^2$,  the solutions of the SPDE are GRFs defined on $\mathbb{S}^2$. Our proposed system of SPDEs approach inherits this property.
The only place has been changed is that the system of SPDEs is directly defined on $\mathbb{S}^2$. Another advantage of our approach is that the GMRF approximation can still be used.
In other words, we can use GMRFs to represent GRFs for computation. For more information about GRFs on manifolds, we refer to \citet[Section $3.1$]{lindgren2011explicit}.

\begin{figure}[tbp]
    \centering
    \subfigure[]{\includegraphics[width=0.48\textwidth,height=0.48\textwidth]{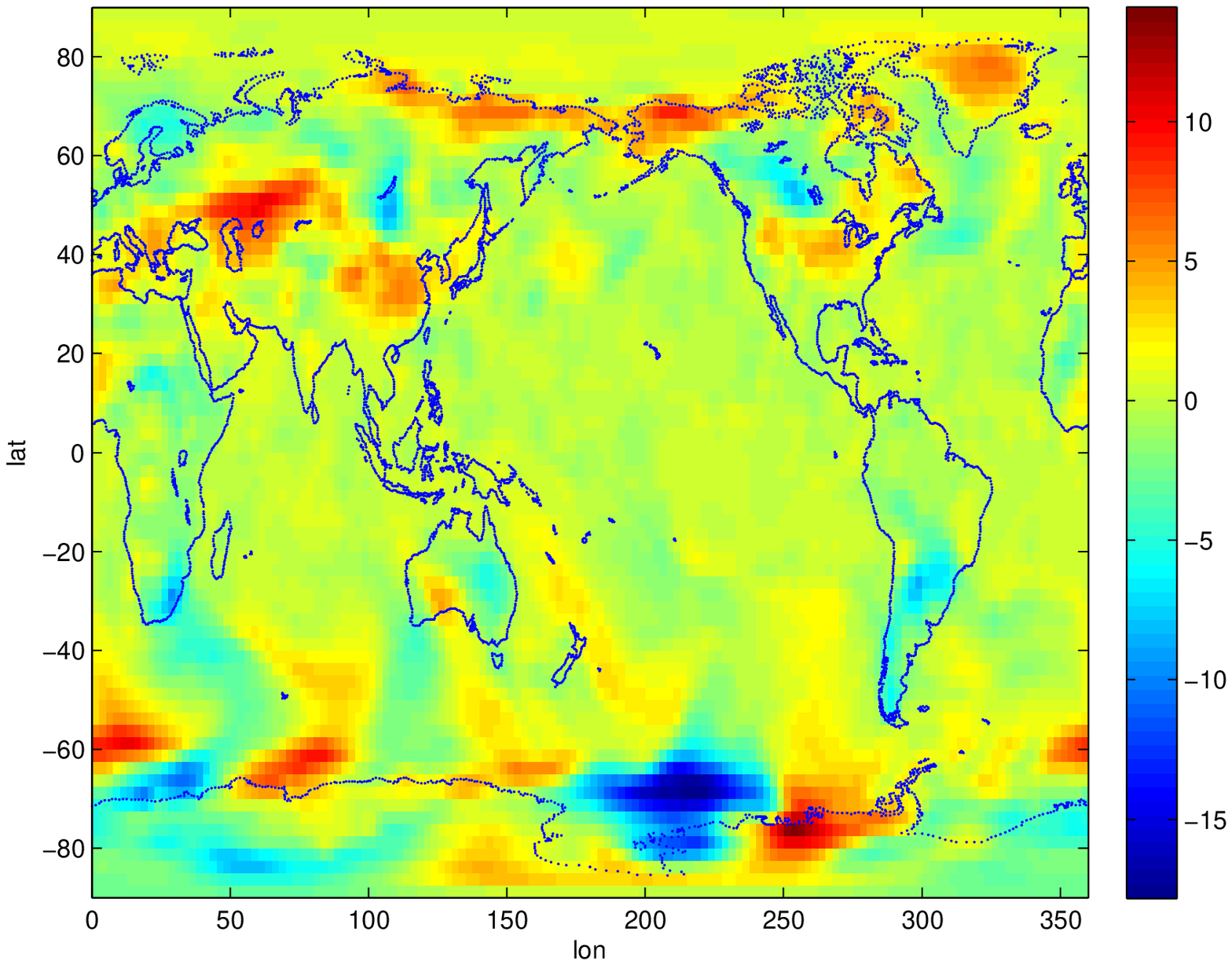} \label{fig: osci_real_data_2d_temp}} 
    \subfigure[]{\includegraphics[width=0.48\textwidth,height=0.48\textwidth]{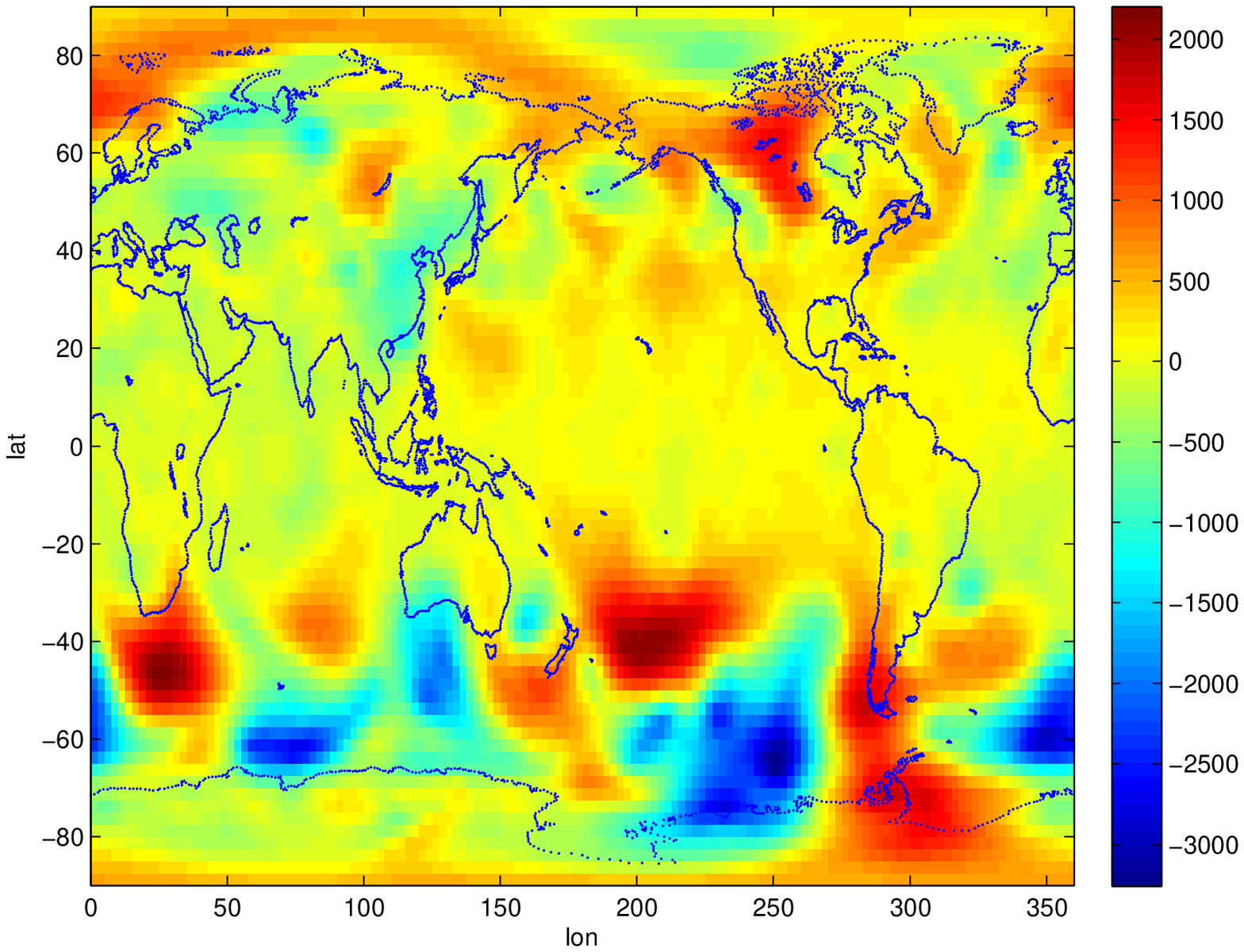} \label{fig: osci_real_data_2d_pres}} 
 %   \subfigure[]{\includegraphics[width=0.8\textwidth,height=0.6\textwidth]{correlation_oscillation_case1} \label{fig: osci_sampling_cov1}}
    \caption{Dataset with temperature (a) and pressure (b) from ERA $40$ database.} 
\label{fig: osci_real_data_2d}
 \end{figure}

\begin{figure}[tbp]
    \centering
    \subfigure[]{\includegraphics[width=0.48\textwidth,height=0.48\textwidth]{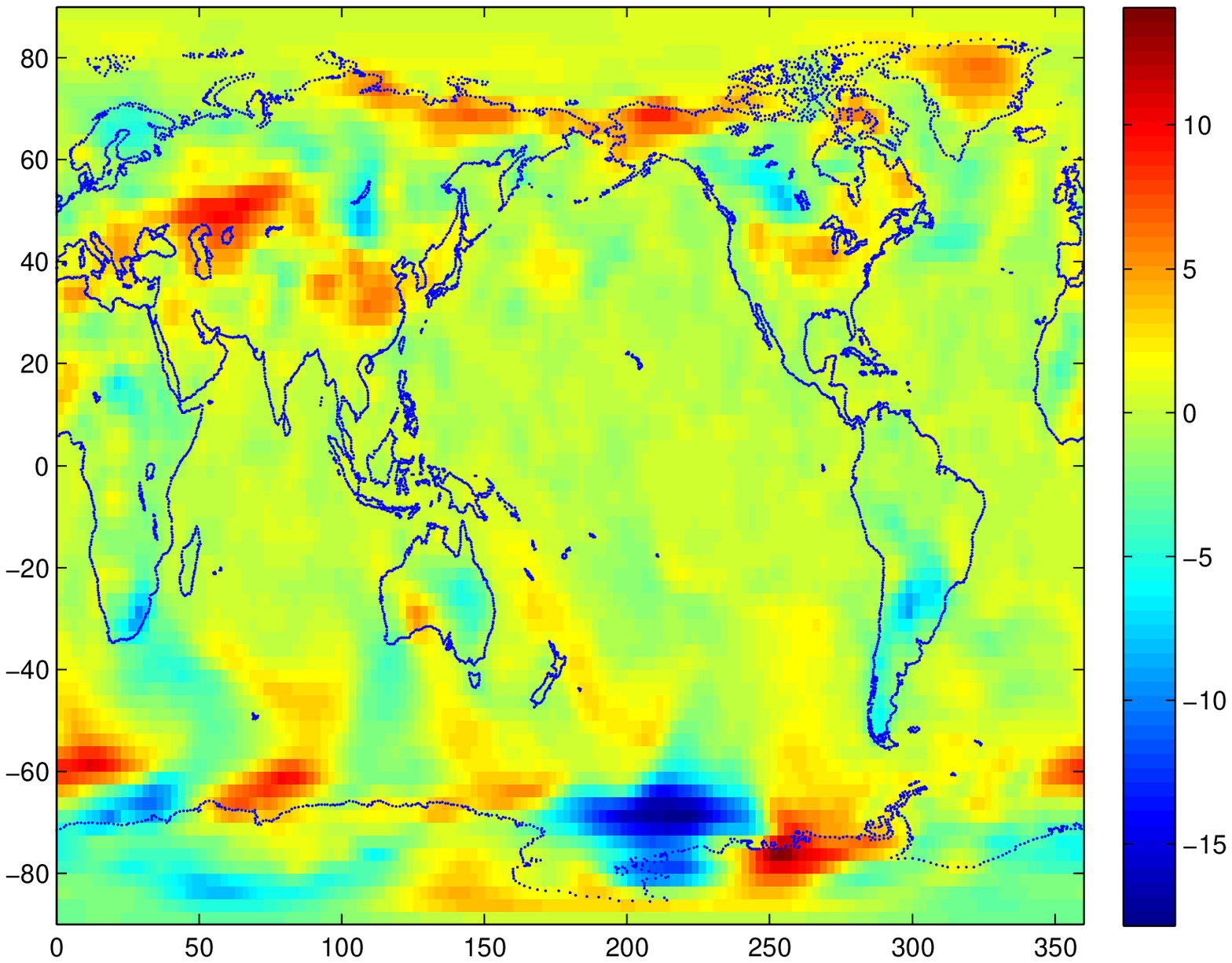} \label{fig: osci_real_data_2d_temp_estimated}} 
    \subfigure[]{\includegraphics[width=0.48\textwidth,height=0.48\textwidth]{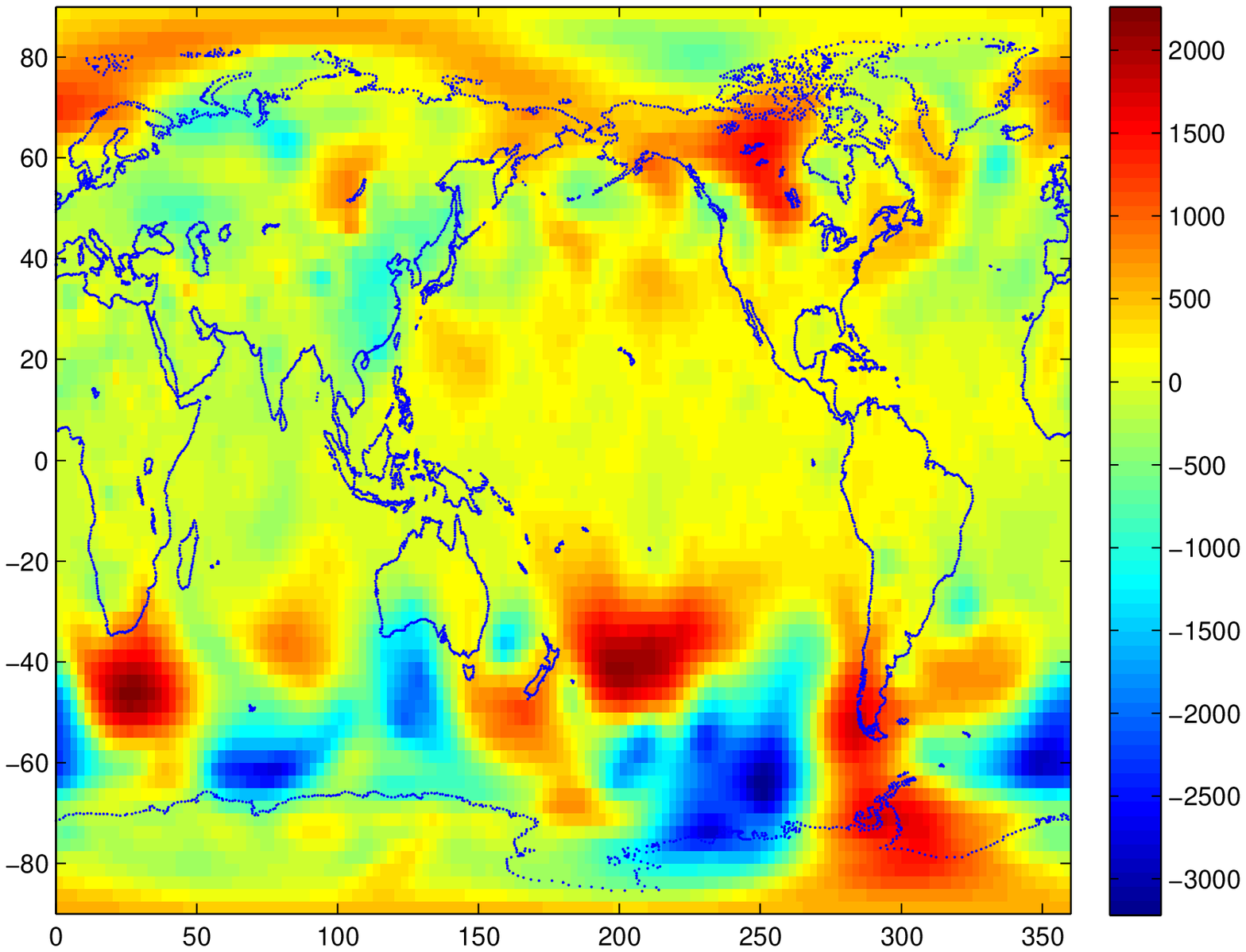} \label{fig: osci_real_data_2d_pres_estimated}}
    \caption{Reconstructed bivariate random fields for temperature (a) and pressure (b).} 
\label{fig: osci_real_data_estimated_2d}
 \end{figure}

\begin{figure}[tbp]
    \centering
    \subfigure[]{\includegraphics[width=0.58\textwidth,height=0.58\textwidth]{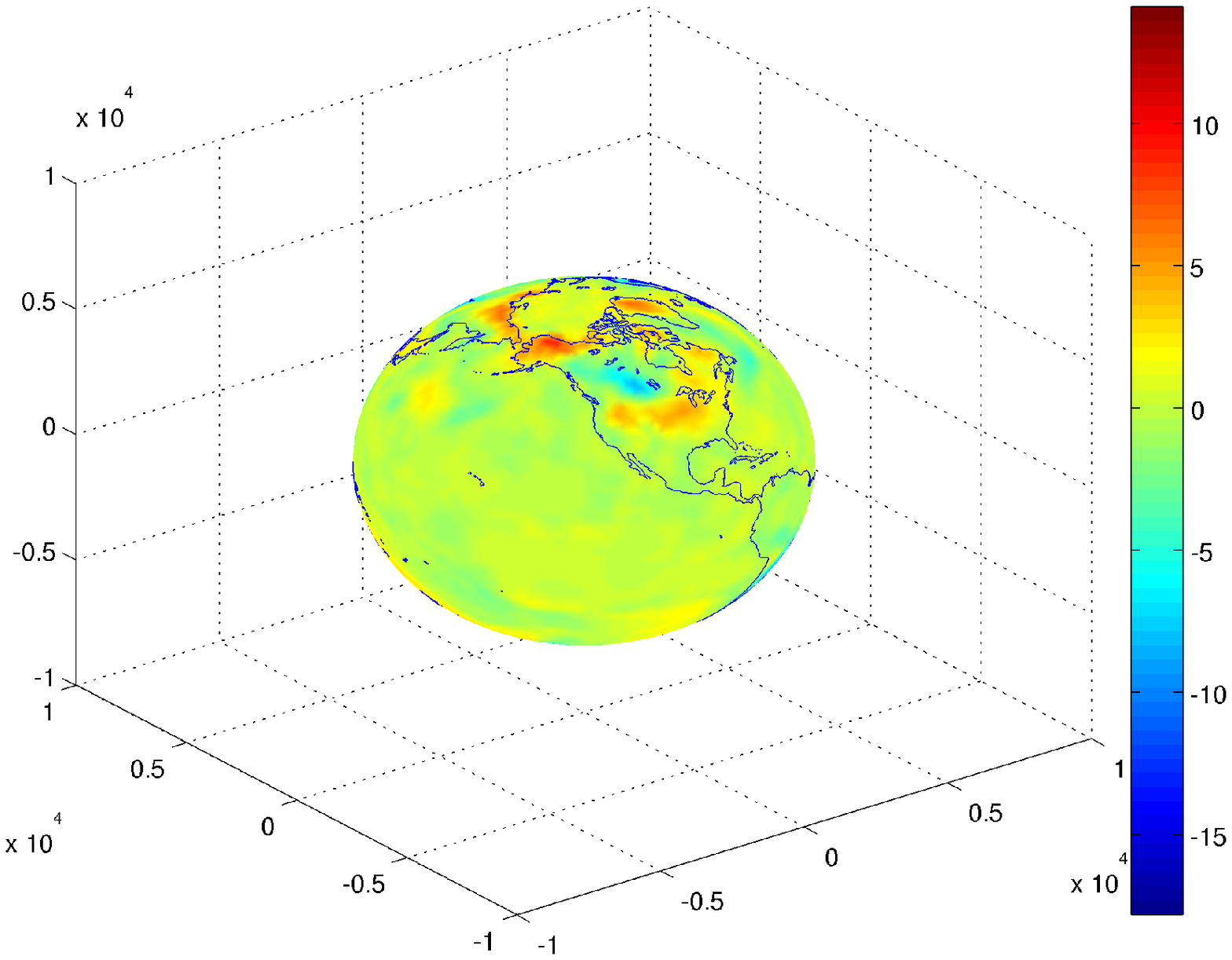} \label{fig: osci_real_data_3d_temp}} 
    \subfigure[]{\includegraphics[width=0.58\textwidth,height=0.58\textwidth]{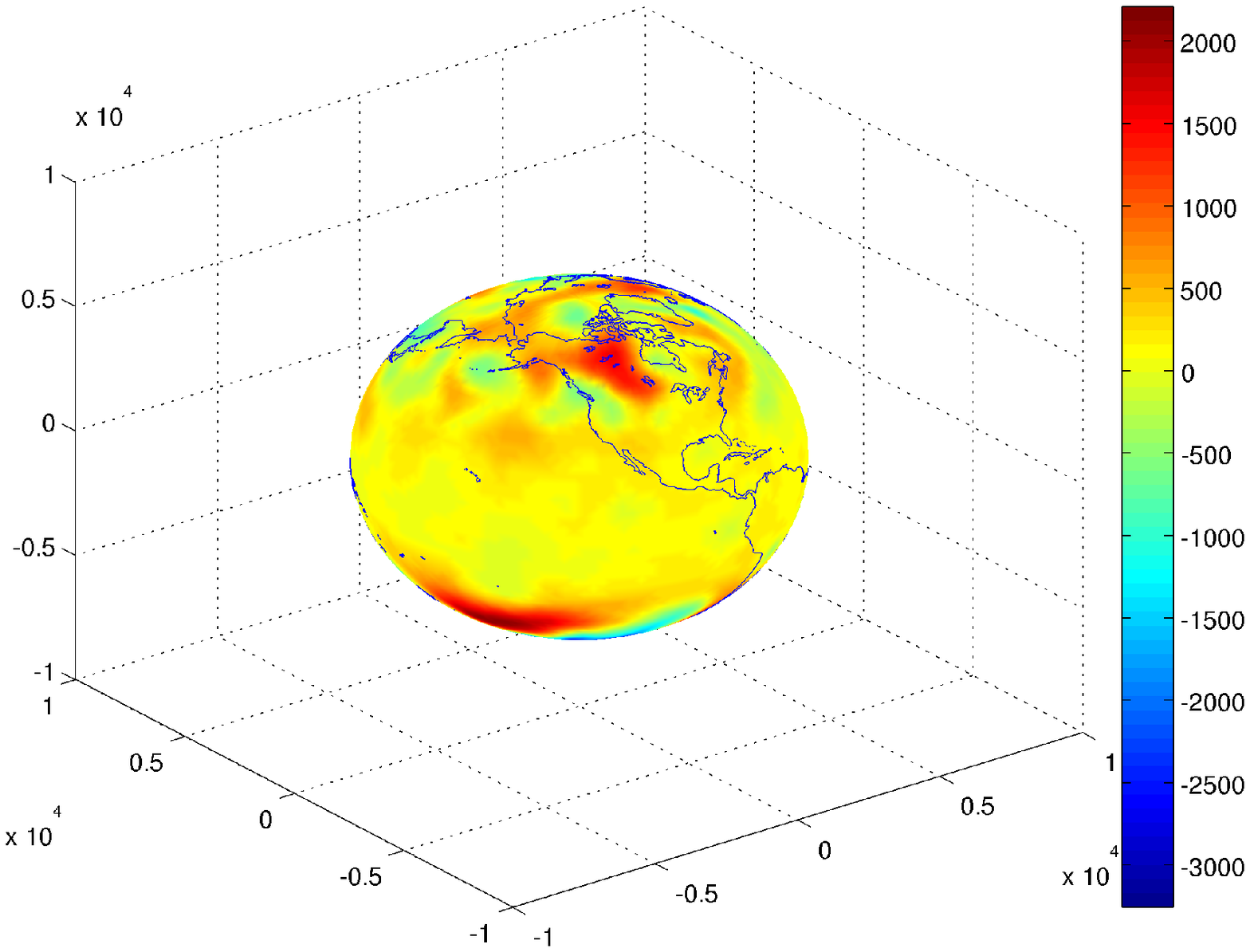} \label{fig: osci_real_data_3d_pres}} 
    \caption{Dataset with temperature (a) and pressure (b) on the sphere from ERA $40$ database.} 
\label{fig: osci_real_data_3d}
 \end{figure}

\begin{figure}[tbp]
    \centering
    \subfigure[]{\includegraphics[width=0.58\textwidth,height=0.58\textwidth]{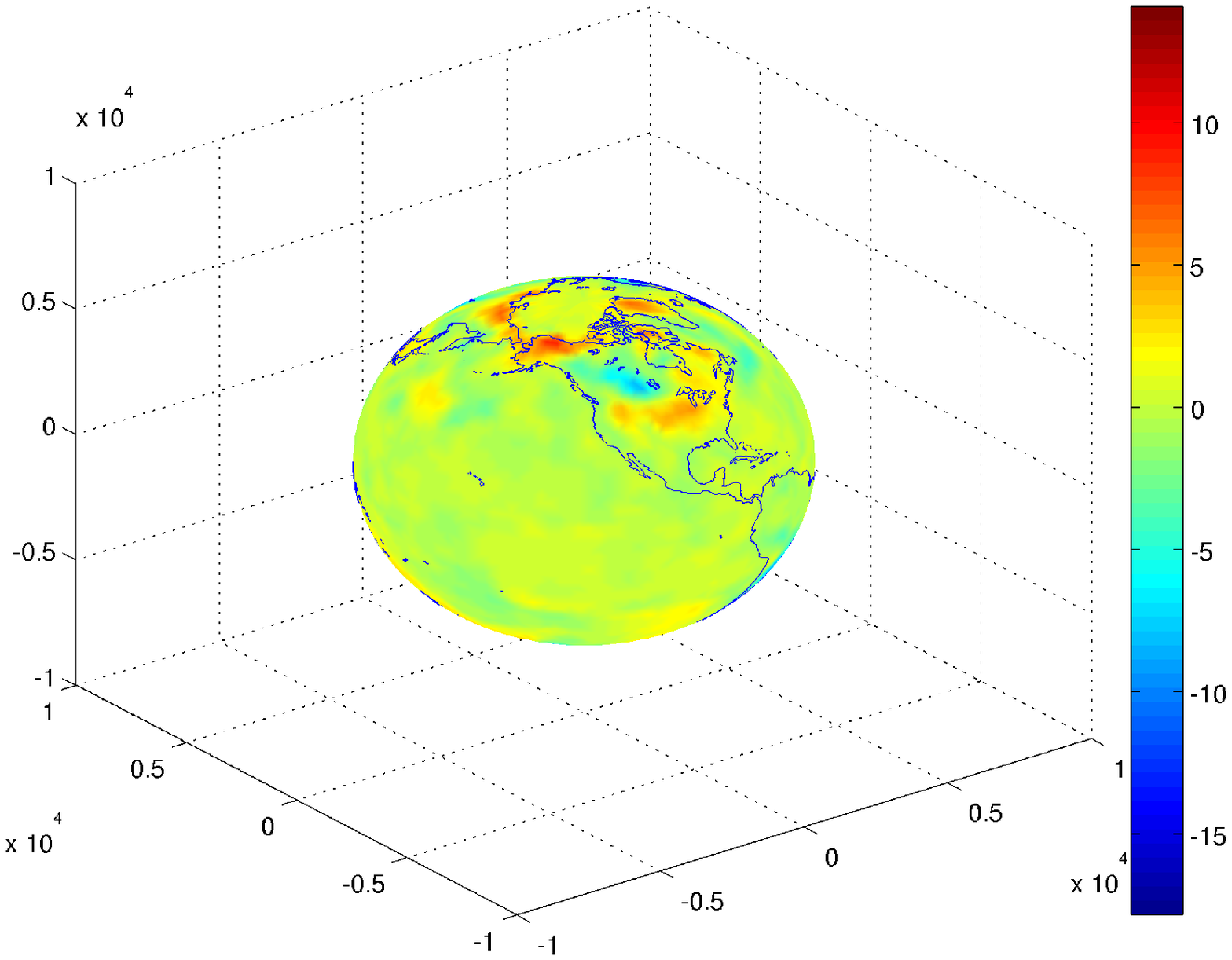} \label{fig: osci_real_data_3d_temp_estimated}} 
    \subfigure[]{\includegraphics[width=0.58\textwidth,height=0.58\textwidth]{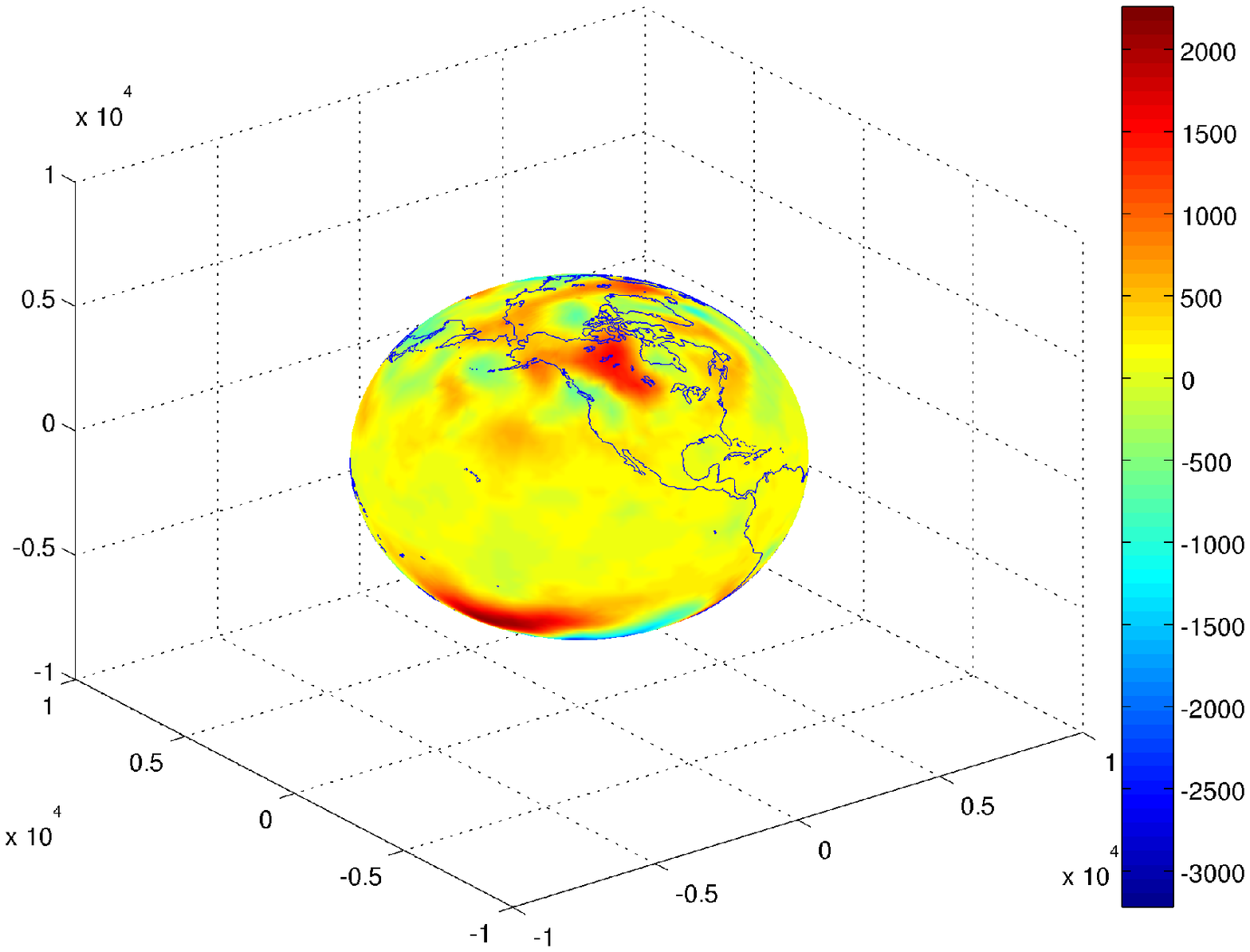} \label{fig: osci_real_data_3d_pres_estimated}} 
    \caption{Reconstructed bivariate random fields for temperature (a) and pressure (b) on the sphere.} 
\label{fig: osci_real_data_estimated_3d}
 \end{figure}

Since it is known that the pressure on the globe has an oscillating covariance function, it is reasonable to set $x_1(\boldsymbol{s})$ as the temperature and $x_2(\boldsymbol{s})$ as the pressure, and let the second noise process
$\varepsilon_2(\boldsymbol{s})$ have an oscillating covariance function, but not the first noise process $\varepsilon_1(\boldsymbol{s})$.
The original dataset is shown in Fig. \ref{fig: osci_real_data_2d_temp} and Fig. \ref{fig: osci_real_data_2d_pres}, and the reconstructed temperature and pressure are shown 
in Fig. \ref{fig: osci_real_data_2d_temp_estimated} and Fig. \ref{fig: osci_real_data_2d_pres_estimated}. 
We also give $3D$ images for the true datasets on the sphere in Fig. \ref{fig: osci_real_data_3d_temp} and Fig. \ref{fig: osci_real_data_3d_pres}, and the reconstructed 
temperature and pressure on the sphere in Fig. \ref{fig: osci_real_data_3d_temp_estimated} and Fig. \ref{fig: osci_real_data_3d_pres_estimated}. 
One thing we want to point out is that we follow the methodology given in \citet{lindgren2011explicit} and construct the GRF on the unit radius sphere $\mathbb{S}^2$. 
Another important point is that we set $\kappa_{n_1}^2 = h_{11}$ to simplify the model and inference. 

\begin{figure}[tbp]
    \centering
     \includegraphics[width=0.8\textwidth,height=0.8\textwidth]{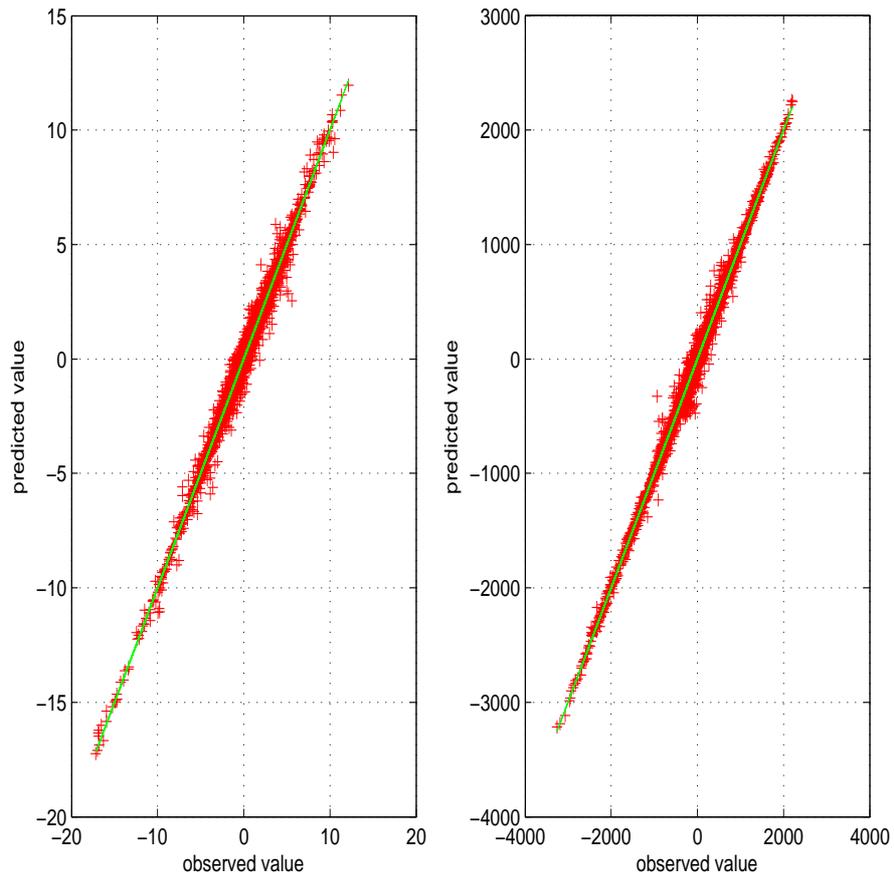} 
    \caption{Prediction for the leaving out $5000$ observations for temperature (left) and pressure (right)} 
\label{fig: osci_real_data_prediction}
 \end{figure}

\begin{figure}[tbp]
    \centering
     \includegraphics[width=0.8\textwidth,height=0.55\textwidth]{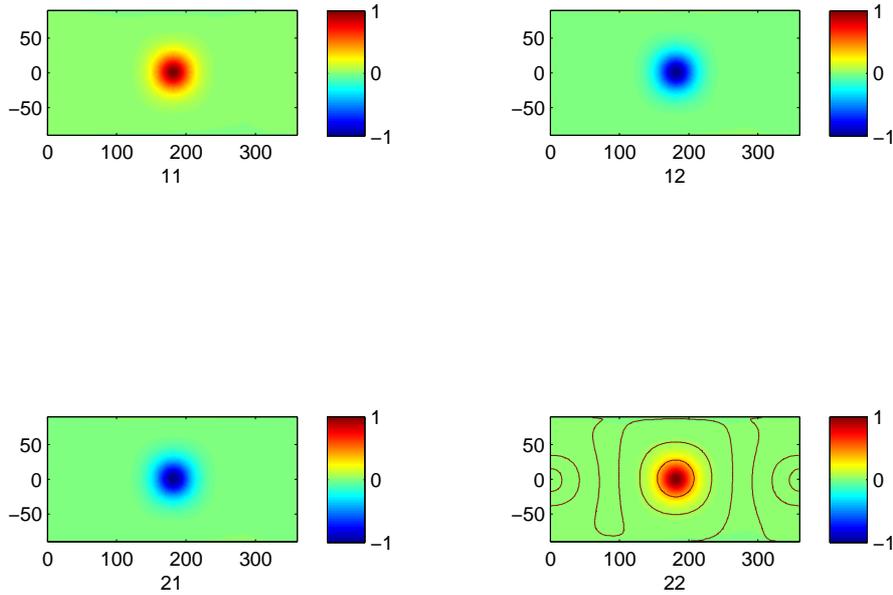} 
    \caption{Correlation functions for the bivariate random fields. $11$ indicates the marginal correlation of the temperature. $12$ and $21$ indicate the cross-correlation between the temperature and pressure. 
             $22$ indicates the marginal correlation of the pressure.} 
\label{fig: osci_real_data_correlation}
 \end{figure}

In order to check the predictive performance of our approach, we have divided the dataset into two subsets. We used a subset containing $5368$ observations for both temperature and pressure for estimating the parameters
and predict the remaining $5000$ observations. The estimates are given in Table \ref{tab: osci_real_data_parameters}.
From the results we notice that the model captures the empirical knowledge that the temperature and pressure are negative correlated since $b_{21} > 0$.
The prediction for the $5000$ observations are given in Fig. \ref{fig: osci_real_data_prediction}. From a prediction point of view, the model works well since most of the prediction are close to the true observed values.
The correlation functions are given in Fig. \ref{fig: osci_real_data_correlation}, and we notice that the covariance function of pressure indeed has oscillation behavior.  

\begin{table}
\centering
\caption{Inference for real dataset}
\begin{tabular}{c|c}
  \hline
  \hline
Parameters     &    Estimated                    \\
\hline
$b_{11}$       &    $8.952 \times 10^5$          \\
$b_{21}$       &       $1.546$                   \\
$b_{22}$       &    $4.714 \times 10^1$          \\ 
$h_{11}$       &    $1.224 \times 10^{-6}$       \\
$h_{22}$       &    $8.089 \times 10^{-6}$       \\
$\kappa_{n_2}$ &    $1.013 \times 10^{-3}$       \\
$\omega$       &      $0.819$                    \\     
\hline
 \end{tabular}
 \label{tab: osci_real_data_parameters} 
\end{table}

\section{Discussion and future work} \label{sec: osci_discussion}
Due to the increasing importance of spatial statistics in applications, new approaches for handling different complex datasets are demanded. The methodologies for dealing with multivariate datasets appear in many areas, 
such as in air quality \citep{brown1994multivariate, schmidt2003bayesian}, weather forecasting \citep{courtier1998ecmwf,reich2007multivariate}, and economics \citep{gelfand2004nonstationary, sain2007spatial}.
Two of the most important issues with these methodologies are how to handle large datasets and how to ensure the nonnegative definiteness constraint for the covariance function. 
\citet{gneitingmatern} gave some theorems in order to construct valid covariance functions for multivariate random fields. In their approach every component in the matrix-valued covariance function
was a Mat\'ern covariance function. \citet{hu2012multivariate} proposed to use the systems of SPDEs to construct multivariate GRFs with isotropic and non-oscillating covariance functions.
The summary paper by \citet{sun2012geostatistics} discussed the approaches for how to handle large datasets.
They discussed several approaches such as separable covariance structures \citep{genton2007separable, fuentes2006testing}, covariance tapering \citep{furrer2006covariance, zhang2008covariance}, 
likelihood approximations \citep{vecchia1988estimation, stein2004approximating}, fixed rank kriging and fixed rank filtering \citep{cressie2008fixed,cressie2010fixed} 
and Gaussian Markov random fields approximation \citep{rue2002fitting, rue2004approximating, rue2005gaussian, rue2009approximate, lindgren2011explicit}. 

This paper is an extension of \citet{lindgren2011explicit} and \citet{hu2012multivariate}. The main contribution of this paper is the proposed approach for 
constructing multivariate random fields with oscillating covariance functions
using systems of SPDEs. The main idea is to use noise processes with oscillating covariance functions in order to introduce oscillation in the covariance functions of the random fields. 
We recommend to use the triangular systems of SPDEs since these models have some advantages. For instance, we have fewer hyper-parameters and
we can locate which random fields have non-oscillating, oscillating, possibly oscillating covariance functions. This approach can construct many models discussed by 
\citet{hu2012multivariate} if we set the oscillation parameter $\omega = 0$. It also inherit most of the advantages of the SPDE approach discussed by \citet{lindgren2011explicit} and systems of SPDEs approach discussed by
\citet{hu2012multivariate}. 

The two main challenges in multivariate random fields mentioned above can be partially solved with our model.
On the theoretical side, the covariance functions of the multivariate random fields fulfill the nonnegative definite constraint automatically. On the computational side, the GMRF representation makes the precision matrices to be sparse. 
Thus numerical algorithms for sparse matrices can be used for fast sampling and inference.
Four simulated datasets and one real dataset have been used to illustrate how to use our approach in different situations. The results have illustrated the effectiveness of the proposed approach. 

There are several possible extensions for further research, such as constructing non-stationary multivariate GRFs from the systems of SPDEs, and spatio-temporal models both in $\mathbb{R}^d$ and on manifolds.
It should also be possible to use the integrated nested Laplace approximation (INLA) framework \citep{rue2009approximate} for doing inference for the multivariate GRFs. More applied work using the proposed approach
is under development. 
 
\appendix
\section*{Appendix A} \label{sec: osci_appendix_a}
There are different kinds of parametrization for the system of SPDEs. The main idea is to change the operators in \eqref{eq: osci_bivariate_SPDE_operator_full} and use a different 
parametrization as discussed in Section \ref{sec: osci_bivariateGRFs}. Two intuitive operator matrices are
\begin{equation} \label{eq: osci_bivariate_SPDE2_operator2}
  \mathscr{L}_2(\boldsymbol{\theta}) = \begin{pmatrix}
 b_{11} \left( h_{11} -\Delta \right) &  0 \\
 b_{21} \left( h_{21} -\Delta \right) & b_{22} \left( h_{22} -\Delta \right)
\end{pmatrix},
\end{equation}

\begin{equation}  \label{eq: osci_bivariate_SPDE3_operator3}
  \mathscr{L}_3(\boldsymbol{\theta}) = \begin{pmatrix}
 b_{11} \left( h_{11} -\Delta \right) &  0 \\
 b_{21}  & b_{22} 
\end{pmatrix}.
\end{equation}
With \eqref{eq: osci_bivariate_SPDE2_operator2} and \eqref{eq: osci_bivariate_SPDE3_operator3} the correlations between the fields will be changed. The operator matrix given in \eqref{eq: osci_bivariate_SPDE2_operator2}
introduces more flexibility since we have one more parameter $h_{21}$ to control the range of cross-correlation.  
However, it might be hard to estimate all the parameters in this case. With operator matrix given in \eqref{eq: osci_bivariate_SPDE3_operator3}, we have fewer parameters in the model,
but the correlation structure between the fields is simplified. The second random field has the same correlation range as the first field.
These two systems can be use in different applications.

\section*{Appendix B}  \label{sec: osci_appendix_b}
When we use the triangular systems of SPDEs, we need to set the constraint $\kappa_{n_1}^2 = h_{11}$ since when the first noise process is generated from Equation \eqref{eq: osci_spde_simple}, they are not identifiable together.
Write the system of equations for the bivariate random field given in \eqref{eq: osci_bivariate_SPDE1_explicit} explicitly as
\begin{equation*} 
 \begin{split}
 b_{11} (h_{11} - \Delta) x_1(\boldsymbol{s}) & = \varepsilon_1(\boldsymbol{s}),   \\
 b_{21} x_1(\boldsymbol{s}) + b_{22} (h_{22} - \Delta) x_2(\boldsymbol{s}) & = \varepsilon_2(\boldsymbol{s}).
 \end{split}
\end{equation*}
Assume the first noise process $\varepsilon_1(\boldsymbol{s})$ is generated by
\begin{equation} \label{eq: osci_univariate_noise} 
 (\kappa_{n_i}^2 - \Delta) \varepsilon_i(\boldsymbol{s}) = \mathcal{W}(\boldsymbol{s}).
\end{equation}
We can now rewrite the first equation in system \eqref{eq: osci_bivariate_SPDE1_explicit} as a system of equations 
\begin{equation} \label{eq: osci_bivariate_indentifiable_kappa_n1}
 \begin{split}
 b_{11} (h_{11} - \Delta) x_1(\boldsymbol{s}) & = \varepsilon_1(\boldsymbol{s}),   \\
(\kappa_{n_1}^2 - \Delta) \varepsilon_1(\boldsymbol{s}) & = \mathcal{W}(\boldsymbol{s}).
 \end{split}
\end{equation}
This system of equations can be rewritten into one equation with white noise as the driving process,
\begin{equation} \label{eq: osci_univariate_x1} 
 (h_{11} - \Delta)(\kappa_{n_1}^2 - \Delta) x_1(\boldsymbol{s}) = \mathcal{W}(\boldsymbol{s}).
\end{equation}
It is obvious that $\kappa_{n_1}^2$ and $h_{11}$ are not identifiable from each other since $(h_{11} - \Delta)$ and $(\kappa_{n_1}^2 - \Delta)$ commute. Therefore, we suggest the constraint $\kappa_{n_1}^2 = h_{11}$.
However, if the first noise process is oscillating and is generated from Equation \eqref{eq: osci_univariate_complex}, we don't need this constraint because they are identifiable.
However, we still recommend to use this setting in order to simplify the inference.
                                                                     
\bibliographystyle{plainnat}
\linespread{0.8}{\small {\bibliography {../Reference/Ref}}}

\begin{thebibliography}{52}
\providecommand{\natexlab}[1]{#1}
\providecommand{\url}[1]{\texttt{#1}}
\expandafter\ifx\csname urlstyle\endcsname\relax
  \providecommand{\doi}[1]{doi: #1}\else
  \providecommand{\doi}{doi: \begingroup \urlstyle{rm}\Url}\fi

\bibitem[Banerjee et~al.(2004)Banerjee, Carlin, and
  Gelfand]{banerjee2004hierarchical}
S.~Banerjee, B.P. Carlin, and A.E. Gelfand.
\newblock \emph{{Hierarchical modeling and analysis for spatial data}}.
\newblock Chapman \& Hall, 2004.
\newblock ISBN 158488410X.

\bibitem[Bolin and Lindgren(2009)]{bolin2009wavelet}
D.~Bolin and F.~Lindgren.
\newblock Wavelet markov models as efficient alternatives to tapering and
  convolution fields.
\newblock Technical report, Mathematical Statistics, Centre for Mathematical
  Sciences, Faculty of Engineering, Lund University, 2009.

\bibitem[Bolin and Lindgren(2011)]{bolin2011spatial}
D.~Bolin and F.~Lindgren.
\newblock Spatial models generated by nested stochastic partial differential
  equations, with an application to global ozone mapping.
\newblock \emph{The Annals of Applied Statistics}, 5\penalty0 (1):\penalty0
  523--550, 2011.

\bibitem[Brenner and Scott(2008)]{brenner2008mathematical}
S.C. Brenner and L.R. Scott.
\newblock \emph{The mathematical theory of finite element methods}, volume~15.
\newblock Springer Verlag, 2008.

\bibitem[Brown et~al.(1994)Brown, Le, and Zidek]{brown1994multivariate}
P.J. Brown, N.D. Le, and J.V. Zidek.
\newblock Multivariate spatial interpolation and exposure to air pollutants.
\newblock \emph{Canadian Journal of Statistics}, 22\penalty0 (4):\penalty0
  489--509, 1994.

\bibitem[Courtier et~al.(1998)Courtier, Andersson, Heckley, Pailleux,
  Vasiljevic, Hamrud, Hollingsworth, Rabier, and Fisher]{courtier1998ecmwf}
P~Courtier, E~Andersson, W~Heckley, J~Pailleux, D~Vasiljevic, M~Hamrud,
  A~Hollingsworth, F~Rabier, and M~Fisher.
\newblock The ecmwf implementation of three-dimensional variational
  assimilation (3d-var). i: Formulation.
\newblock \emph{Quarterly Journal of the Royal Meteorological Society},
  124:\penalty0 1783--1807, 1998.

\bibitem[Cressie and Johannesson(2008)]{cressie2008fixed}
N.~Cressie and G.~Johannesson.
\newblock Fixed rank kriging for very large spatial data sets.
\newblock \emph{Journal of the Royal Statistical Society: Series B (Statistical
  Methodology)}, 70\penalty0 (1):\penalty0 209--226, 2008.

\bibitem[Cressie et~al.(2010)Cressie, Shi, and Kang]{cressie2010fixed}
N.~Cressie, T.~Shi, and E.L. Kang.
\newblock Fixed rank filtering for spatio-temporal data.
\newblock \emph{Journal of Computational and Graphical Statistics}, 19\penalty0
  (3):\penalty0 724--745, 2010.

\bibitem[Cressie(1993)]{cressie1993statistics}
N.A.C. Cressie.
\newblock \emph{Statistics for spatial data}, volume 298.
\newblock Wiley-Interscience, 1993.

\bibitem[Diggle and Ribeiro~Jr(2006)]{diggle2006model}
P.J. Diggle and P.J. Ribeiro~Jr.
\newblock \emph{Model-based Geostatistics}.
\newblock Springer, 2006.

\bibitem[Diggle et~al.(1998)Diggle, Tawn, and Moyeed]{diggle1998model}
P.J. Diggle, JA~Tawn, and RA~Moyeed.
\newblock {Model-based geostatistics}.
\newblock \emph{Journal of the Royal Statistical Society: Series C (Applied
  Statistics)}, 47\penalty0 (3):\penalty0 299--350, 1998.
\newblock ISSN 1467-9876.

\bibitem[Fuentes(2006)]{fuentes2006testing}
M.~Fuentes.
\newblock Testing for separability of spatial--temporal covariance functions.
\newblock \emph{Journal of Statistical Planning and Inference}, 136\penalty0
  (2):\penalty0 447--466, 2006.

\bibitem[Fuglstad(2010)]{fuglstad2010approximating}
G.A. Fuglstad.
\newblock Approximating solutions of stochastic differential equations with
  gussian markov random fields.
\newblock Technical report, Department of Mathematical Science, Norwegian
  University of Science and Technology, 2010.

\bibitem[Fuglstad(2011)]{fuglstad2011spatial}
G.A. Fuglstad.
\newblock Spatial modelling and inference with spde-based gmrfs.
\newblock Master's thesis, Department of Mathematical Sciences, Norwegian
  University of Science and Technology, 2011.

\bibitem[Furrer et~al.(2006)Furrer, Genton, and Nychka]{furrer2006covariance}
R.~Furrer, M.G. Genton, and D.~Nychka.
\newblock Covariance tapering for interpolation of large spatial datasets.
\newblock \emph{Journal of Computational and Graphical Statistics}, 15\penalty0
  (3):\penalty0 502--523, 2006.

\bibitem[Gelfand et~al.(2004)Gelfand, Schmidt, Banerjee, and
  Sirmans]{gelfand2004nonstationary}
A.E. Gelfand, A.M. Schmidt, S.~Banerjee, and CF~Sirmans.
\newblock Nonstationary multivariate process modeling through spatially varying
  coregionalization.
\newblock \emph{Test}, 13\penalty0 (2):\penalty0 263--312, 2004.

\bibitem[Gelfand et~al.(2010)Gelfand, Diggle, Fuentes, and
  Guttorp]{gelfand2010handbook}
A.E. Gelfand, P.J. Diggle, M.~Fuentes, and P.~Guttorp.
\newblock \emph{Handbook of spatial statistics}.
\newblock CRC Press, 2010.

\bibitem[Genton(2007)]{genton2007separable}
M.G. Genton.
\newblock Separable approximations of space-time covariance matrices.
\newblock \emph{Environmetrics}, 18\penalty0 (7):\penalty0 681--695, 2007.

\bibitem[Gneiting(1998)]{gneiting1998simple}
T.~Gneiting.
\newblock Simple tests for the validity of correlation function models on the
  circle.
\newblock \emph{Statistics \& probability letters}, 39\penalty0 (2):\penalty0
  119--122, 1998.

\bibitem[Gneiting et~al.(2010)Gneiting, Kleiber, and Schlather]{gneitingmatern}
T.~Gneiting, W.~Kleiber, and M.~Schlather.
\newblock {Mat{\'e}rn Cross-Covariance Functions for Multivariate Random
  Fields}.
\newblock \emph{Journal of the American Statistical Association}, 105\penalty0
  (491):\penalty0 1167--1177, 2010.
\newblock ISSN 0162-1459.

\bibitem[Goff and Jordan(1988)]{goff1988stochastic}
J.A. Goff and T.H. Jordan.
\newblock {Stochastic modeling of seafloor morphology: Inversion of sea beam
  data for second-order statistics}.
\newblock \emph{Journal of Geophysical Research}, 93\penalty0 (B11):\penalty0
  13589--13, 1988.
\newblock ISSN 0148-0227.

\bibitem[Handcock and Stein(1993)]{handcock1993bayesian}
M.S. Handcock and M.L. Stein.
\newblock {A Bayesian analysis of kriging}.
\newblock \emph{Technometrics}, 35\penalty0 (4):\penalty0 403--410, 1993.
\newblock ISSN 0040-1706.

\bibitem[Hartman and H{\"o}ssjer(2008)]{hartman2008fast}
L.~Hartman and O.~H{\"o}ssjer.
\newblock {Fast kriging of large data sets with Gaussian Markov random fields}.
\newblock \emph{Computational Statistics \& Data Analysis}, 52\penalty0
  (5):\penalty0 2331--2349, 2008.
\newblock ISSN 0167-9473.

\bibitem[Hjelle and D{\ae}hlen(2006)]{hjelle2006triangulations}
{\O}.~Hjelle and M.~D{\ae}hlen.
\newblock \emph{Triangulations and applications}.
\newblock Springer Verlag, 2006.

\bibitem[Hu et~al.(2012{\natexlab{a}})Hu, Simpson, Lindgren, and
  Rue]{hu2012multivariate}
X.~Hu, D.P. Simpson, F.~Lindgren, and H.~Rue.
\newblock Multivariate gaussian random fields using systems of stochastic
  partial differential equations.
\newblock \emph{statistical report, Norwegian University of Science and
  Technology}, 2012{\natexlab{a}}.

\bibitem[Hu et~al.(2012{\natexlab{b}})Hu, Simpson, and Rue]{hu2012specifying}
X.~Hu, D.P. Simpson, and H.~Rue.
\newblock Specifying gaussian markov random fields with incomplete orthogonal
  factorization using givens rotations.
\newblock Technical report, Department of Mathematical Science, norwegian
  University of Science and Technology, 2012{\natexlab{b}}.

\bibitem[Jones(1963)]{jones1963stochastic}
R.H. Jones.
\newblock Stochastic processes on a sphere.
\newblock \emph{The Annals of mathematical statistics}, 34\penalty0
  (1):\penalty0 213--218, 1963.

\bibitem[Jun and Stein(2007)]{jun2007approach}
M.~Jun and M.L. Stein.
\newblock An approach to producing space--time covariance functions on spheres.
\newblock \emph{Technometrics}, 49\penalty0 (4):\penalty0 468--479, 2007.

\bibitem[Kaufman et~al.(2008)Kaufman, Schervish, and
  Nychka]{kaufman2008covariance}
C.G. Kaufman, M.J. Schervish, and D.W. Nychka.
\newblock Covariance tapering for likelihood-based estimation in large spatial
  data sets.
\newblock \emph{Journal of the American Statistical Association}, 103\penalty0
  (484):\penalty0 1545--1555, 2008.

\bibitem[Kloeden and Platen(1999)]{kloeden1992numerical}
P.E. Kloeden and E.~Platen.
\newblock \emph{{Numerical solution of stochastic differential equations}}.
\newblock Springer, 3rd edition, 1999.
\newblock ISBN 3540540628.

\bibitem[Lindgren et~al.(2011)Lindgren, Rue, and
  Lindstr{\"o}m]{lindgren2011explicit}
F.~Lindgren, H.~Rue, and J.~Lindstr{\"o}m.
\newblock An explicit link between gaussian fields and gaussian markov random
  fields: the stochastic partial differential equation approach.
\newblock \emph{Journal of the Royal Statistical Society: Series B (Statistical
  Methodology)}, 73\penalty0 (4):\penalty0 423--498, 2011.

\bibitem[Lindgren(2010)]{georg2010secondstationary}
G.~Lindgren.
\newblock \emph{A second course on stationary stochastic processes}.
\newblock Center for Mathematical Sciences, Lund University, December 2010.

\bibitem[Mat{\'e}rn(1986)]{matern1986spatial}
B.~Mat{\'e}rn.
\newblock \emph{{Spatial variation}}.
\newblock Springer-Verlag Berlin, 1986.

\bibitem[Reich and Fuentes(2007)]{reich2007multivariate}
B.J. Reich and M.~Fuentes.
\newblock A multivariate semiparametric bayesian spatial modeling framework for
  hurricane surface wind fields.
\newblock \emph{The Annals of Applied Statistics}, 1\penalty0 (1):\penalty0
  249--264, 2007.

\bibitem[Robert(2007)]{robert2007bayesian}
C.~Robert.
\newblock \emph{The Bayesian choice: from decision-theoretic foundations to
  computational implementation}.
\newblock Springer Verlag, 2007.

\bibitem[Rue(2001)]{rue2001fast}
H.~Rue.
\newblock {Fast sampling of Gaussian Markov random fields}.
\newblock \emph{Journal of the Royal Statistical Society: Series B (Statistical
  Methodology)}, 63\penalty0 (2):\penalty0 325--338, 2001.
\newblock ISSN 1467-9868.

\bibitem[Rue and Held(2005)]{rue2005gaussian}
H.~Rue and L.~Held.
\newblock \emph{{Gaussian Markov random fields: theory and applications}}.
\newblock Chapman \& Hall, 2005.
\newblock ISBN 1584884320.

\bibitem[Rue and Tjelmeland(2002)]{rue2002fitting}
H.~Rue and H.~Tjelmeland.
\newblock {Fitting Gaussian Markov random fields to Gaussian fields}.
\newblock \emph{Scandinavian Journal of Statistics}, 29\penalty0 (1):\penalty0
  31--49, 2002.
\newblock ISSN 1467-9469.

\bibitem[Rue et~al.(2004)Rue, Steinsland, and Erland]{rue2004approximating}
H.~Rue, I.~Steinsland, and S.~Erland.
\newblock {Approximating hidden Gaussian Markov random fields}.
\newblock \emph{Journal of the Royal Statistical Society: Series B (Statistical
  Methodology)}, 66\penalty0 (4):\penalty0 877--892, 2004.
\newblock ISSN 1467-9868.

\bibitem[Rue et~al.(2009)Rue, Martino, and Chopin]{rue2009approximate}
H.~Rue, S.~Martino, and N.~Chopin.
\newblock {Approximate Bayesian inference for latent Gaussian models by using
  integrated nested Laplace approximations}.
\newblock \emph{Journal of the Royal Statistical Society: Series B (Statistical
  Methodology)}, 71\penalty0 (2):\penalty0 319--392, 2009.
\newblock ISSN 1467-9868.

\bibitem[Sain and Cressie(2007)]{sain2007spatial}
S.R. Sain and N.~Cressie.
\newblock A spatial model for multivariate lattice data.
\newblock \emph{Journal of Econometrics}, 140\penalty0 (1):\penalty0 226--259,
  2007.

\bibitem[Schmidt and Gelfand(2003)]{schmidt2003bayesian}
A.M. Schmidt and A.E. Gelfand.
\newblock A bayesian coregionalization approach for multivariate pollutant
  data.
\newblock \emph{Journal of Geophysical Research}, 108\penalty0 (D24):\penalty0
  8783, 2003.

\bibitem[Shaby and Ruppert(2012)]{shaby2012tapered}
B.~Shaby and D.~Ruppert.
\newblock Tapered covariance: Bayesian estimation and asymptotics.
\newblock \emph{Journal of Computational and Graphical Statistics}, 21\penalty0
  (2):\penalty0 433--452, 2012.

\bibitem[Stein(1999)]{stein1999interpolation}
M.L. Stein.
\newblock \emph{{Interpolation of Spatial Data: some theory for kriging}}.
\newblock Springer Verlag, 1999.
\newblock ISBN 0387986294.

\bibitem[Stein et~al.(2004)Stein, Chi, and Welty]{stein2004approximating}
M.L. Stein, Z.~Chi, and L.J. Welty.
\newblock Approximating likelihoods for large spatial data sets.
\newblock \emph{Journal of the Royal Statistical Society: Series B (Statistical
  Methodology)}, 66\penalty0 (2):\penalty0 275--296, 2004.

\bibitem[Sun et~al.(2012)Sun, Li, and Genton]{sun2012geostatistics}
Y.~Sun, B.~Li, and M.G. Genton.
\newblock Geostatistics for large datasets.
\newblock \emph{Advances and challenges in space-time modelling of natural
  events}, pages 55--77, 2012.

\bibitem[Vecchia(1988)]{vecchia1988estimation}
A.V. Vecchia.
\newblock Estimation and model identification for continuous spatial processes.
\newblock \emph{Journal of the Royal Statistical Society. Series B
  (Methodological)}, pages 297--312, 1988.

\bibitem[Wei(2006)]{wei2006time}
W.W.S. Wei.
\newblock \emph{Time series analysis: univariate and multivariate methods}.
\newblock Addison-Wesley, 2006.

\bibitem[Whittle(1954)]{whittle1954stationary}
P.~Whittle.
\newblock {On stationary processes in the plane}.
\newblock \emph{Biometrika}, 41\penalty0 (3-4):\penalty0 434--449, 1954.
\newblock ISSN 0006-3444.

\bibitem[Whittle(1963)]{whittle1963stochastic}
P.~Whittle.
\newblock Stochastic processes in several dimensions.
\newblock \emph{Bull. Int. Statist. Inst.}, 40:\penalty0 974--994, 1963.

\bibitem[Zhang and Du(2008)]{zhang2008covariance}
H.~Zhang and J.~Du.
\newblock Covariance tapering in spatial statistics.
\newblock \emph{Positive definite functions: From Schoenberg to space-time
  challenges}, pages 181--196, 2008.

\bibitem[Zienkiewicz et~al.(2005)Zienkiewicz, Taylor, Taylor, and
  Zhu]{zienkiewicz2005finite}
O.C. Zienkiewicz, R.L. Taylor, R.L. Taylor, and JZ~Zhu.
\newblock \emph{The finite element method: its basis and fundamentals},
  volume~1.
\newblock Butterworth-heinemann, 2005.

\end{thebibliography}

% \newpage
% \thispagestyle{empty}
% \mbox{}
\end{document}